\documentclass{aa}   %% Astronomy & Astrophysics style class v8.2

\usepackage{graphicx,natbib,twoopt}
\usepackage[varg]{txfonts}           %% A&A font choice
\usepackage{hyperref}
\usepackage{acronym} 

\hypersetup{
  colorlinks=true,   %% links colored instead of frames
  urlcolor=blue,     %% external hyperlinks
  linkcolor=red,     %% internal latex links (eg Fig)
}

\bibpunct{(}{)}{;}{a}{}{,}    %% natbib cite format used by A&A and ApJ

\makeatletter
\newcommand{\bibnote}[2]{\global\@namedef{#1note}{#2}}
\newcommand{\biblink}[2]{\global\@namedef{#1link}{#2}}

\makeatother

\makeatletter
 \newcommandtwoopt{\citeads}[3][][]{%
   \nonstopmode%              %% fix to not stop at error message in latex
   \href{http://adsabs.harvard.edu/abs/#3}%
        {\def\hyper@linkstart##1##2{}%
         \let\hyper@linkend\@empty\citealp[#1][#2]{#3}}%   %% Rutten, 2000
   \biblink{#3}{\href{http://adsabs.harvard.edu/abs/#3}{ADS}}%
   \errorstopmode}            %% fix to resume stopping at error messages 
 \newcommandtwoopt{\citepads}[3][][]{%
   \nonstopmode%              %% fix to not stop at error message in latex
   \href{http://adsabs.harvard.edu/abs/#3}%
        {\def\hyper@linkstart##1##2{}%
         \let\hyper@linkend\@empty\citep[#1][#2]{#3}}%     %% (Rutten 2000)
   \biblink{#3}{\href{http://adsabs.harvard.edu/abs/#3}{ADS}}%
   \errorstopmode}            %% fix to resume stopping at error messages
 \newcommandtwoopt{\citetads}[3][][]{%
   \nonstopmode%              %% fix to not stop at error message in latex
   \href{http://adsabs.harvard.edu/abs/#3}%
        {\def\hyper@linkstart##1##2{}%
         \let\hyper@linkend\@empty\citet[#1][#2]{#3}}%     %% Rutten (2000)
   \biblink{#3}{\href{http://adsabs.harvard.edu/abs/#3}{ADS}}%
   \errorstopmode}            %% fix to resume stopping at error messages 
 \newcommandtwoopt{\citeyearads}[3][][]{%
   \nonstopmode%              %% fix to not stop at error message in latex
   \href{http://adsabs.harvard.edu/abs/#3}%
        {\def\hyper@linkstart##1##2{}%
         \let\hyper@linkend\@empty\citeyear[#1][#2]{#3}}%  %% 2000
   \biblink{#3}{\href{http://adsabs.harvard.edu/abs/#3}{ADS}}%
   \errorstopmode}            %% fix to resume stopping at error messages 
\makeatother

\newacro{ADS}{Astrophysics Data System}
\newacro{NLTE}{non-local thermodynamic equilibrium}
\newacro{NASA}{National Aeronautics and Space Administration}

\begin{document}

%% simple header.  Change into A&A or ApJ commands for those journals

%\twocolumn[{%
%\vspace*{4ex}
%\begin{center}
 % {\Large \bf Anisotropy of the galaxy cluster X-ray luminosity-temperature relation}\\[4ex]       
  %{\large \bf Konstantinos Migkas$^{1}$ and 
   %           Thomas H. Reiprich$^1$}\\[4ex]
              
  %\begin{minipage}[t]{15cm}
   %     $^1$ Argelander-Institut f{\"u}r Astronomie, Universit{\"a}t Bonn, Auf dem H{\"u}gel 71, 53121 Bonn, Germany\\ \email{kmigkas@astro.uni-bonn.de}\\

  %\vspace*{2ex}
  %\end{minipage}
%\end{center}
%}] 
\title{The SRG/eROSITA All-Sky Survey}
\subtitle{SRG/eROSITA cross-calibration with \textit{Chandra} and \textit{\textit{XMM-Newton}} using galaxy cluster gas temperatures}

\author{K. Migkas$^{1,2,3}$, D. Kox$^2$, G. Schellenberger$^4$, A. Veronica$^2$, F. Pacaud$^2$, T. H. Reiprich$^2$, Y. E. Bahar$^5$, F. Balzer$^5$, E. Bulbul$^5$, J. Comparat$^5$, K. Dennerl$^5$, M. Freyberg$^5$, C. Garrel$^5$, V. Ghirardini$^5$, S. Grandis$^6$, M. Kluge$^5$, A. Liu$^5$, M. E. Ramos-Ceja$^5$, J. Sanders$^5$, X. Zhang$^5$}

\institute{$^1$ Leiden Observatory, Leiden University, PO Box 9513, 2300 RA Leiden, the Netherlands  \\ 
$^2$Argelander-Institut f{\"u}r Astronomie, Universit{\"a}t Bonn, Auf dem H{\"u}gel 71, 53121 Bonn, Germany\\
$^3$SRON Netherlands Institute for Space Research, Niels Bohrweg 4, NL-2333 CA Leiden, the Netherlands\\
$^4$Center for Astrophysics | Harvard \& Smithsonian, 60 Garden Street, Cambridge, MA 02138, USA \\
$^5$Max Planck Institute for Extraterrestrial Physics, Gie\ss enbachstra\ss e 1, 85748 Garching bei M{\"u}nchen, Germany\\
$^6$ Universit{\"a}t Innsbruck, Institut f{\"u}r Astro- und Teilchenphysik, Technikerstr. 25/8, 6020 Innsbruck, Austria\\
\email{kmigkas@strw.leidenuniv.nl}
}

\date{Received date} 
\abstract{Galaxy cluster gas temperatures ($T$) play a crucial role in many cosmological and astrophysical studies. However, it has been shown that $T$ measurements can significantly vary between different X-ray telescopes. These $T$ biases can propagate to several cluster applications in which $T$ can be used, such as measuring hydrostatic cluster masses and constraining the angular variation of cosmological parameters. Thus, it is important to accurately cross-calibrate X-ray instruments to account for systematic biases. In this work, we present the cross-calibration between Spectrum Roentgen Gamma/eROSITA (SRG/eROSITA) and \textit{Chandra}/ACIS and between SRG/eROSITA and \textit{XMM-Newton}/EPIC using for the first time a large sample of galaxy cluster $T$. To do so, we used the first eROSITA All-Sky Survey data and the preliminary extremely expanded HIgh FLUx Galaxy Cluster Sample, a large X-ray flux-limited cluster catalog. We spectroscopically measured X-ray $T$ for 186 independent cluster regions with both SRG/eROSITA and \textit{Chandra}/ACIS in a self-consistent way for three energy bands: 0.7-7 keV (full), 0.5-4 keV (soft), and 1.5-7 keV (hard). We did the same with SRG/eROSITA and \textit{XMM-Newton}/EPIC for 71 different cluster regions and all three bands. We find that SRG/eROSITA measures systematically lower $T$ than the other two instruments, with hotter clusters deviating more than cooler ones. For the full band, SRG/eROSITA returns 20$\%$ and 14$\%$ lower $T$ than \textit{Chandra}/ACIS and \textit{XMM-Newton}/EPIC, respectively, when the two other instruments each measure $k_{\text{B}}T\approx 3$ keV. The discrepancy respectively increases to 38\% and 32\% when \textit{Chandra}/ACIS and \textit{XMM-Newton}/EPIC each measure $k_{\text{B}}T\approx 10$ keV. On the other hand, the discrepancy becomes milder for low-$T$ galaxy groups. Moreover, a broken power law fit demonstrated that there is a break at the SRG/eROSITA-\textit{Chandra}/ACIS scaling relation at $k_{\text{B}}T\approx 1.7-2.7$ keV, depending on the energy band. The soft band shows a marginally lower discrepancy compared to the full band. In the hard band, the cross-calibration of SRG/eROSITA and the other instruments show very strong differences. We tested several possible systematic biases (such as multiphase cluster gas, Galactic absorption, non-Gaussian scatter, and selection effects) to identify the reason behind the cross-calibration discrepancies, but none could significantly alleviate the tension. For now, it is most likely that the systematically lower SRG/eROSITA $T$ can be attributed to systematic effective area calibration uncertainties; however, the exact role of multiphase cluster gas in the observed $T$ discrepancies needs to be further investigated. Furthermore, we provide conversion factors between SRG/eROSITA, \textit{Chandra}/ACIS, and \textit{XMM-Newton}/EPIC $T$ that will be beneficial for future cluster studies that combine SRG/eROSITA $T$ with data from other X-ray instruments. Finally, we also provide conversion functions between the official eRASS1 cluster catalog $T$ and the equivalent core and core-excised \textit{Chandra}/ACIS and \textit{XMM-Newton}/EPIC $T$.}

%Galaxy cluster gas temperatures are used for a broad range of cosmological and astrophysical applications. Therefore, an 


\titlerunning{eROSITA cross-calibration with \textit{\textit{Chandra}} and \textit{XMM-Newton}}
\authorrunning{K. Migkas et al. }

\maketitle

\section{Introduction} \label{intro}

Galaxy clusters, the largest virialized systems in the Universe, play a critical role in our understanding of various astrophysical and cosmological phenomena. X-ray observations of their most massive baryonic component, the hot intracluster medium, offer a wealth of information on a wide range of cluster properties. Among these, the intracluster medium temperature ($T$) is particularly valuable, as it is a key factor in many cosmological applications of galaxy clusters. 

Widespread cosmological use of galaxy clusters comes through the halo mass function, which obtains tight cosmological constraints by modeling the number of cluster halos per mass and redshift. For such a test, an unbiased modeling of cluster masses is of immense significance. Accurate measurements of $T$ profiles are used to determine the total hydrostatic mass of clusters, which can then be utilized to constrain cosmology \citep[e.g.,][]{Gerrit2017}.\footnote{We note here that $T$ is not used for the primary cosmological constraints of the eROSITA survey \citep{ghirardini}.} Moreover, since $T$ is a low-scattered proxy of the total cluster mass, the latter can also be directly estimated through the mass-temperature scaling relation $M-T$ \citep[e.g.,][]{lovisari,lovisari20,bulbul}. Another vital need for high-quality $T$ measurements arises from the studies of cosmic isotropy with cluster scaling relations \citep{migkas21}. Measuring $T$ does not require any strong cosmological assumptions. As such, the directionality of its correlation with other cosmology-dependent cluster properties constitutes a powerful test of the isotropy of cosmic expansion and the existence of bulk flows in the local Universe. Furthermore, X-ray gas $T$ measurements are an essential component in a substantial number of astrophysical studies, including assessments of the impact of baryonic feedback on cluster scaling relations \citep[e.g.,][]{mittal} as well as the detection of the warm-hot intergalactic medium \citep[e.g.,][]{werner,eckert_fil}, and many others. It is therefore evident that an unbiased determination of $T$ is of utter importance for cluster science. 

However, it is well established that discrepancies exist between cluster $T$ measurements obtained by various X-ray telescopes. \citet[][S15 hereafter]{schellenberger} showed that \textit{XMM-Newton}/EPIC (hereafter, \textit{XMM-Newton}) returned systematically lower $T$ values than \textit{Chandra}/ACIS (hereafter, \textit{Chandra}) using a sample of 64 galaxy clusters observed by both telescopes. This discrepancy was found to increase as a function of $T$ from consistency for $T\approx 1$ keV clusters\footnote{Here, we ignore the Boltzmann constant $k_{\text{B}}$ multiplied by the temperature to obtain energy units. Thus, we adopted the notation $k_{\text{B}}T\equiv T$.} to a $\approx 29\%$ higher \textit{Chandra} $T$ ($T_{\text{Chandra}}$ hereafter) for $T\approx10$ keV clusters. Furthermore, the temperature difference was stronger in the soft X-ray than in the hard X-ray band. Other studies that used only eight to 11 clusters found similar, often milder, cluster temperature discrepancies between \textit{XMM-Newton}, \textit{Suzaku}, \textit{Chandra}, and \textit{NuSTAR} \citep{nevalainen,kettula,wallbank}. 

If $T$ disagreements between different X-ray telescopes are not properly taken into account, they can lead to several systematic biases. For instance, S15 showed that using $T$ values only from \textit{XMM-Newton} or \textit{Chandra} without accounting for their cross-calibration issue results in an 8\% difference in the estimated $\Omega_{\text{m}}$ when using hydrostatic masses and the cluster mass function. Moreover, using the \citet{lovisari20} $M-T$ relation, a 10\% bias in $T$ shifts the total cluster masses by 16\%. Finally, when testing cosmic isotropy with cluster scaling relations, a 10\% systematic difference in $T$ causes a 12\% shift in $H_0$, inducing artificial anisotropies \citep{migkas21}. 

%From all the above it becomes evident that a proper cross-calibration of $T$ measurements across different X-ray telescopes is a necessity for cluster cosmology.

Cluster science will significantly progress in the coming years thanks to the X-ray cluster catalogs provided by the extended ROentgen Survey with an Imaging Telescope Array \citep[eROSITA;][]{merloni,predehl21}. On board the Spectrum-Roentgen-Gamma (SRG) space observatory \citep{sunyaevSRG} launched in July 2019, eROSITA consists of seven telescope modules (TMs), each with 54 nested mirror shells. SRG/eROSITA (hereafter, eROSITA) will conduct a four-year survey, mapping the entire X-ray sky for the first time in the 21st century. Thanks to its large effective area and broad X-ray energy range coverage, eROSITA is expected to detect all massive galaxy clusters ($M\geq 3\times 10^{14}\ M_{\odot}$) in the observable Universe away from the Galactic plane \citep{merloni,pillepich}. For many thousands of these clusters, precise spectroscopic $T$ measurements will become possible \citep{borm}, significantly increasing the cluster sample size with available $T$ values. Thus, a cross-calibration of eROSITA-measured cluster $T$ with those from \textit{XMM-Newton} and \textit{Chandra} is of utter importance for studies that wish to jointly analyze data coming from these X-ray telescopes. 

Recently, \citet{turner} used eight clusters from the eROSITA Final Equatorial Depth Survey (eFEDS) cluster catalog \citep{Liu} and found that eROSITA returns $25\pm 9 \%$ lower $T$ than \textit{XMM-Newton} in the broad X-ray band, although different energy bands were used for the two instruments. Due to the limited data, the dependence of the discrepancy on cluster temperature could not be constrained. That work provided a first indication of the cross-calibration between these two telescopes but lacked the statistical power for a more precise comparison and a deeper investigation of the discrepancy. Other studies that focused on the eROSITA analysis of single clusters also found temperature discrepancies between eROSITA, \textit{XMM-Newton}, and \textit{Chandra}. \citet{sanders22} and \citet{whelan22} found that eROSITA showed higher $T$ than \textit{XMM-Newton} and only slightly lower $T$ than \textit{Chandra} for their respective clusters. On the other hand, \citet{veronica22} found mildly lower $T$ with eROSITA than \textit{XMM-Newton} both in the broad and soft bands for another single cluster. \citet{liu23} found that, in the full band, eROSITA measures $\sim 60\%$ and $\sim 45\%$ lower $T$ than \textit{Chandra} and \textit{XMM-Newton}, respectively. The discrepancy was alleviated when they measured the eROSITA $T$ ($T_{\text{eROSITA}}$ hereafter) in the 2-8 keV band, which agreed with the $T_{\text{Chandra}}$ from the broad X-ray band. These results, which are based on single cluster comparisons, often contradict each other, and this indicates that such individual comparisons are probably not sufficient to characterize the cross-calibration of eROSITA with other X-ray instruments. Due to the non-negligible scatter of these relations, single cluster comparisons are subject to noise fluctuations; thus, one needs large cluster samples to robustly quantify any systematic $T$ discrepancy. Moreover, the above studies often used different metal abundances and Galactic absorption values for different instruments when comparing their respective $T$. To obtain a clearer picture of the eROSITA-\textit{Chandra}-\textit{XMM-Newton} cross-calibration, one needs to perform spectral fits with self-consistent "nuisance" parameters across different telescopes so an artificial systematic bias in the $T$ estimation can be avoided. 

In this work, we use the data products of the first eROSITA All-Sky survey (eRASS1) to accurately assess the cross-calibration between eROSITA, \textit{XMM-Newton}, and \textit{Chandra}. To do so, we measured the cluster gas $T$ using eRASS1 data for (nearly) all systems included in the preliminary extremely expanded HIgh FLux Galaxy Cluster Sample (eeHIFLUGCS) catalog, as presented in \citet[][M20 hereafter]{migkas20}, and lying in the western Galactic hemisphere. We measured the core and core-excised $T$ and metallicity $Z$ for 120 and 51 clusters with \textit{Chandra} and \textit{XMM-Newton} $T$ measurements, respectively, from the same cluster areas and for the 0.7-7 keV energy band. For the majority of these clusters, we also compared the spectral fits for the soft (0.5-4 keV) and hard (1.5-7 keV) X-ray bands. Our results provide clear conversion functions between cluster $T$ measurements coming from different X-ray telescopes, which are of great value for any joint cluster analysis of eROSITA, \textit{XMM-Newton}, and \textit{Chandra}.

The paper is organized as follows: In Sect. \ref{Sample}, we describe the cluster sample used in this work. In Sect. \ref{data_red_anal}, we describe the data reduction and spectral analysis of the eROSITA, \textit{XMM-Newton}, and \textit{Chandra} data. In Sect. \ref{statistics}, we describe the statistical methods followed in order to compare the results across different telescopes and quantify the significance of any differences. In Sect. \ref{mainresults}, we present the cross-calibration results between all telescopes and energy bands. In Sect. \ref{broken_pow}, we present the broken power law fits for the eROSITA-\textit{Chandra} scaling relations. In Sect. \ref{discussion}, we discuss the possible reasons behind the observed $T$ discrepancies and the impact of this work. Finally, in Sect. \ref{conclusions}, the conclusions of this work are given. 

\section{Sample}\label{Sample}

The galaxy cluster sample used for this work was presented and described in M20. In a nutshell, it is a subsample of the eeHIFLUGCS catalog (Pacaud, et al. in prep.), which is a nearly complete, X-ray flux-limited cluster sample. eeHIFLUGCS was selected based on a re-analysis of ROSAT data of the clusters included in the Meta-Catalogue of X-ray detected Clusters of galaxies \citep[MCXC;][]{mcxc}. After masking the Galactic plane ($|b|\leq 20^{^{\circ}}$) and the Magellanic clouds, the only selection criterion is an unabsorbed X-ray flux of $f_{\text{X}}\geq 5\times 10^{-12}$ ergs/s/cm$^2$ in the 0.1-2.4 keV band. Only clusters with sufficient \textit{XMM-Newton} or \textit{Chandra} data were kept in the M20 sample. That allowed for a precise $T$ measurement within the $0.2-0.5\ R_{500}$\footnote{$R_{500}$ refers to the cluster radius within which the average cluster density is 500$\times$ the critical density of the Universe.} cluster annulus in the 0.7-7 keV band. Moreover, multiple cluster systems and clusters with strong AGN contamination were discarded. This resulted in 313 clusters in M20. In \citet{migkas21} we measured more properties for this sample, including the core ($<0.2\ R_{500}$) $T$, which is also utilized here. The M20 $R_{500}$ values used in this work were taken from MCXC and they were calculated through the X-ray luminosity-total mass scaling relation of \citet{arnaud2} for a flat $\Lambda$CDM cosmology with $H_0=70$ km/s/Mpc and $\Omega_{\text{m}}=0.3$. For very few clusters, the MCXC $R_{500}$ were further corrected in M20 due to the additional cleaning of the MCXC X-ray luminosity (see M20 for details).

In this work, we use the western Galactic hemisphere eRASS1 data to which the German eROSITA Consortium holds proprietary rights. This results in 155 different clusters from M20, which is the sample used in this work. Out of these 155, 111 are analyzed with \textit{Chandra} and 53 with \textit{XMM-Newton} data (with nine clusters analyzed with both instruments). All cluster data used in this work are publicly available.\footnote{\url{https://github.com/kmigkas/eROSITA-cross-calibr-data}.}

\section{Data analysis and cluster measurements}\label{data_red_anal}

\subsection{eROSITA}

\subsubsection{Data reduction}
We used the first eROSITA-All Sky Survey data (eRASS:1) with processing version 010. The products of the eRASS are divided into $4,700$ sky tiles, each of the size of slightly overlapping $3.6\times3.6^\circ$ area \citep{predehl21}. We downloaded all sky tiles where the clusters of our sample are located, as well as some surrounding sky tiles for sky background estimation. The calibration and data reduction steps of the eRASS:1 data were performed with the extended Science Analysis Software (eSASS) version 211214 \citep{Brunner2022,merloni24}. The first step of the data reduction was to generate clean event lists and photon images. We ran the \texttt{evtool} tasks routine in the energy band $0.2-10.0$ keV and set the parameters \texttt{flag=0xc00fff30} to remove bad pixels and the strongly vignetted corners of the square CCDs, and \texttt{pattern=15} to include all patterns (single, double, triple, and quadruple). The exposure maps of the corresponding event files were generated using the \texttt{expmap} task. If multiple eRASS sky tiles were to be merged, for instance, in the case of very extended clusters, the \texttt{radec2xy} task was used to first align the sky tiles before merging them with \texttt{evtool} task. Count-rate images were generated by dividing the photon images and their corresponding exposure maps.

\subsubsection{eSASS source detection chain}\label{eSASS}
We used the eSASS source detection chain to identify and exclude point sources from the eROSITA data.\footnote{At the time this analysis was performed, the official eRASS1 point source catalog \citep{merloni24} was not finalized. The detected point sources in our work largely overlap with the official eRASS1 point source catalog.} The eSASS source detection chain consists of four consecutive tasks, relying on the results already obtained during image creation. The first step is to run the \texttt{erbox} task in local mode, which is based
on a sliding box algorithm and detects peaks in the image by estimating a background. It supplies a first list of possible sources in the image as well as an updated detection mask excluding the source positions (cheese map). Next, the task \texttt{erbackmap} creates a background map from an image by masking the sources found by \texttt{erbox} and smoothing. Afterward, another iteration of the task \texttt{erbox} is performed in map mode. It uses the background map to create a more accurate list of sources. Finally, the \texttt{ermldet} task is used to characterize the sources as point or extended sources. It supplies a final source list and a source image. Additionally, the task \texttt{catprep} can be used to convert the source list into a
source catalog fits file. All point and extended sources were masked with a radius large enough so the immediate surroundings of the mask converged to the local background level. There are very few cases were extended sources needed masking, due to the way the M20 sample was constructed (excluding clusters with nearby extended sources, e.g., double clusters). After the completion of the eSASS source detection chain, a visual inspection was performed to manually mask apparent sources that were missed by the algorithm. 

\subsubsection{Spectral analysis}\label{spectral}
To extract the source and background spectra, and their response matrix files (RMFs) and ancillary response files (ARFs), the task \texttt{srctool} was used. We extracted spectra of all seven telescope modules combined (referred to as TM0), as well as the combinations of TM1,2,3,4,6 (TM8), and TM5,7 (TM9). The latter are the cameras affected by the optical light leak at the very soft X-ray bands \citep{predehl21}. To compare the results between eROSITA, \textit{XMM-Newton}, and \textit{Chandra}, one wants to keep the temperature measurement method as similar as possible. For this purpose, we use the exact same spectra extraction regions for the clusters as used in M20. These are the $0.2-0.5R_{500}$ annuli centered at the X-ray emission peak as seen by \textit{XMM-Newton} or \textit{Chandra}, as well as the cluster core region, $<0.2R_{500}$. 

For the masking, two different approaches were used. In the first case, the masking process described in Sect. \ref{eSASS} was followed, based solely on the eROSITA point source detection and manual masking after visual inspection. As a second approach, the same masks as in the \textit{XMM-Newton} and \textit{Chandra} data were used. For the eROSITA-\textit{XMM-Newton} clusters, we used the exact same masks as in M20, obtained from the \textit{XMM-Newton} observations. For the eROSITA-\textit{Chandra} clusters, we masked the same point sources as in the \textit{Chandra} data, but with a $30"$ radius mask to account for the larger eROSITA point spread function (PSF). The two approaches returned nearly indistinguishable temperature results for nearly all clusters. This is due to the fact that we focus on central cluster regions ($<0.5 R_{500}$) with high surface brightness, where minor changes in the background treatment usually have a negligible impact. Therefore, we adopted the first approach as the default one, to better "simulate" the measurements of eROSITA temperatures that future studies will wish to convert to \textit{XMM-Newton} or \textit{Chandra} temperatures based on our findings.

The background region of each cluster was defined based on two criteria. Firstly, the inner radius should correspond to the $1.6R_{500}$ of the respective cluster (as in M20) so that the background spectra do not include bright residual emission from the cluster. The outer radius is calculated such that the background area is at least four times as large as the source region. In some cases, the background spectra had very few counts. In those cases, the outer radii were scaled up so that each background spectrum is expected to contain at least 300 counts, and the source detection chain was reiterated.
\par
The spectral fitting was performed with \texttt{XSPEC} \citep{XSPEC} version 12.12.0. The model can be described as
\begin{equation}
\begin{split}
\mathtt{Model =} &\quad\mathtt{constant\times[apec_1 + phabs\times(apec_2 +}\\
&\quad\mathtt{powerlaw)] + phabs\times apec_3+PIB}.\\
\end{split}
\label{eq:spectral_model}
\end{equation}

The \texttt{constant} (arcmin$^2$) term represents the cosmic X-ray background (CXB) components scaled to the areas of the source regions. The components consist of the unabsorbed thermal emission from the Local Hot Bubble (LHB; \texttt{apec$\mathtt{_1}$}), the absorbed Milky Way Halo (MWH; \texttt{apec$\mathtt{_2}$}), and the cosmic X-ray background from the unresolved sources (\texttt{powerlaw}). The absorption along the line of sight by the Milky Way is modeled by a \texttt{phabs} model and its parameter is set to the $N_{\text{H,tot}}$ values from \cite{willingale} (same as in M20). The temperatures for LHB and MWH were fixed at 0.099 and 0.225 keV respectively \citep{McCammon_2002}, while the photon index of the \texttt{powerlaw} was fixed at 1.46 \citep[e.g.;][]{Luo_2017}. The second term is for the source spectra, which is an absorbed thermal emission component (\texttt{phabs$\times$apec$\mathtt{_3}$}). The last term (PIB) describes the particle-induced (instrumental) background, which is modeled as in \citet[][see that paper for details]{veronica23}. For this, we adapted the results of the eROSITA EDR filter wheel closed (FWC)\footnote{\href{https://erosita.mpe.mpg.de/edr/eROSITAObservations/EDRFWC/}{https://erosita.mpe.mpg.de/edr/eROSITAObservations/EDRFWC/}} data analysis. The FWC spectra were rescaled to the observed spectra using the $7-9$ keV band, where the PIB dominates. The normalizations of all PIB components were left to vary. The best-fit results were then used as starting points in the subsequent spectral fittings, where the PIB normalizations were left free to vary within the $3\sigma$ range of the best-fits. 

The spectral fitting was performed in three different bands: the $0.7-7.0$ keV (full, same band used in M20), $0.5-4.0$ keV (soft), and $1.5-7.0$ keV (hard) bands.\footnote{Due to the limited number of counts, the soft and hard bands inevitably overlap between $1.5-4$ keV. Restricting the range of either of these bands significantly reduces the quality of $T_{\text{eROSITA}}$ constraints and the value of the cross-instrument comparison. For instance, by using the 2-7 keV band we reduce the number of available $T_{\text{eROSITA}}$ by $\sim 40\%$ compared to the 1.5-7 keV band.} All spectra (CXB+source) were fitted simultaneously for individual TMs for each band. Moreover, the TM8 and TM9 modules were first fitted separately. As discussed in Sect. \ref{TM8-TM9}, the $T$ results of the two modules agree within $\lesssim 2 \sigma$.\footnote{For the TM9 modules that suffer from the light leak, the $T$ estimation was rather insensitive on the exact lower energy limit that was adopted ($0.7$ or $0.8$ keV). Thus, the same energy bands with TM8 were used.} Therefore, we fit simultaneously both modules, that is, we use TM0 as the default for the main analysis. Additionally, we performed all fits with the metal abundance ($Z$) both fixed to the M20 value and free to vary. To determine the best-fit spectral parameters, we used C-statistics \citep{Cash_1979}, which is a more suitable estimator for Poissonian counts. The Solar metal abundance table from \cite{aspl} was also used.

Finally, due to its low sensitivity at high energies ($\gtrsim 2.5$ keV), eROSITA was unable to constrain $T$\footnote{This means the symmetrized $T$ uncertainty $(\sigma_{\text{T+}}+\sigma_{\text{T-}})/2$ is larger than the best-fit $T$ value (i.e., $T/{\sigma_T}<1$), the spectral fit failed to converge, or the returned $T$ were irrational ($>25$ keV or negative values).} in the full band for $11\%$ (23) and $33\%$ (21) of the cluster regions for which we have \textit{Chandra} and \textit{XMM-Newton} $T$ respectively. These fractions are similar for the soft band $T$ as well. For the hard band, this fraction increases to $47\%$ (96) and $60\%$ (56) respectively. All $T/\sigma_T<1$  clusters were excluded from the analysis (the negligible effects of this selection are explored in Sect. \ref{selection-bias}). For the eROSITA-\textit{Chandra} cross-calibration, this resulted in 185 (full band), 179 (soft band), and 108 (hard band) available cluster temperatures. The median $T/\sigma_T$ is 4.4, 4.2, and 2.3 respectively. For the eROSITA-\textit{XMM-Newton} cross-calibration, this resulted in 71 (full band), 68 (soft band), and 38 (hard band) cluster temperatures. The median $T/\sigma_T$ is 3.2, 3.0, and 1.9 respectively.

\subsection{\textit{Chandra} and \textit{XMM-Newton}}

\subsubsection{Data reduction}
%For the \textit{Chandra} and \textit{XMM-Newton} analysis we use the ACIS I+S and EPIC (MOS1+MOS2+PN) instruments respectively. 
The data reduction and analysis of the \textit{Chandra} and \textit{XMM-Newton} observations are described in detail in \citet{gerrit2} and \citet{miriam} respectively. For \textit{Chandra} the ACIS I/S instruments were used, with no grating (no HETG/LETG). For \textit{XMM-Newton}, the EPIC (MOS1/MOS2/PN) instrument was used. In short, pointed observations of both \textit{Chandra} and \textit{XMM-Newton} were treated for solar flare contamination, bad pixels, anomalous CCD state \citep[for \textit{XMM-Newton} only;][]{kuntz} and out-of-time events, masking of point sources and extended emission sources not related to the cluster, vignetting, exposure time correction, and instrumental background. The HEASOFT 6.20, XMMSAS v16.0.0, and CIAO v4.8 with CALDB 4.7.6 software packages were used in the M20 analysis and the full band results. Any additional spectral fitting in this work was performed using the already available spectra from M20, produced by the same software packages. For the additional soft and hard band \textit{XMM-Newton} spectral analysis, we use the same software packages. For the additional \textit{Chandra} spectral fitting, the CIAO v4.13 software package was used with CALDB 4.9.4.\footnote{The full band $T_{\text{Chandra}}$ fits were repeated with CIAO v4.13 and insignificant, non-systematic $T$ changes were found compared to the default M20 values. Moreover, several \textit{XMM-Newton} clusters were reanalyzed with HEASOFT 6.29 and XMMSAS v18.0.0. and again, no significant, systematic changes were found in the measured $T$. Any small $T$ changes were significantly smaller than the scatter of the scaling relations in Sect. \ref{mainresults}. Hence, the exact used software package versions do not affect the results and conclusions of this work.}

\subsubsection{Spectral analysis}\label{spectral-Chandra-xmm}

The spectral analysis methodology for both \textit{Chandra} and \textit{XMM-Newton} is described in M20. In brief, we extracted two independent spectra per cluster, from within the $\leq 0.2\ R_{500}$ and $0.2-0.5\ R_{500}$ regions. For \textit{Chandra}, the PIB was obtained from the stow event files \citep{gerrit2}. It was renormalized to the 9.5-12 keV count-rate of the cluster observation before being subtracted from the source spectra. The  CXB was extracted from within $1^{\circ}-2^{\circ}$ around the cluster using the seven ROSAT All-Sky Survey (RASS) bands \citep{snowden97}, as done during the M20 measurements. For \textit{XMM-Newton}, the PIB was measured utilizing filter wheel closed observations. It was then rescaled to the cluster observation count-rates of the unexposed EPIC corners ($>925"$ from the pointing's center) using the $2.5-5$ keV (MOS and PN) and $8-9$ keV (MOS) energy bands. The CXB was extracted from the \textit{XMM-Newton} field of view (FOV), from regions at $\gtrsim 1.6\times R_{500}$ from the cluster's center. When this was not possible, the CXB was extracted from the outer $1'$ width annulus of the FOV. In this case, an extra, free-to-vary \texttt{apec} component was added to the CXB model to account for residual cluster emission. 

The fitted model is the same as for eROSITA (Sect. \ref{spectral}), without the PIB component. For both \textit{Chandra} and \textit{XMM-Newton}, the rescaled PIB was subtracted from the source spectra. To account for an imperfect PIB subtraction, additional Gaussian lines are added to the model, representing fluorescence line residuals \citep[for details see][M20]{gerrit2,miriam}. The CXB-only spectra were first fitted alone. The best-fit CXB values are then used as starting points and the full model (cluster emission$+$CXB) is left free to vary when fitting the source spectra. All properties of each model component (e.g., normalizations, temperatures, and metallicities) were linked across the three EPIC detectors during the \textit{XMM-Newton} fits. For each spectrum, different fits are performed for three energy bands, as in eROSITA. The $0.7-7$ keV results were taken from M20. For this work, we additionally performed the $0.5-4$ keV and $1.5-7$ keV spectral fits for all spectra, leaving $Z$ free to vary and fixing $N_{\text{H}}$ to the M20 value. For consistency with M20, we used XSPEC v12.9.1 for both telescopes. A $\chi^2-$statistic was used for both \textit{Chandra} and \textit{XMM-Newton} to determine the best-fit spectral parameters, as in M20.\footnote{The number of counts and bins (every 25 counts) per fitted spectrum is very high (several tens of thousand counts) and the $\chi^2-$statistic gives very similar results to a c-statistic. The lack of significant bias due to the use of the $\chi^2$ instead of the c-statistic is discussed in Sect. \ref{T-shift-sect}.}

\section{Statistical methodology}\label{statistics}
For the comparison of different temperature measurements, we adopt a linear relation in the logarithmic space of the form:
\begin{equation}
\log_{10}{\frac{T_{\text{Y}}}{T_{\text{piv}}}}=A+B\times \log_{10}{\frac{T_{\text{X}}}{T_{\text{piv}}}}, 
\label{scal_rel}
\end{equation}
where $T_{\text{Y}}$ and $T_{\text{X}}$ are the compared temperature distributions and $A$ and $B$ are the intercept and slope of the relation respectively. Moreover, $T_{\text{piv}}=4.5$ keV for eROSITA-\textit{Chandra} comparisons and $T_{\text{piv}}=3$ keV for eROSITA-\textit{XMM-Newton} comparisons (i.e., rounded median values of \textit{Chandra} and \textit{XMM-Newton} $T$ distributions). Eq. \ref{scal_rel} corresponds to a single power law in linear space. A broken (double) power law fit was also tested. However, for all temperature comparisons, it was "strongly disfavored" or "disfavored" by the Bayesian and Akaike Information Criteria (BIC and AIC respectively) since it did not improve the fit significantly to justify the introduction of extra free parameters. Nevertheless, we present the results for a broken power law fit in Sect. \ref{broken_pow} for the eROSITA-\textit{Chandra} scaling relations for which a broken power law fit showed the greatest improvement compared to a single power law fit.

For the linear regression of Eq. \ref{scal_rel} we use a likelihood maximization method (LMM hereafter) as the default. Specifically, we maximize the log-likelihood function 
\begin{equation}
\begin{aligned}
\ln{\mathcal{L}}&=\\
=&-\dfrac{1}{2}\sum^N_{i=1}\left[\dfrac{(\log{T^{'}_{\text{Y},i}}-A-B\times \log{T^{'}_{\text{X},i}})^2}{\sigma_i^2+\sigma_\text{intr}^2}+\ln{(\sigma_i^2+\sigma_\text{intr}^2)}\right],
\label{likelihood}
\end{aligned}
\end{equation}
where 
\begin{equation}
T^{'}_{\text{Y},i}=\dfrac{T_{\text{Y},i}}{T_{\text{piv}}},\ \  T^{'}_{\text{X},i}=\dfrac{T_{\text{X},i}}{T_{\text{piv}}},\ \  \text{and}\ \ \sigma^2_i=\sigma^2_{\log{T_{\text{Y},i}}i}+B^2\times \sigma^2_{{\log{T_{\text{X},i}}}},
\label{uncertainties-sum}
\end{equation}
and $\sigma_{\text{intr}}$ is the intrinsic scatter of the relation. Moreover, $\sigma_{\log{T}}=\log{e}\times \dfrac{T_{\text{max}}-T_{\text{min}}}{2T}$ are the $T$ measurement uncertainties in log-space, symmetrized from the linear-space $68.3\%\ T$ uncertainties. The fitted parameters here are $A$, $B$, and $\sigma_{\text{intr}}$. The total scatter $\sigma_{\text{tot}}$ is given by the mean quadratic sum of $\sigma_i$ and $\sigma_{\text{intr}}$. 

Additionally to LMM we also utilize the Bivariate Correlated Errors and intrinsic Scatter (BCES) package by \citet{akritas}. For every scaling relation, the best-fit values of $A$ and $B$ were obtained using the BCES(orth) method. The latter minimizes the orthogonal distance of the data points compared to the best-fit line, as opposed to LMM, which mostly considers the residuals on the y-axis. Due to the much larger measurement uncertainties of $T_{\text{eROSITA}}$, a minimization in the y-axis (as in LMM) is more appropriate than an orthogonal distance regression. Nevertheless, BCES(orth) serves as a measure of how sensitive our results are on the exact fitting method.

Similarly to \citet{pratt}, the total scatter in the y-axis $\sigma_{\text{tot}}$ is measured using the error-weighted vertical distance to the best-fit regression line:

\begin{equation}
\sigma^2_{\text{tot}}= \frac{1}{N-2}\sum^N_{i=1}w_i\left(Y_i-A_{\text{orth}}-B_{\text{orth}}\times X_i\right)^2 
\label{scatterYX},
\end{equation}
where $Y=\log_{10}{\dfrac{T_{\text{Y},i}}{T_{\text{piv}}}}$ and $X=\log_{10}{\dfrac{T_{\text{X},i}}{T_{\text{piv}}}}$. Furthermore, 
\begin{equation}
\begin{aligned}
  &w_i=\dfrac{1/\sigma_i^2}{(1/N)\sum^N_{i=1}1/\sigma^2_i}\ \  \text{and}\ \ \sigma^2_i=\sigma^2_{Y_i}+B^2\times \sigma^2_{X_i}. 
  \end{aligned}
  \label{weight}
\end{equation}
Here, $\sigma_{\text{intr}}$ is then given by the quadratic difference between $\sigma_{\text{tot}}$ and $\sigma_i$. Although we use BCES(orth), we measure the scatter in the y-axis direction (instead of the orthogonal scatter) to make a direct comparison with the results of LMM.

For both LMM and BCES, the 68.3\% ($1\sigma$) credible intervals of fitted model parameters are estimated by performing 10,000 bootstrap resamplings with replacement. From the posterior distribution of the best-fit parameters, we determine the $16_{\text{th}}$ and $84_{\text{th}}$ percentiles. In the 2-dimensional planes, the 1, 2, and 3$\sigma$ ellipses are drawn by finding the ellipse with the smallest area that encompasses the 68.3\%, 95\%, and 99.7\% of the 10,000 bootstrap data points.

Finally, following S15, we define a measure $\xi$ of the statistical deviation between the two compared $T$ distributions in linear space,
\begin{equation}
    \xi=\text{med}\dfrac{T_{\text{Y}}-T_{\text{X}}}{\sqrt{\sigma_{T_{\text{Y}}}^2+\sigma_{T_{\text{X}}}^2}},
\end{equation}
where the median of $\xi$ for all clusters is considered. The term $\xi$ can take both positive and negative values, depending on which $T$ distribution is systematically higher or lower. If there is no systematic difference between the two compared instruments, one expects $\xi\approx 0$.

\section{Single power law fits}\label{mainresults} 

In this section we present the results from the cluster $T$ comparison between eROSITA, \textit{Chandra}, and \textit{XMM-Newton}. All scaling relation results are shown in Table \ref{tab:comparison2}.
\subsection{eROSITA versus \textit{Chandra} temperatures}\label{Chandra-single-pow-results}
eROSITA returns systematically lower cluster $T$ than \textit{Chandra} for all three energy bands. The discrepancy is a function of cluster $T$, with hot, massive clusters deviating more than low $T$ groups. For the latter, the two instruments seem to return similar $T$ across all energy bands. Overall, the discrepancy between eROSITA and \textit{Chandra} cluster $T$ is the strongest for the hard energy band. This is unlike what was found in previous studies \citep[e.g, S15;][]{kettula} where X-ray instruments tend to return consistent $T$ in the hard band ($>2$ keV) while showing stronger disagreement in the soft band ($<2$ keV). However, the band definition in this work is different than past studies and the soft and hard bands overlap at $1.5-4$ keV; therefore, a direct comparison is challenging.

\subsubsection{Full band}
The $T$ comparison for the full band between the two instruments is displayed in the left panel of Fig. \ref{Chandra-scaling-rel}. The relation shows a slope of $B=0.781^{+0.020}_{-0.023}$, deviating from the equality line by $>10\sigma$ (Fig. \ref{ero-Chandra}). The total scatter of the relation ($\approx 30\%$) is dominated by the $T_{\text{eROSITA}}$ uncertainties of high-$T$ clusters, rather than the intrinsic scatter ($\approx 10\%$). The latter remains rather constant with increasing $T_{\text{Chandra}}$ (see Appendix \ref{scatter-dependence} for the details of scatter dependency on $T$). Due to the energy dependence of eROSITA's effective area, $T_{\text{eROSITA}}$ uncertainties generally increase with increasing $T$. Similarly, for a fixed $T_{\text{Chandra}}$, clusters seen as high$-T$ systems by eROSITA (i.e., upscattered) show larger uncertainties than downscattered clusters that seem to have lower $T$.

eROSITA returns $\approx 25\%$ lower $T$ than \textit{Chandra} for clusters with $T_{\text{Chandra}}=4.5$ keV. For more massive clusters with $T_{\text{Chandra}}=10$ keV, the deviation rises to $\approx 38\%$. On the contrary, galaxy groups with $T_{\text{Chandra}}\lesssim 2$ keV do not show any systematic $T$ difference between the two telescopes. The value of $\xi=-1.14$ shows that individual clusters do not deviate significantly from the equality line due to their large $T_{\text{eROSITA}}$ uncertainties. Therefore, the large cluster sample used here is crucial to compensate for these individual uncertainties and robustly determine the cross-calibration between eROSITA and \textit{Chandra} cluster $T$.

The single power law fit seems to slightly overestimate the expected $T_{\text{eROSITA}}$ for $T_{\text{Chandra}}\approx 1$ keV clusters by $\approx 5\%$ for LMM. Most galaxy groups at this $T$ range seem to lie closer to the equality line. This overestimation is partially caused by a single, upscattered cluster region (the $0.2-0.5\ R_{500}$ annulus of S0805). For BCES(orth), the expected $T_{\text{eROSITA}}$ is overestimated by $18\%$. This strongly demonstrates that LMM provides a better fit than BCES in this case. The systematic non-Gaussian scatter at low $T$ may suggest the need for a broken power law. This is presented in Sect. \ref{broken_pow}.

\subsubsection{Soft band}

The soft band $T$ comparison is displayed in the middle panel of Fig. \ref{Chandra-scaling-rel} and it is very similar to the full band $T$ comparison. All scaling relation parameters are consistent within $1\sigma$ with the full band results, as shown in Fig. \ref{ero-Chandra}. Once again, LMM shows a steeper slope than BCES(orth), representing the low $T$ systems better. The systematic residuals at low $T$ persist in the soft band as well, while $\xi$ shows that the median $T$ discrepancy between the two instruments is only slightly larger than in the full band. Overall, no significant change is observed in the eROSITA-\textit{Chandra} cross-calibration by limiting the spectroscopic analysis to softer energies.

\subsubsection{Hard band}
\begin{figure*}[hbtp]
               \includegraphics[width=0.33\textwidth]{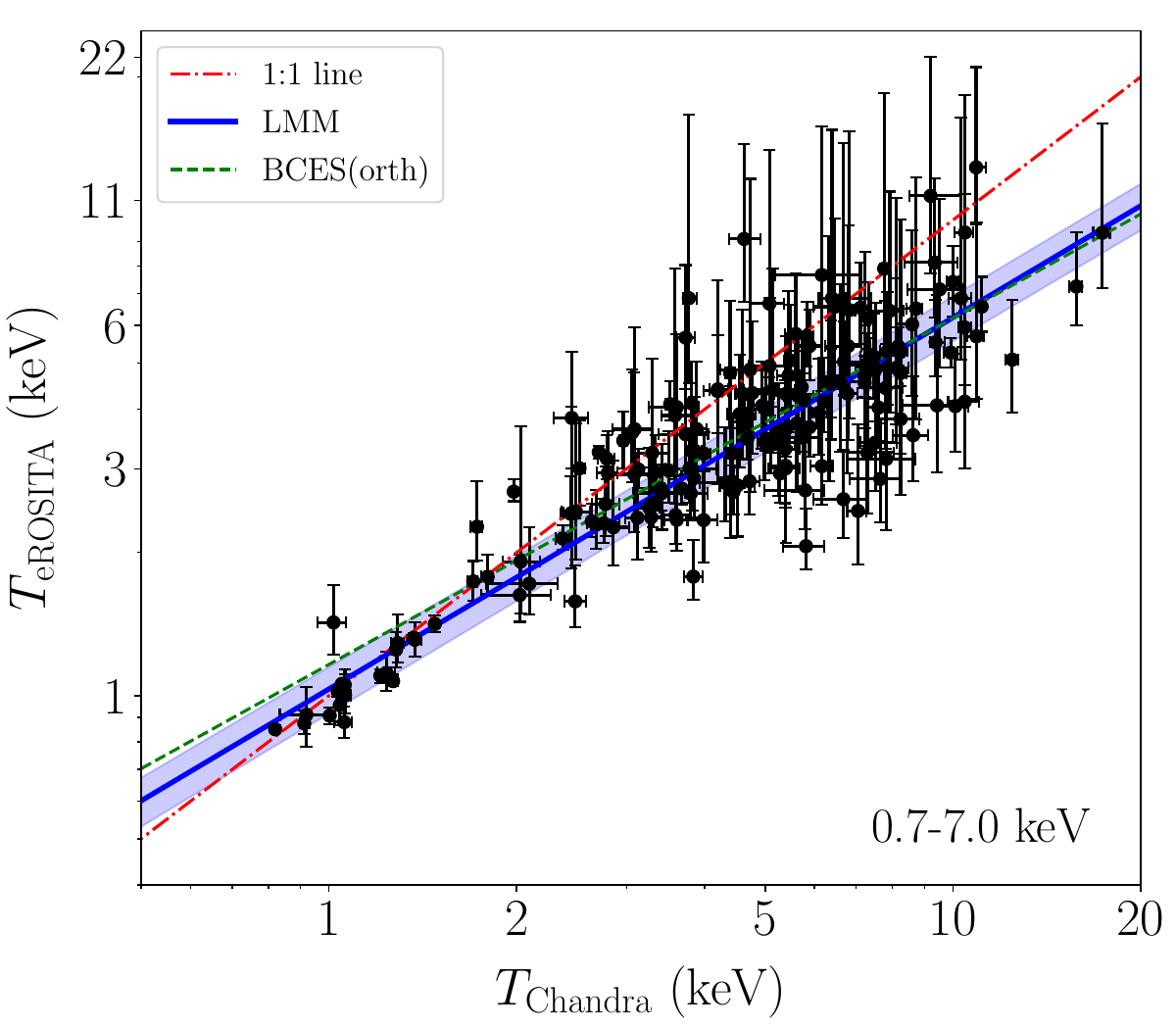}
               \includegraphics[width=0.33\textwidth]{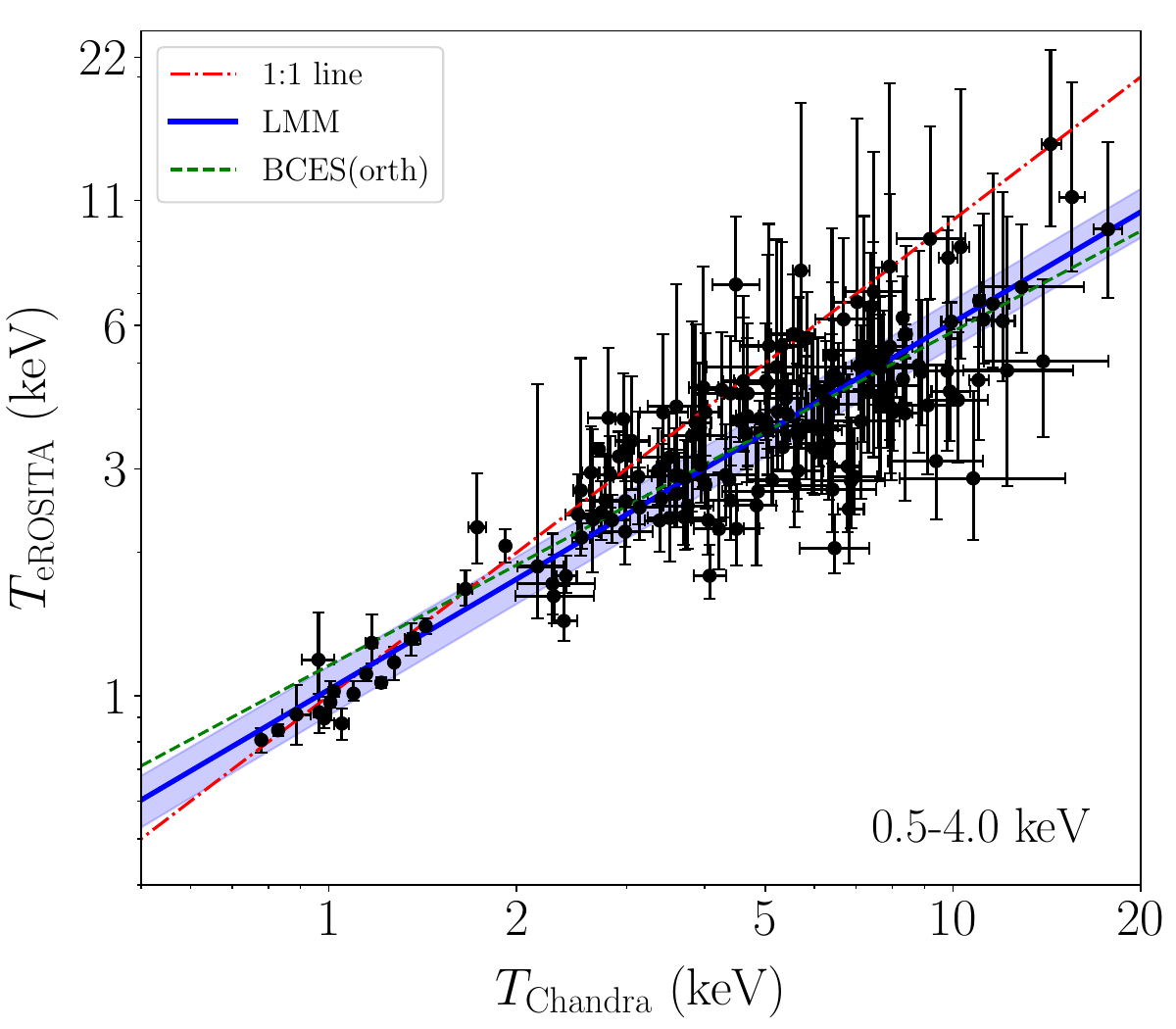}
               \includegraphics[width=0.33\textwidth]{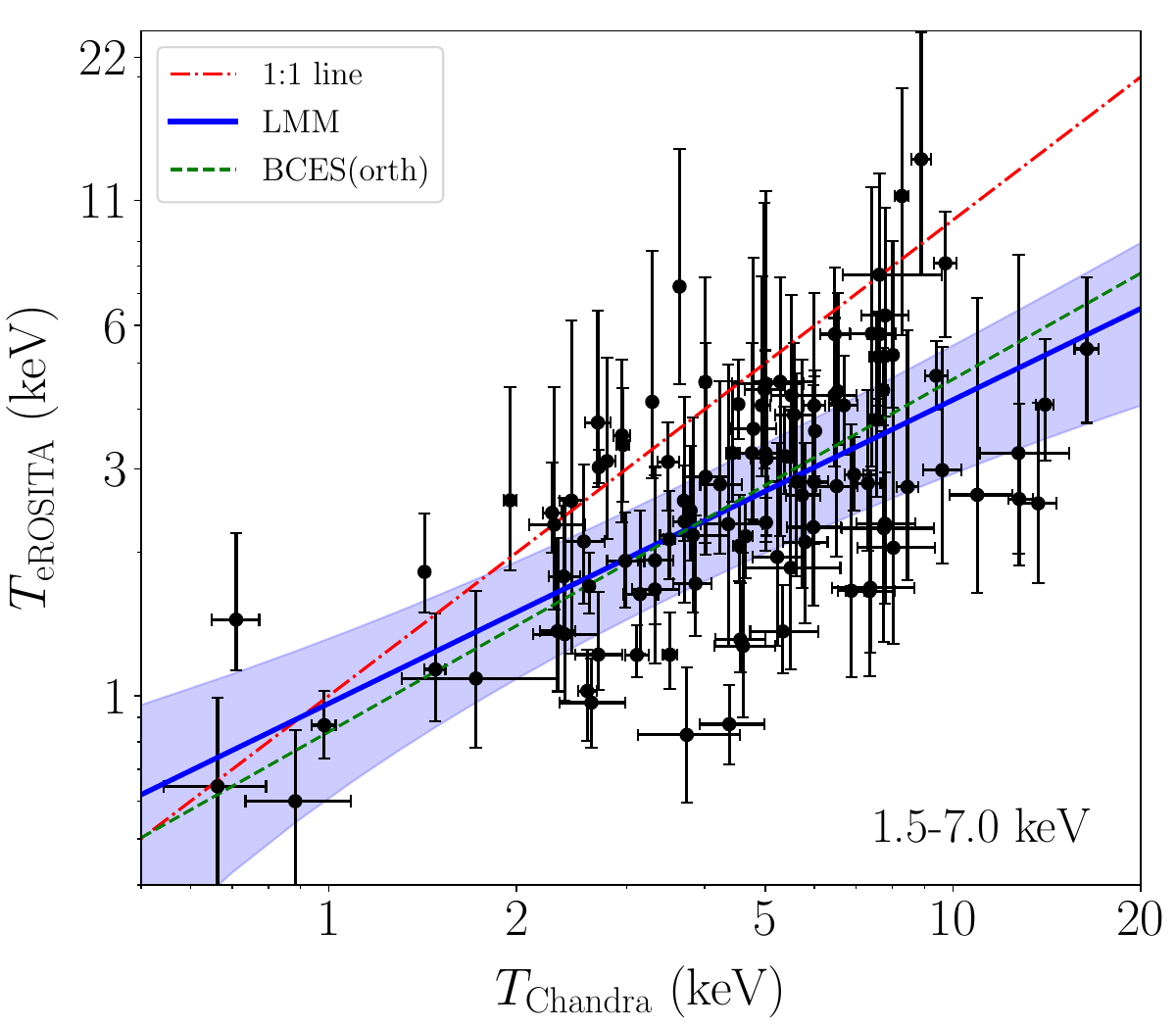}
               \caption{Comparison between eROSITA and \textit{Chandra} temperatures for the full (left), soft (middle), and hard (right) bands. The best-fit scaling relation line by LMM (blue) and BCES(orth) (green dashed) are displayed. The equality 1:1 line is shown in red (dash dot). The blue shaded area represents the LMM statistical error plus the intrinsic scatter.}
        \label{Chandra-scaling-rel}
\end{figure*}

\begin{figure}[hbtp]
               \includegraphics[width=0.49\textwidth]{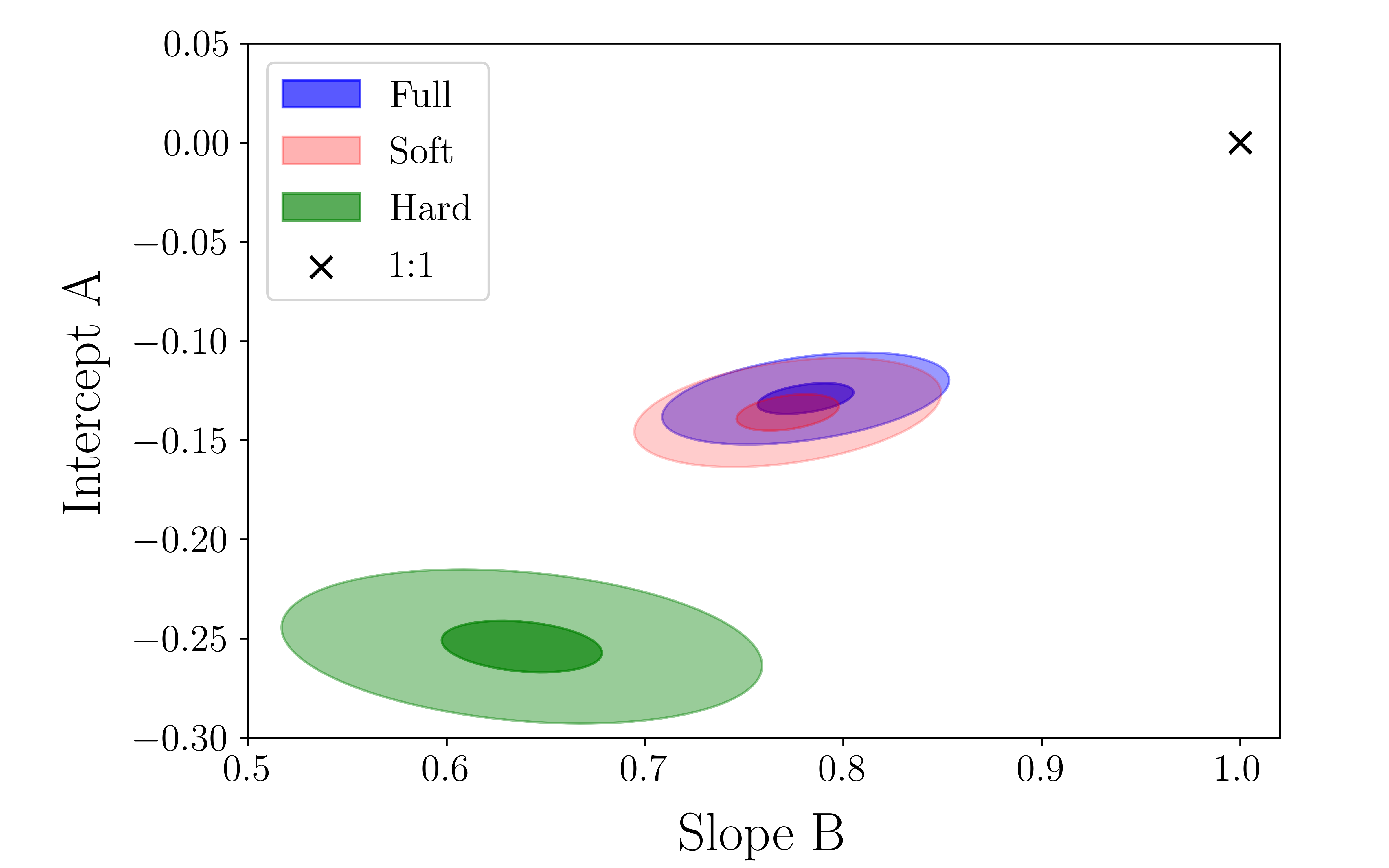}
               \caption{$1\sigma$ (68.3$\%$) and $3\sigma$ (99.7$\%$) confidence levels for the eROSITA-\textit{Chandra} scaling relations for the full (blue), soft (pink), and hard (green) bands. The 1:1 line is represented by the black cross. The band best-fit lines for all bands deviate by $\gtrsim 10\sigma$ from the 1:1 line.}
        \label{ero-Chandra}
\end{figure}

The $T$ comparison for the hard band between eROSITA and \textit{Chandra} is displayed in the right panel of Fig. \ref{Chandra-scaling-rel}. Overall, the offset in the measured $T$ between the two instruments is stronger in the hard band than in the soft and full bands, as shown in Fig. \ref{ero-Chandra}. The dependency of the offset on $T$ is also stronger with a slope of $B=0.638\pm 0.076$, $1.8\sigma$ lower than in the full band. Low $T$ systems seem consistent between the two telescopes for the hard band as well, while high-$T$ clusters deviate more than in the full and soft bands. Specifically, $T_{\text{Chandra}}=1$ keV clusters return the same $T$ for both telescopes, while eROSITA measures $\approx 44\%$ lower $T_{\text{eROSITA}}$ for $T_{\text{Chandra}}=4.5$ keV systems and $58\%$ lower $T_{\text{eROSITA}}$ for $T_{\text{Chandra}}=10$ keV. $\xi$ also has a higher value than for the other bands. The average intrinsic scatter is $\approx 3$ times larger than in the other bands. Its high value is mostly driven by $T_{\text{Chandra}}\lesssim 5$ keV, and it is rather stable within this $T_{\text{Chandra}}$ range. For hotter clusters the scatter reduces by $\approx 50\%$, although the statistical uncertainties are too large to robustly support a $T$-dependent scatter behavior (Fig. \ref{scatter-dependency}.) Finally, there is no obvious need for a broken power law in this case.

\subsection{eROSITA versus \textit{XMM-Newton} temperatures}\label{xmm-single-pow-results}
eROSITA shows systematically lower cluster $T$ than \textit{XMM-Newton} for all three energy bands, with the discrepancy being a function of cluster $T$, similar to the eROSITA-\textit{Chandra} comparison. However, the discrepancy is milder this time for the full and soft energy bands. For the hard band, the relation is very loosely constrained due to the large scatter. There is no obvious need for a broken power law fit for any band. 

\subsubsection{Full band}
The $T$ comparison for the full band between eROSITA and \textit{XMM-Newton} is displayed in the left panel of Fig. \ref{xmm-scaling-rel}. The relation shows a slope of $B=0.825^{+0.074}_{-0.066}$, supporting a $T-$dependence of the discrepancy between the two instruments at $2.4\sigma$. Considering also the intercept $A=-0.078^{+0.016}_{-0.017}$, which deviates more strongly from 1:1 than the slope, the overall relation deviates from the equality line by $\approx 5\sigma$ (Fig. \ref{xmm-contours})\footnote{Naively the reader might think this is not obvious from the eye test in Fig. \ref{xmm-scaling-rel}. However, the clusters close to the 1:1 line are the ones with large $T_{\text{eROSITA}}$ uncertainty, and hence, they carry lower statistical weight than clusters that show lower $T_{\text{eROSITA}}$.}, where the bulk of the tension comes from the intercept $A=-0.078^{+0.016}_{-0.017}$. The intrinsic scatter is once again low ($\approx 9\%$) and relatively constant for $T_{\text{XMM}}\lesssim 4$ keV clusters. For hotter systems, it drops to $<1\%$, although it remains within $<2\sigma$ from the average scatter. Nevertheless, this might indicate a $T-$dependent intrinsic scatter behavior that will be explored when the statistical uncertainties of $T_{\text{eROSITA}}$ reduce with future data. The total scatter is dominated again by the $T_{\text{eROSITA}}$ uncertainties.

For galaxy groups with $T\lesssim 1.5$ keV, the two instruments agree within $5.5\%$. For average-sized clusters with $T_{\text{XMM}}=3$ keV, eROSITA measures $\approx 16\%$ lower $T$ than \textit{XMM-Newton}. For more massive clusters with $T_{\text{XMM}}=7$ keV, the deviation rises to $\approx 28\%$. The value of $\xi=-0.41$ is mostly affected by the clusters with  $T_{\text{eROSITA}}\geq T_{\text{XMM}}$, which are the ones with large $T_{\text{eROSITA}}$ uncertainties and not much statistical weight. 

\subsubsection{Soft band}
The eROSITA-\textit{XMM-Newton} cross-calibration in the soft band is shown in the middle panel of Fig. \ref{xmm-scaling-rel}. In general, the soft band is slightly closer to the equality line than the full band (Fig. \ref{xmm-contours}), although the two bands give consistent results within $1\sigma$. The slightly better agreement is also shown from the lower $\xi=-0.25$ value. Overall, the soft band scaling relation deviates by $4.2\sigma$ by the equality line, with the main source of tension being the intercept value $A=-0.065^{+0.018}_{-0.019}$, which implies that $T_{\text{XMM}}=3$ keV clusters show a $14\%$ lower $T_{\text{eROSITA}}$. Furthermore, eROSITA measures a $23\%$ lower $T_{\text{eROSITA}}$ for $T_{\text{XMM}}=7$ keV clusters, while at low-T there is no meaningful difference with the full band, with the two instruments agreeing with each other. As in the eROSITA-\textit{Chandra} cross-calibration, LMM fitting returns a steeper slope than BCES(orth), representing the low $T$ systems better. Overall, a marginally better agreement between eROSITA and \textit{XMM-Newton} is observed in softer X-ray energies than in the full band.

\subsubsection{Hard band}\label{xmm-hard-band}
\begin{figure*}[hbtp]
               \includegraphics[width=0.33\textwidth]{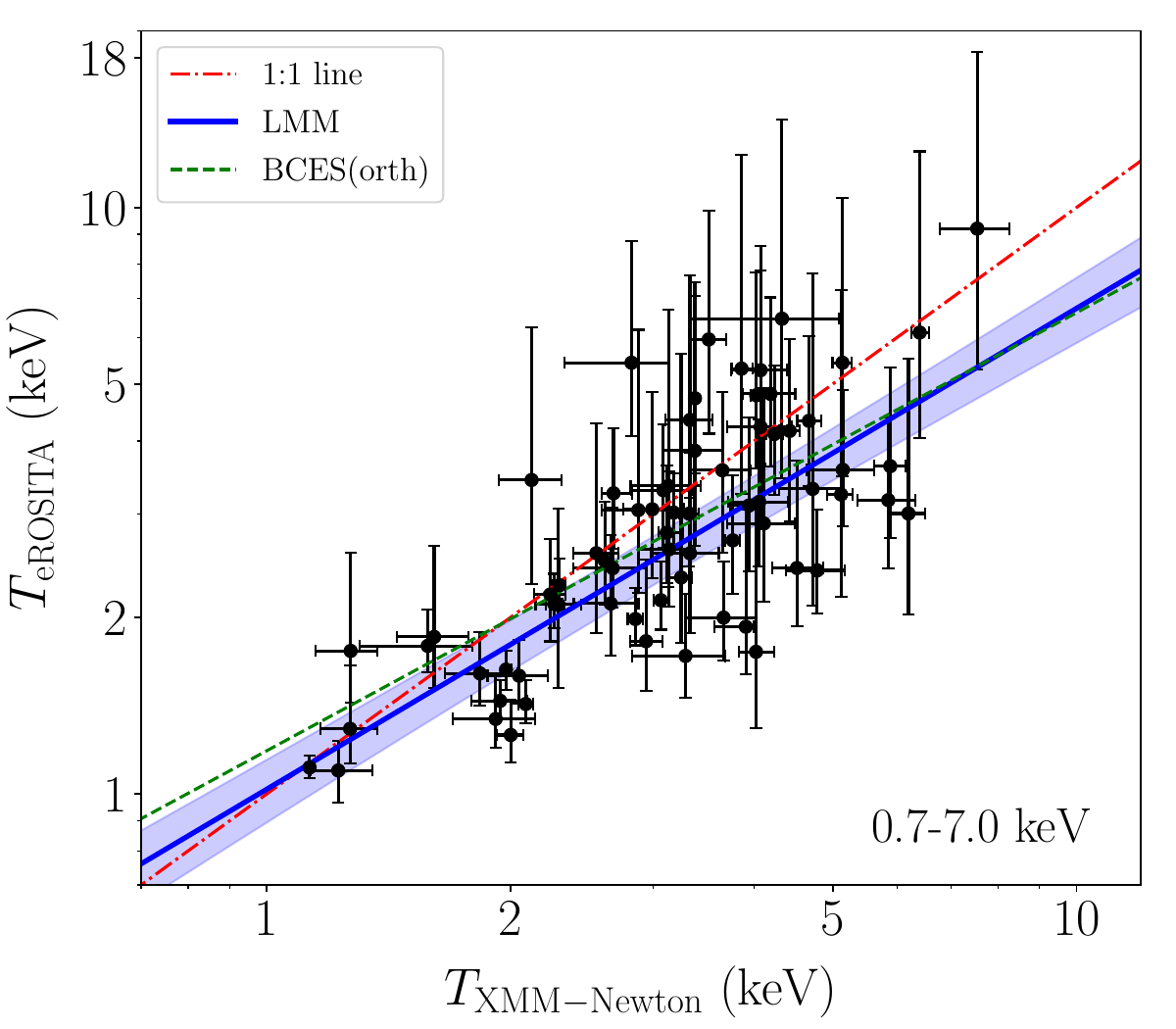}
               \includegraphics[width=0.33\textwidth]{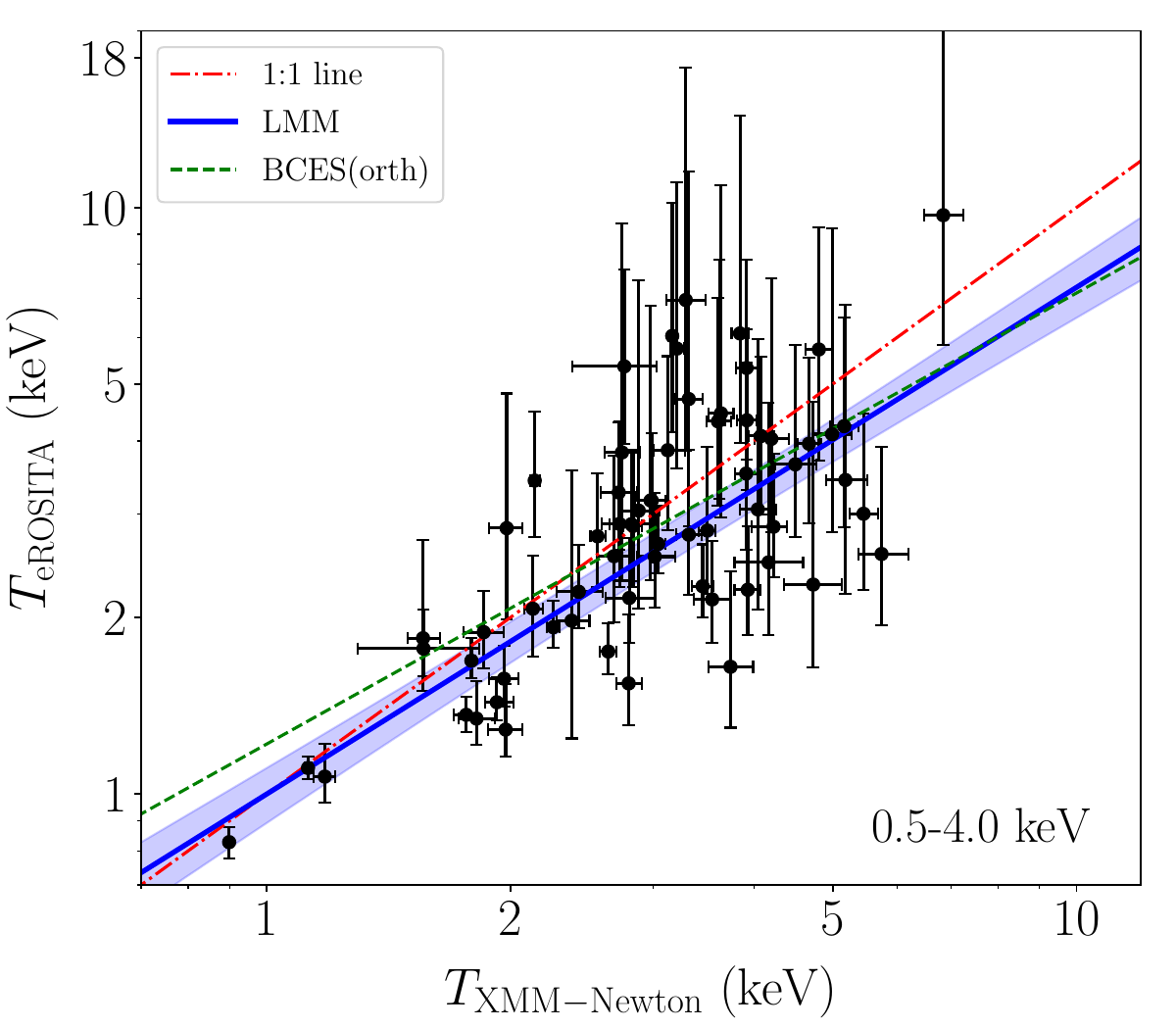}
                \includegraphics[width=0.33\textwidth]{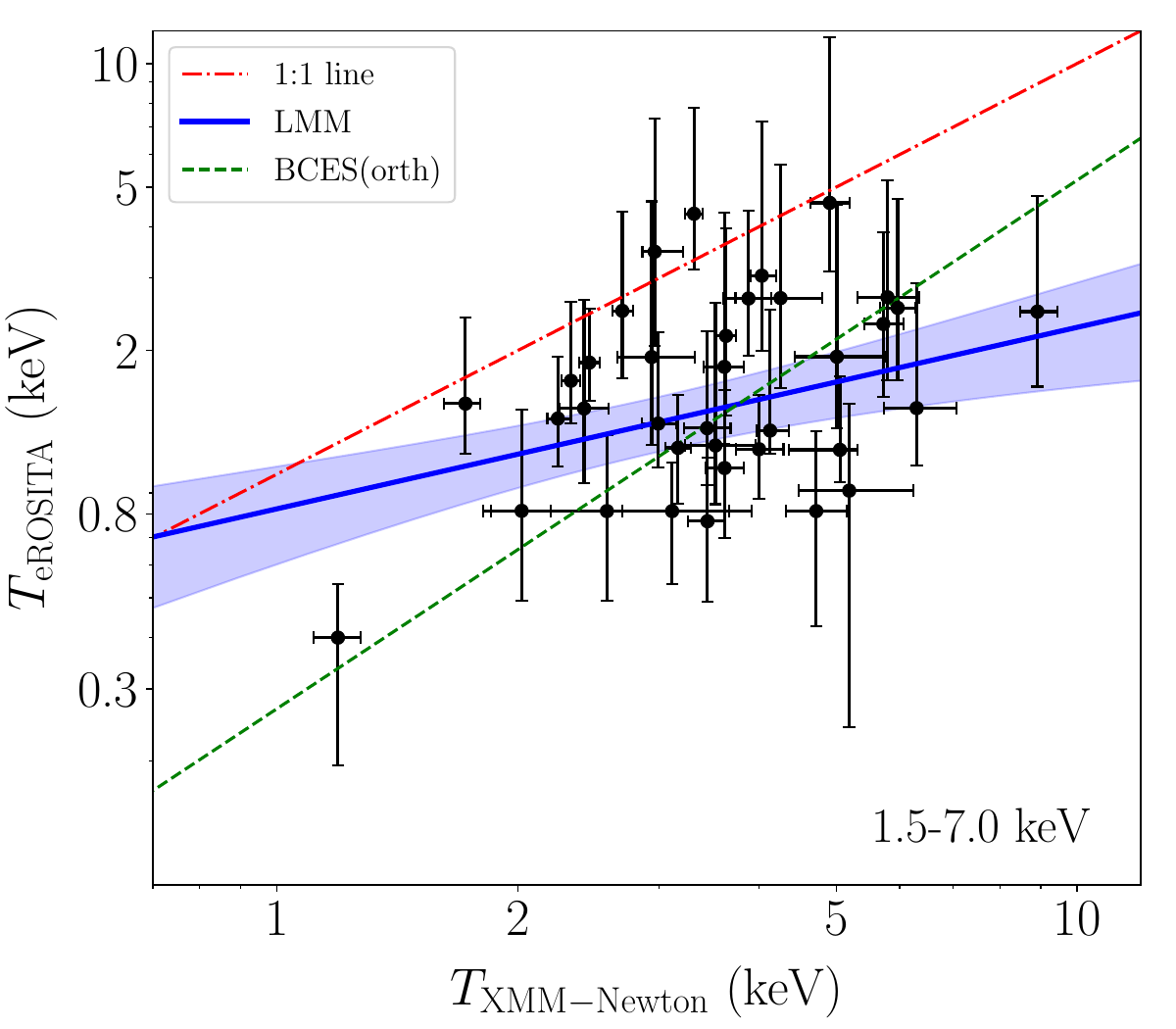}
               \caption{Same as in Fig. \ref{Chandra-scaling-rel} but for the comparison between eROSITA and \textit{XMM-Newton} temperatures.}
        \label{xmm-scaling-rel}
\end{figure*}

\begin{figure}[hbtp]
               \includegraphics[width=0.49\textwidth]{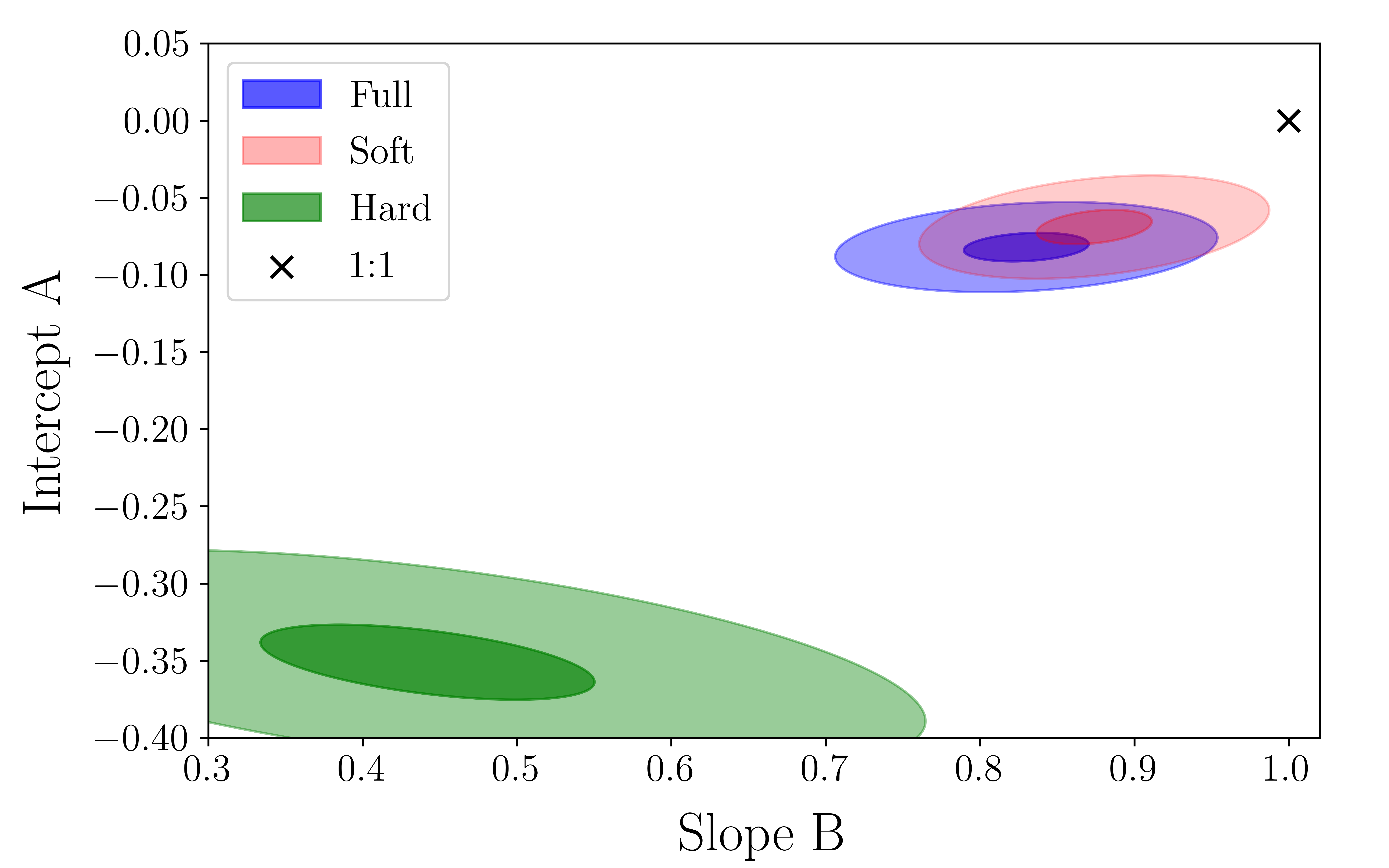}
               \caption{$1\sigma$ (68.3$\%$) and $3\sigma$ (99.7$\%$) confidence levels for the eROSITA-\textit{XMM-Newton} scaling relations for the full (blue), soft (pink), and hard (green) bands. The 1:1 line is represented by the black cross. The full and soft band best-fit lines deviate by $5\sigma$ and $4.2\sigma$ respectively from the 1:1 line.}
        \label{xmm-contours}
\end{figure}

The $T$ comparison for the hard band between eROSITA and \textit{XMM-Newton} is displayed in the right panel of Fig. \ref{xmm-scaling-rel}. Due to the limited number of available $T_{\text{eROSITA}}$ in the hard band, and their large uncertainties, the relation is not as robustly constrained as for the other bands. This is also clear from the large slope difference that LMM and BCES(orth) return. Nevertheless, one can safely conclude that eROSITA measures much lower $T$ than \textit{XMM-Newton} in the hard band. The average $T_{\text{eROSITA}}$ underestimation for a $T_{\text{XMM}}=3$ keV cluster is $55\%$. The slope is flatter (but highly uncertain) while the average intrinsic scatter is $\approx 1.5-2$ times larger than the respective full and soft band values. The intrinsic scatter also remains stable for $T_{\text{XMM}}\lesssim 4$ keV, while it decreases to almost zero for hotter clusters, as in the full band (but still within $\lesssim 1.5\sigma$ from the average). We need to note that before we constrain the scaling relation, two strong outliers were excluded despite fulfilling the criteria to be in the sample. More details can be found in Sect. \ref{Xmm-hard-outliers}. 

\begin{table*}[htbp]
\centering
\caption{Best-fit parameters for the eROSITA-\textit{Chandra} and eROSITA-\textit{XMM-Newton} scaling relations for all energy bands and fitting methods using the parametrization in Eq. \ref{scal_rel} and Eq. \ref{scatterYX}. }
\label{tab:comparison2}
\begin{tabular}{c | c c c c c c | c}
\hline
\hline
& & & & & & & \\
Comparison & Band & Method & $A$ & $B$ & $\sigma_{\text{intr}}$ & $\sigma_{\text{tot}}$ & $\xi$ \\
\hline
\hline\\[-0.25cm]
 &  & LMM & $-0.129_{-0.007}^{+0.008}$ & $0.781_{-0.023}^{+0.020}$ & $0.046\pm 0.015$ & $0.129\pm 0.008$ & \\
 & Full &  &  &  &  &  & -1.14 \\
 &  & BCES(orth) & $-0.112_{-0.008}^{+0.009}$ & $0.728_{-0.029}^{+0.026}$ & $0.040\pm 0.006$ & $0.116\pm 0.007$ & \\
\cline{2-8}
\\[-0.25cm]
 & & LMM & $-0.136\pm 0.008$ & $0.772\pm 0.025$ & $0.048\pm0.018$ & $0.131\pm 0.011$ & \\
eROSITA-\textit{Chandra} & Soft &  &  &  &  & & -1.27\\
 &  & BCES(orth) & $-0.131\pm 0.009$ & $0.702_{-0.030}^{+0.027}$ & $0.041\pm 0.006$ & $0.114\pm 0.006$ & \\
\cline{2-8}
\\[-0.25cm]
 &  & LMM & $-0.254\pm 0.021$ & $0.638\pm 0.076 $ & $0.143\pm 0.024$ & $0.271\pm 0.015$ & \\
 & Hard &  &  &  &  & & -1.68\\
 &  & BCES(orth) & $-0.245\pm 0.038$ & $0.741_{-0.151}^{+0.184}$ & $0.095\pm 0.012$ & $0.231_{-0.016}^{+0.012}$ & \\
\hline
\\[-0.25cm]
 & & LMM & $-0.078_{-0.017}^{+0.016}$ & $0.825_{-0.066}^{+0.074}$ & $0.038\pm 0.019$ & $0.164\pm 0.013$ & \\
 & Full &  &  &  &  & & -0.41 \\
 &  & BCES(orth) & $-0.047\pm 0.017$ & $0.747_{-0.094}^{+0.082}$ & $0.067\pm 0.007$ & $0.144\pm0.009$ & \\
\cline{2-8}
\\[-0.25cm]
& & LMM & $-0.065_{-0.019}^{+0.018}$ & $0.869_{-0.067}^{+0.073}$ & $0.027\pm 0.016$ & $0.174\pm 0.014$ & \\
eROSITA-\textit{XMM-Newton} & Soft &  &  &  &  & & -0.25\\
 &  & BCES(orth) & $-0.025\pm 0.020$ & $0.769_{-0.116}^{+0.094}$ & $0.074\pm 0.009$ & $0.156\pm 0.011$ & \\
\cline{2-8}
\\[-0.25cm]
 & & LMM & $-0.351_{-0.036}^{+0.038}$ & $0.442_{-0.182}^{+0.205}$ & $0.055\pm 0.039$ & $0.263\pm 0.024$ & \\
 & Hard  &  &  &  &  & & -2.31 \\
 &  & BCES(orth) & $-0.434_{-0.041}^{+0.043}$  & $1.287_{-0.400}^{+0.698}$ & $0.107\pm 0.041$ & $0.279\pm 0.033$ & \\
\hline
\end{tabular}
\end{table*}

\section{Broken power law fits}\label{broken_pow} 
From Fig. \ref{Chandra-scaling-rel} and the systematic residuals at $T\lesssim 2$ keV, it seems that the full and soft band eROSITA-\textit{Chandra} $T$ scaling relations might be better described by a broken power law at low $T$ instead of a single power law. To explore this, we fit a broken power law
\begin{equation}
\log_{10}\left(\frac{T_{\text{Y}}}{T_{\text{piv}}}\right) = A_{\text{bpl}} + \left\{
\begin{array}{l}
B_1 \times \log_{10}\left(\dfrac{T_{\text{X}}}{T_{\text{break}}}\right), \quad \text{for } T_{\text{X}} \leq T_{\text{break}} \\
\\[1ex]
B_2 \times \log_{10}\left(\dfrac{T_{\text{X}}}{T_{\text{break}}}\right), \quad \text{for } T_{\text{X}} > T_{\text{break}},
\end{array}
\right.
\label{broken_pow_eq}
\end{equation}
where $T_{\text{break}}$ is the $T$ where the power law changes (the so-called power law knee) and $B_1$ and $B_2$ are the slopes of the power laws before and after $T_{\text{break}}$. We note that the intercept $A_{\text{bpl}}$ here corresponds to $T_{\text{break}}$ and not $T_{\text{piv}}$, as in the single power law case. Therefore, $A$ and $A_{\text{bpl}}$ should not be compared. Moreover, the two power law parts have their own $\sigma_{\text{intr}}$ fitted. The fitting is performed only with LMM since BCES(orth) does not allow for a broken power law fit. Finally, the $T_{\text{break}}$ uncertainty is rather large and degenerates the fit of all other parameters as well, making the fit non-informative. Therefore, we fix $T_{\text{break}}$ to its best-fit value.

We stress again that both the BIC and AIC criteria disfavor the broken power law compared to the single power law fit. This is because the improvement of the fit (i.e., an increase of model likelihood $\ln{\mathcal{L}}$) is not enough to justify the extra, fitted model parameters.\footnote{This conclusion does not change even if we use a single $\sigma_{\text{intr}}$ parameter for the broken power law} Nevertheless, this is probably due to the limited number of low-$T$ clusters to which the first part of the broken power law is fit and not due to the lack of a real break in the scaling relation. Therefore, the broken power law results might be more accurate for converting eROSITA and \textit{Chandra} $T$ in the full and soft bands (i.e., reduced scatter compared to single power law), even if this is not strictly justified from BIC and AIC. Finally, all broken power law scaling relation results are shown in Table \ref{broken_pow_table}.

\subsection{Full band}
\begin{figure}[hbtp]
               \includegraphics[width=0.45\textwidth]{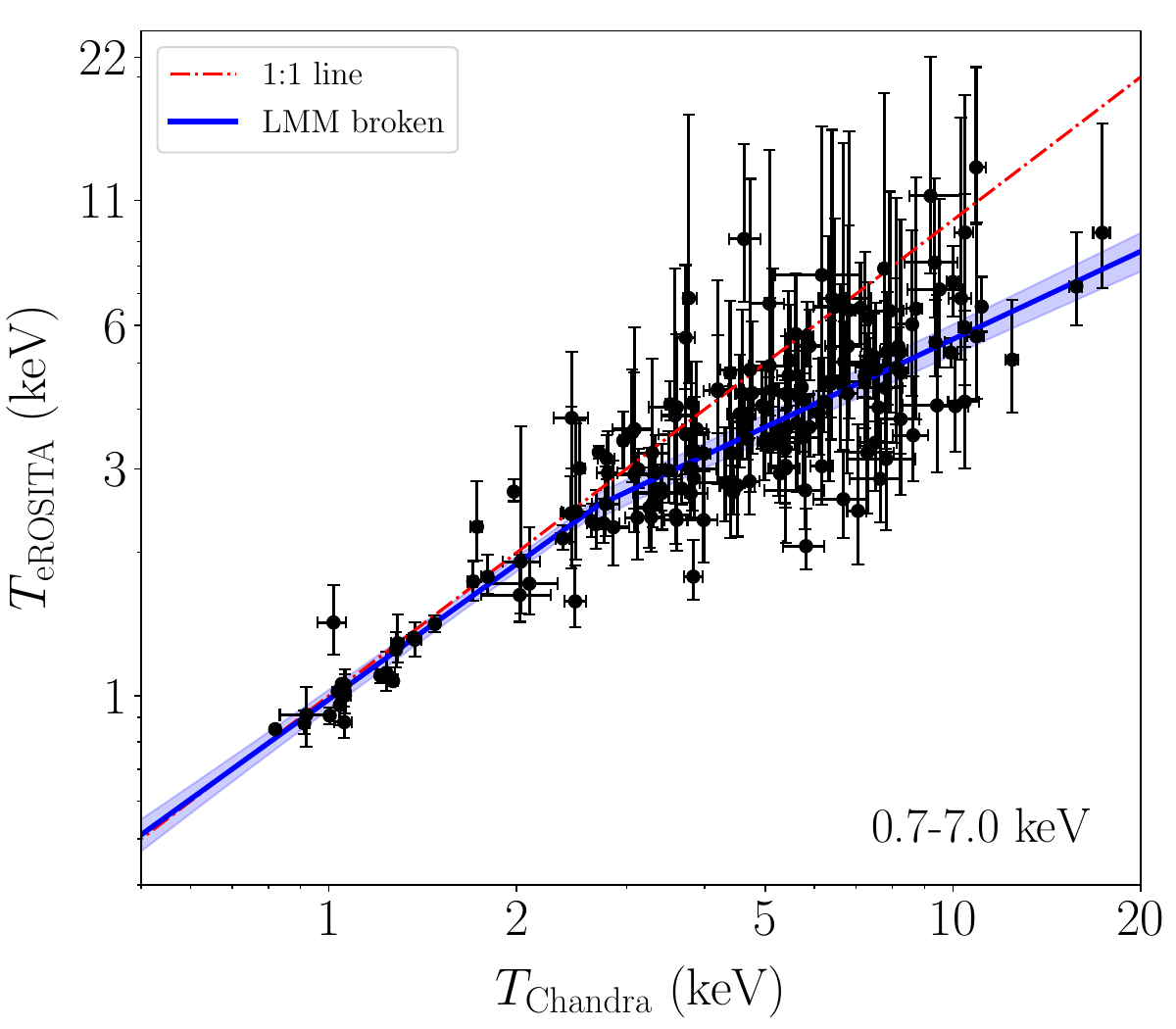}
               \includegraphics[width=0.45\textwidth]{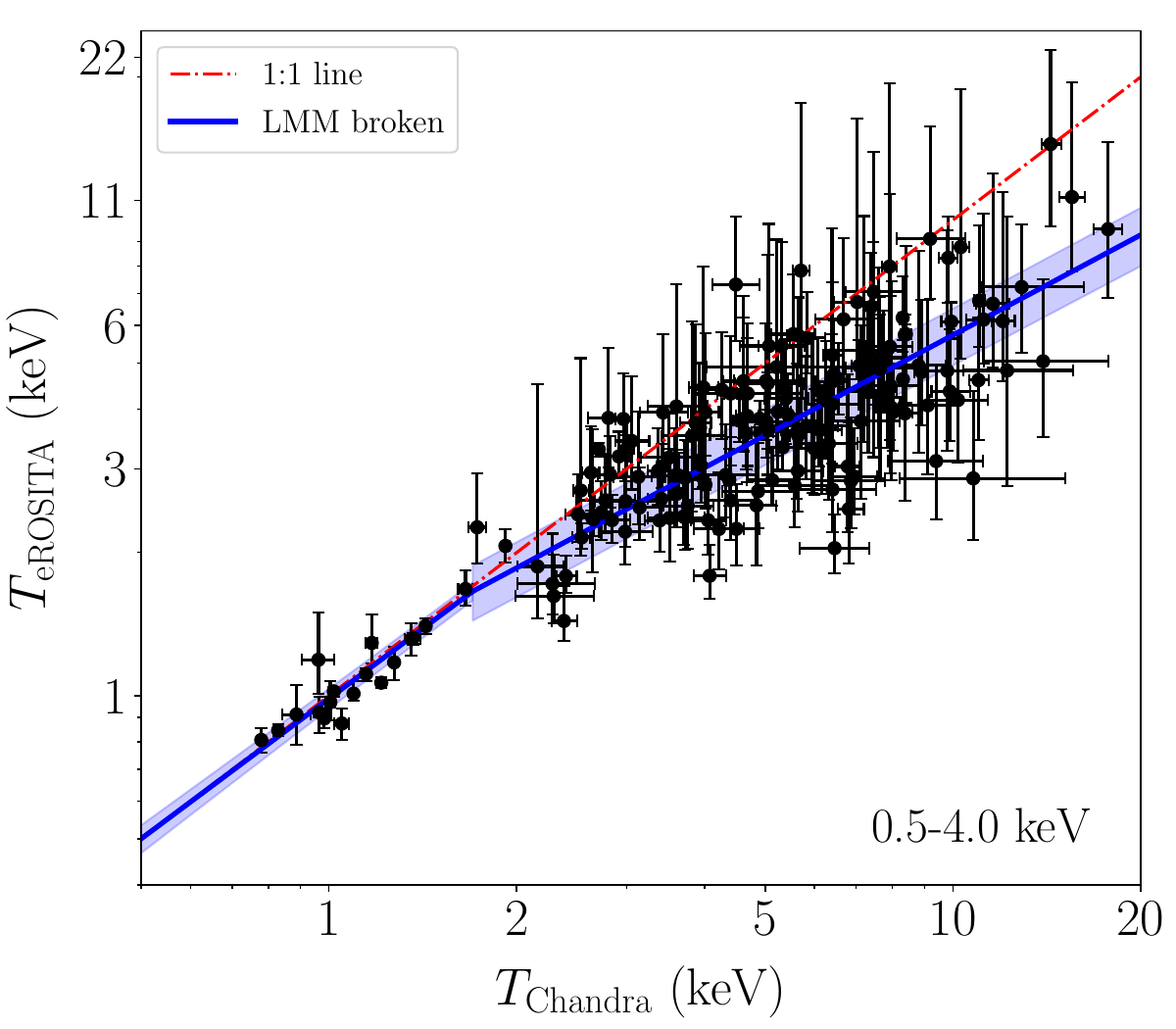}
               \caption{Same as in Fig. \ref{Chandra-scaling-rel} but with a broken power law fit with LMM for the full (top) and soft (bottom) bands, as shown in Eq. \ref{broken_pow_eq}. The break point of the power law is found at $T_{\text{break}}=2.7$ keV and $1.7$ keV, respectively. The low-$T$ clusters are much better fit than the single power law fit, showing consistency between instruments. At high-$T$, the best-fit line deviates more from the 1:1 line than the single power law fit.}
        \label{Chandra-broken}
\end{figure}

The fit results for eROSITA-\textit{Chandra} full band comparison are shown in the top panel of Fig. \ref{Chandra-broken}. The break in the power law is detected at 2.7 keV, where eROSITA measures $7\%$ lower $T$ than \textit{Chandra}. For lower $T$, the slope is $B_1=0.946\pm0.022$, very close to unity. Combined with the intercept $A_{\text{bpl}}=-0.031\pm 0.008$, one sees that the low-$T$ part is similar to the 1:1 line as $T$ decreases, although the statistical deviation from the equality line is still $\approx 4\sigma$. More specifically, eROSITA shows only $\approx 2-5\%$ lower $T$ for $T_{\text{Chandra}}\approx 1-2$ keV clusters, lower than typical measurement uncertainties. The intrinsic scatter is also very small; $3.7\%$ for $T\leq 2.7$ keV, consistent with zero at $2\sigma$.

For $T>2.7$ keV the slope $B_2=0.614\pm 0.024$, strongly demonstrating the more highly evolving discrepancy between the two instruments as $T$ increases. $B_2$ is $24\%$ flatter than the single power law slope and $43\%$ lower than $B_1$. The intrinsic scatter is $\approx 2.8$ times larger than the respective low-$T$ value, at $10\%$. Overall, the broken power law fit brings closer the two instruments in terms of measured $T$ compared to the single power law fit for $T\leq 5.3$ keV clusters, while it increases their discrepancy for $T>5.3$ keV systems.

\subsection{Soft band}
The soft band fit results for the eROSITA-\textit{Chandra} comparison are displayed in the bottom panel of Fig. \ref{Chandra-broken}. The break in the power law is detected at 1.7 keV. For $T\leq 1.7$ keV, $T_{\text{eROSITA}}$ and $T_{\text{Chandra}}$ are consistent within $\leq 2.5\%$. The low-$T$ scaling relation with $B_1=0.979\pm 0.042$ and $A_{\text{bpl}}=-0.011\pm 0.009$ is consistent with the 1:1 line at only $\approx 1\sigma$, closer than the full band fit. The intrinsic scatter is again small and relatively consistent with zero. Thus, no significant deviation from the 1:1 line is observed at the soft band for low-$T$ clusters.

For $T>1.7$ keV, the relation deviates from 1:1 as $T$ rises, with $B_2=0.700\pm 0.019$. However, it returns a less discrepant result than the full band broken power law at high $T$. The intrinsic scatter is $>3$ times larger than the respective low-$T$ value, at $13\%$. Compared to the single power law fit, the broken power law brings closer the measured $T$ of the two instruments at $T\leq 4.3$ keV, while it increases their tension for $T>4.3$ keV clusters.

\begin{table*}[htbp]
\centering
\caption{Best-fit parameters for the eROSITA-\textit{Chandra} broken power law scaling relations in the full and soft bands using the LMM method and the parametrization in Eq. \ref{broken_pow_eq}.}
\label{broken_pow_table}
\begin{tabular}{c c c c c}
\hline
\hline
& & &  \\
$T$ range & $A$ & $B$ & $\sigma_{\text{intr}}$ & $\sigma_{\text{tot}}$  \\
\hline
& & &   \\
& &  Full band &   \\
\hline
$T_{\text{Chandra}}\leq 2.7$ keV &  & $0.946\pm 0.022$ & $0.016\pm 0.008$ & $0.062\pm 0.009$ \\
& $-0.031\pm 0.008$ & & & \\
$T_{\text{Chandra}}>2.7$ keV &  & $0.614\pm 0.024$ & $0.045\pm 0.017$ & $0.151\pm 0.008$ \\
\hline
& & &  \\
& &  Soft band &   \\
\hline
$T_{\text{Chandra}}\leq 1.7$ keV &  & $0.979\pm 0.042$ & $0.017\pm 0.008$ & $0.041\pm 0.006$ \\
& $-0.011\pm 0.009$ & &  \\
$T_{\text{Chandra}}>1.7$ keV &  & $0.700\pm 0.019$ & $0.057\pm 0.018$ & $0.162\pm 0.011$\\
\hline
\end{tabular}
\end{table*}

\section{Discussion}\label{discussion} 
The accurate characterization of the cross-calibration of different X-ray telescopes is crucial to understand the systematic biases and uncertainties of the effective area calibration of X-ray instruments. Furthermore, if the cross-calibration factor between two instruments is not well-known, it hinders the joint analysis of data sets coming from different telescopes. 

In the previous sections, we used a large galaxy cluster $T$ sample for the first time to establish that eROSITA shows systematically lower cluster $T$ compared to \textit{Chandra} and \textit{XMM-Newton}. This discrepancy was found to be stronger the hotter a cluster is, with $T\lesssim 2$ keV clusters not showing significant differences between eROSITA and the other instruments. In general, the soft energy band $0.5-4.0$ keV showed slightly better agreement between different instruments than the full $0.7-7.0$ keV band. At the same time, restricting the spectral fitting to hard X-ray energies ($1.5-7.0$ keV) significantly increased the difference between $T_{\text{eROSITA}}$ and the other $T$ measurements. All the above indicates that the cross-calibration between eROSITA, \textit{Chandra}, and \textit{XMM-Newton}, is a function of spectral shape and energy range used, with the softer spectra showing better agreement. 

Opposite to the eROSITA cross-calibration results, most past studies found that the cluster $T$ differences between other X-ray instruments seem to be stronger at soft X-ray energies and more consistent at harder X-ray bands. In general, these studies failed to pinpoint a specific systematic causing these cross-calibration differences, although a wide range of possible causes was investigated (e.g., S15). The observed cluster $T$ differences were attributed to systematic effective area calibration uncertainties. In this section, we discuss and examine the possible systematic biases behind the established discrepancy between eROSITA $T$ with \textit{Chandra} and \textit{XMM-Newton} $T$.

\subsection{Metal abundance degeneracy with temperature}\label{free-z-section}

The determination of $T$ through spectral fitting is partially influenced by the (mildly) correlated metal abundance $Z$. When converting $T$ values between different X-ray instruments, accounting for the unknown $Z$ value of the second instrument is not feasible; thus one assumes the $Z$ from the first instrument. For this reason, we adopted the $Z$ values from \textit{Chandra} and \textit{XMM-Newton} measurements to use during the eROSITA spectral fitting. However, since the best-fit $T$ and $Z$ values are correlated, fixing $Z$ could potentially bias the $T_{\text{eROSITA}}$ values. In other words, if eROSITA measured systematically different $Z$ and consistent $T$ values, fixing $Z$ might erroneously interpret this as a $T$ discrepancy. Furthermore, due to the so-called "Fe bias" \citep[e.g.;][]{buote,mernier18,riva}, forcing $Z$ to be the same between instruments with different soft- and hard-band sensitivity ratios (e.g., eROSITA and \textit{Chandra}) might lead to systematic $T$ discrepancies if multiphase gas is present. To test if fixing eROSITA $Z$ to the \textit{Chandra} and \textit{XMM-Newton} values causes a systematic bias, we repeated the entire analysis, this time leaving $Z$ free to vary during the eROSITA spectral fits. 

The best-fit results can be found in Table \ref{Free-Z-table}. No scaling relation changes significantly when the new $T_{\text{eROSITA}}$ are used. In fact, the full and soft band scaling relations for both \textit{Chandra} and \textit{XMM-Newton} shift further from the 1:1 line for $T\gtrsim 3$ keV clusters, while the low-$T$ end maintains the relative consistency that showed for the fixed $Z$ case. In general, the best-fit $A$ and $B$ slightly decrease by a few percent, while the scatter increases. The hard band is the most insensitive one to the varying $Z$ change. To conclude, the discrepancy between eROSITA and other telescopes remains the same or increases for all comparisons when $Z$ is left free to vary, disproving the treatment of $Z$ as the cause of the discrepancy. This is consistent with S15, who also did not find a significant improvement in the \textit{Chandra}-\textit{XMM-Newton} cross-calibration when using the same or different $Z$ for the two telescopes.

\subsection{Different effective area sensitivity on soft X-ray and hard X-ray bands}

\subsubsection{Multi-temperature gas structure}\label{multi-T}

\begin{figure}[h]
               \includegraphics[width=0.45\textwidth]{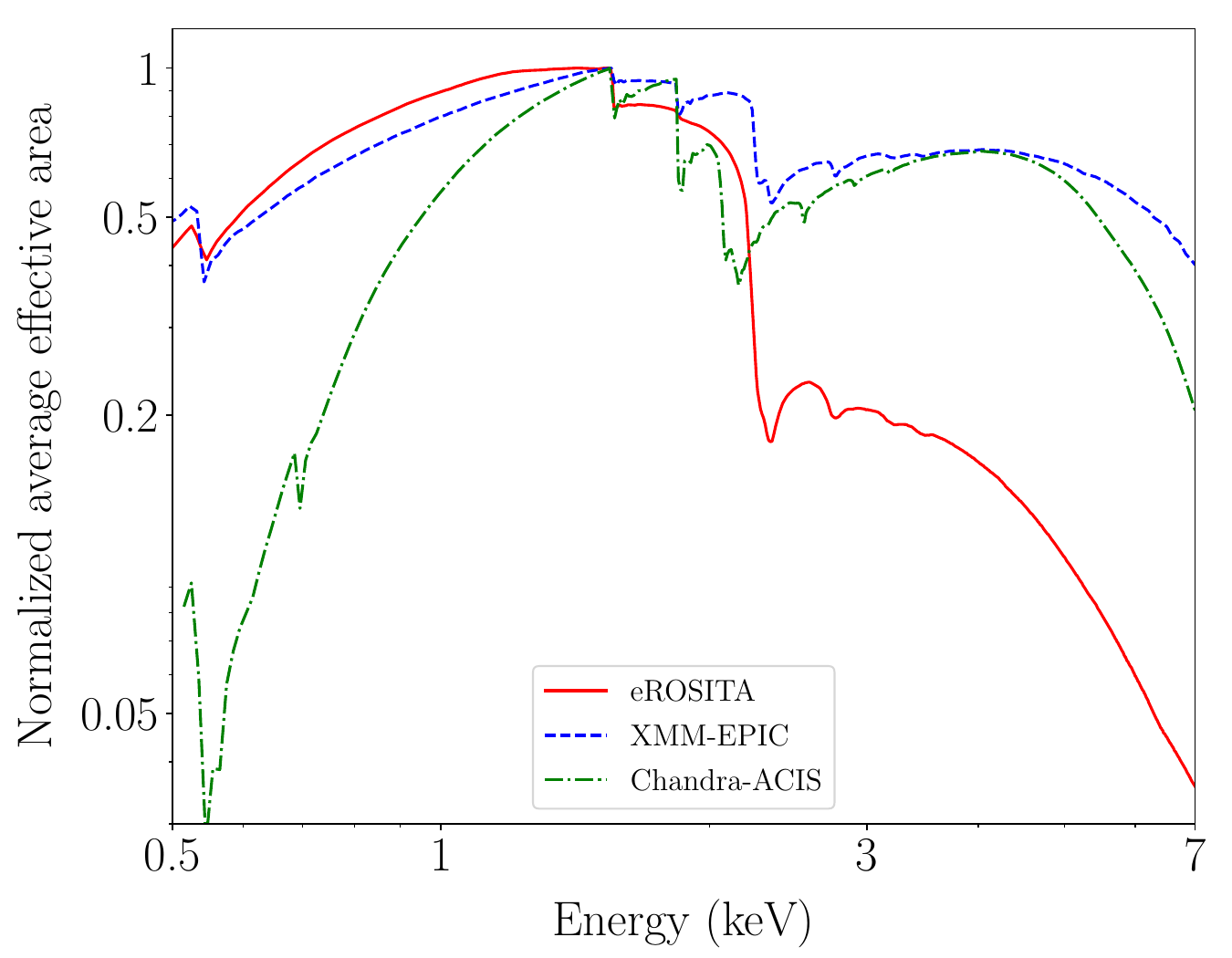}
               \caption{Normalized average effective area for eROSITA TM0 (red, solid), \textit{XMM-Newton}-EPIC (blue, dashed), and \textit{Chandra}-ACIS (green, dot dashed), as derived from the spectra used in this work.}
        \label{Effective-areas}
\end{figure}

The intracluster medium has a multi-temperature structure, which is typically a function of the distance from the cluster center. The observed spectra we use come from 2-dimensional cluster circles and annuli, which include the projection of outer cluster regions in the same line of sight. Hence, the observed spectra include several $T$ gas components that are eventually fit by a single-$T$ model. In addition, multiphase gas might be present at the same cluster radius (e.g, cold and hot gas clouds). In both of these cases, the different sensitivity of the eROSITA effective area at different energies compared to \textit{Chandra} and \textit{XMM-Newton} could result in a systematic bias in the measured single-$T$. As shown in Fig. \ref{Effective-areas}, eROSITA's effective area drops more rapidly from soft to hard X-ray energies compared to \textit{Chandra} and \textit{XMM-Newton}. Consequently, it will potentially assign more weight to the low-$T$ components of multiphase cluster gas, leading to systematically lower measured single-$T$ than the other two telescopes. Such a discrepancy could be also enhanced due to the Fe bias. The latter tends to underestimate single-$T$ measurements when multiphase gas is present, with the effect being stronger for instruments more sensitive to the soft band. The bias introduced to single-$T$ fits due to multiphase gas was discussed in detail in several past studies \citep[e.g.;][]{reiprich13}. In past cross-calibration studies, this was disfavored as the reason behind the observed calibration discrepancy between \textit{XMM-Newton} and \textit{Chandra} (e.g., S15) and other instruments. Nevertheless, it is important to also test this for the eROSITA calibration to fully understand the importance of this possible systematic.

\paragraph{Observational test}

To do so, we first divided our sample into two subsamples: cluster core $T$ and cluster annulus $T$. The former are expected to show stronger discrepancies between eROSITA and the other instruments. This is due to a higher level of multi-temperature structure in the core spectra caused by projection effects and the presence of the (often complicated) cluster core. The comparisons are performed in the full band since the broader band is expected to be more affected by the soft-hard ratio differences of the instruments' effective areas.  The $T$ distributions of the two subsamples are very similar (see Fig. \ref{T-hist}).

\begin{figure}[hbtp]
               \includegraphics[width=0.49\textwidth]{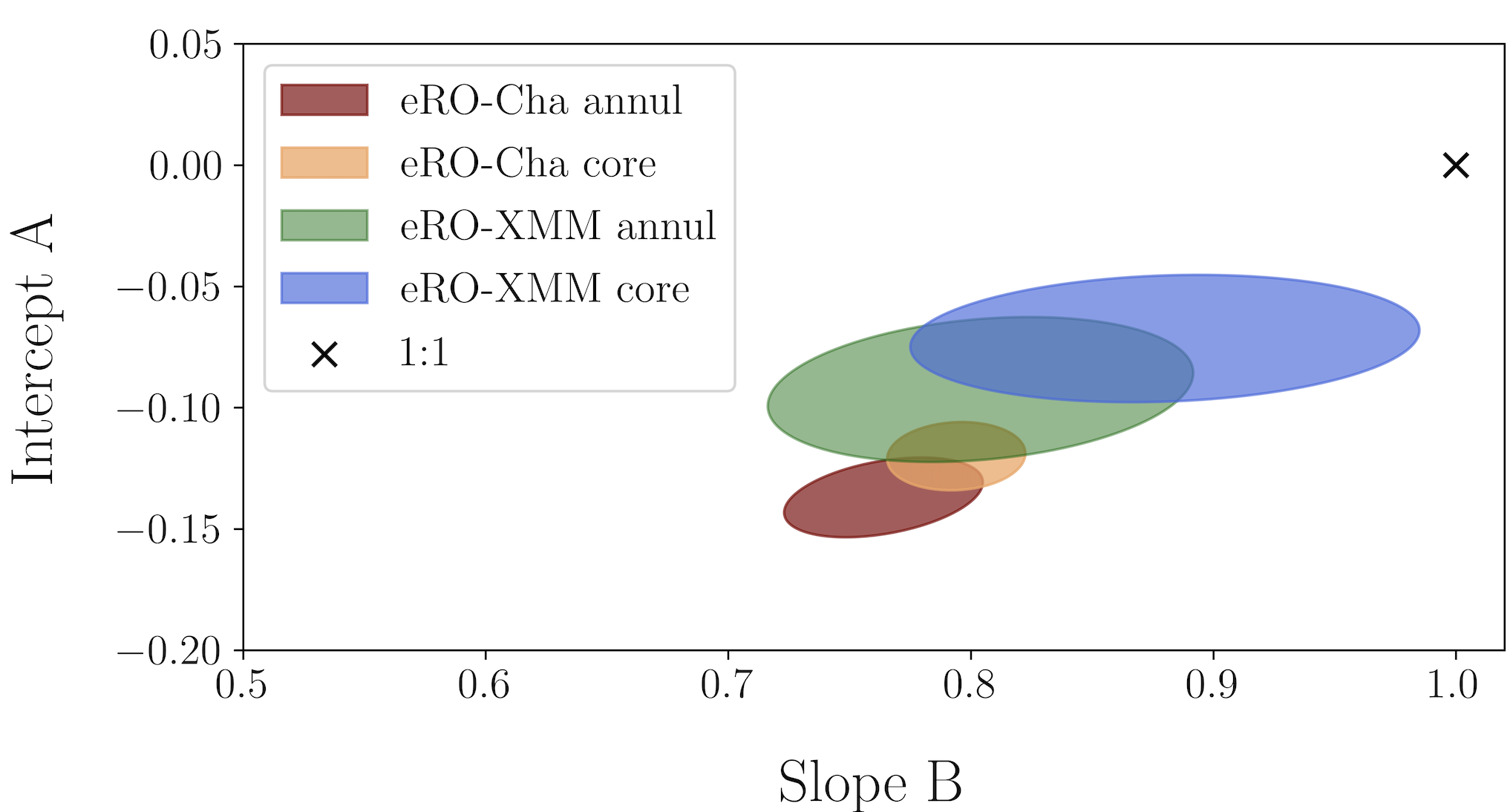}
               \includegraphics[width=0.49\textwidth]{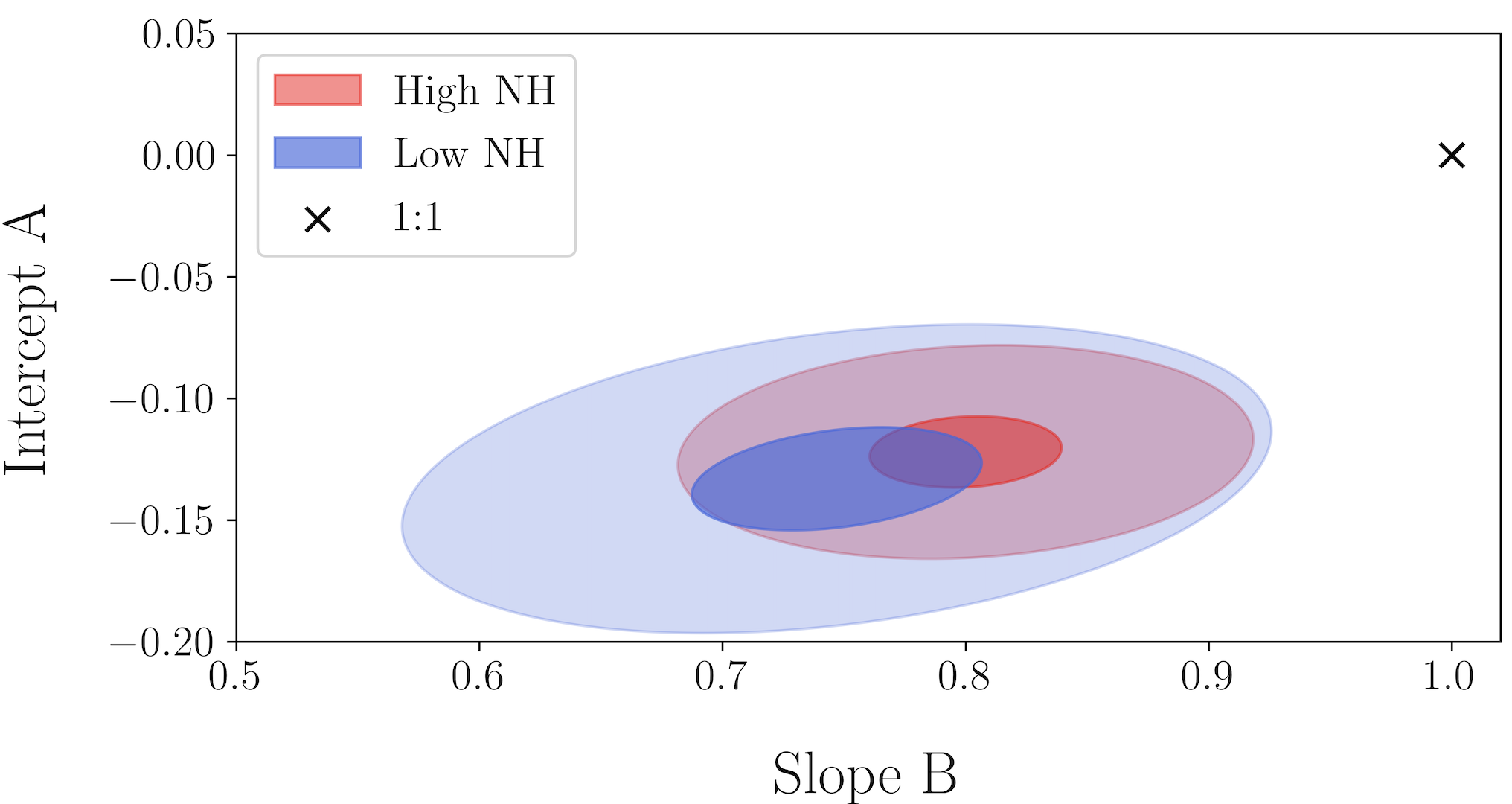}
               \caption{Confidence levels for the eROSITA-\textit{Chandra} and eROSITA-\textit{XMM-Newton} scaling relations for different subsamples, defined based on their cluster properties. \textit{Top}: $1\sigma$ (68.3$\%$) contours for the core-only and annulus-only eROSITA-\textit{Chandra} and eROSITA-\textit{XMM-Newton} scaling relations for the full band. The colors are explained in the figure legend. The best-fit results do not differ significantly for core and annuli cluster $T$ for any instrument comparison. \textit{Bottom}: $1\sigma$ (68.3$\%$) and $3\sigma$ (99.7$\%$) confidence levels for the eROSITA-\textit{Chandra} scaling relations for high $N_{\text{H}}>5.59/$cm$^2$ (pink) and low $N_{\text{H}}<2.45/$cm$^2$ (blue) clusters. The median values of the subsamples are $N_{\text{H}}=1.72/$cm$^2$ and $N_{\text{H}}=7.96/$cm$^2$, respectively. The contours correspond to the full band. The 1:1 line is represented by the black cross. The best-fit results are very similar between clusters with low and high Galactic absorption.}
        \label{core-nh-contours}
\end{figure}

Comparing the $T$ cross-calibration scaling relations for core and annulus spectra, one sees there are no statistically significant differences. As shown in the top panel of Fig. \ref{core-nh-contours}, the core versus annulus scaling relations deviate by $\lesssim 1\sigma$ for both the eROSITA-\textit{Chandra} and eROSITA-\textit{XMM-Newton} comparisons. Interestingly, the core $T$ scaling relations for both instruments are slightly shifted toward the 1:1 line (though still significantly far away) despite their more intense bias caused by the multiphase cluster gas. One would need more data to better understand if this shift is a statistical fluctuation or a true effect.

\paragraph{Simulations from past studies}
Moreover, \citet{zuhone} used hydrodynamical simulations to assess the bias that a multi-temperature structure that resembles realistic cluster temperature profiles introduces to single-$T$ fits of eROSITA. They found that spectroscopic single-thermal models underestimate $T$ by $\sim 10-15\%$ compared to the mass-weighted simulated $T$, with no strong $T$ dependence of this discrepancy. This underestimation is not enough to explain the observed $T$ differences between eROSITA and the other telescopes. On the other hand, compared to the X-ray emission-weighted simulated $T$, \citet{zuhone} found that a spectroscopic single-thermal model underestimates $T$ by $\sim 25-30\%$. If there were no such bias for \textit{Chandra} and \textit{XMM-Newton} $T$, this discrepancy would be enough to alleviate most of the tension between $T_{\text{eROSITA}}$ and the rest. However, both \textit{Chandra} and \textit{XMM-Newton} are expected to also return underestimated $T$ from single-thermal fits, but likely at a lesser degree. Until a similar analysis is performed for these two X-ray instruments, it is unclear exactly how the $T$ comparison between them and eROSITA would be affected.

A relevant test was performed by \citet{reiprich13} where they fit a single-$T$ model to a two-temperature simulated gas with a hot ($T_{\text{hot}}=8$ keV) and a cold ($T_{\text{cold}}=0.5$ keV) component. They found that \textit{Chandra}, \textit{XMM-Newton}, and eROSITA are all expected to return a best-fit $T$ lying between of the two $T$ components, but eROSITA returns the lowest $T$. The reported discrepancy is similar to the $T$ differences we observe in this work. However, to illustrate the effect, \citet{reiprich13} used extreme $T$ differences that are not present in the data we used, where the expected differences, given the measured core and core-excised $T$ and typical cluster $T$ profiles, are much smaller. Also, as noted in S15, such extreme $T$ differences of a two-phase plasma would result in high reduced chi-square values from the single-$T$ fits; this is not observed in our spectral fits. Finally, in that case one would also expect the eROSITA-\textit{Chandra} core-$T$ comparison to deviate more than the annulus-$T$ comparison, which is not the case (Fig. \ref{core-nh-contours}).

\paragraph{Simulations in this work}\label{2T-full-section}
To test the effects of multi-temperature structure on eROSITA $T$ more extensively, we used \texttt{XSPEC} to simulate multi-temperature plasma spectra, following the same approach as in \citet{reiprich13} and S15. We simulated spectra with two temperature components and fit them with a single temperature model, as for the real data. We used $T_{\text{cold}}=0.5,0.75,1$ keV and $T_{\text{hot}}=1,2,4,6$ keV, covering a wide and realistic range of $T$ combinations. The redshift and metallicity were kept fixed to $z=0.05$ and $Z=0.3\ Z_{\odot}$ respectively, which are typical values for the cluster sample we use. Moreover, we used four emission measure ratios (EM) between the cold and hot $T$ components, EM$=0.01,0.05,0.10,0.15$ (as explained later, higher EM returned unrealistic results). For every combination, we simulated 125 random spectra, with a typical number of counts for every instrument, based on the real spectra. In total, we simulated and fit 6000 spectra with a single $T$ model for eROSITA (TM0), \textit{XMM-Newton} (combined EPIC and PN-only), and \textit{Chandra} (ACIS). We assumed that the calibration of each instrument is perfect and, consequently, any differences between instruments should originate due to their different energy dependence of their effective areas.

\begin{figure*}[hbtp]
               \includegraphics[width=0.45\textwidth]{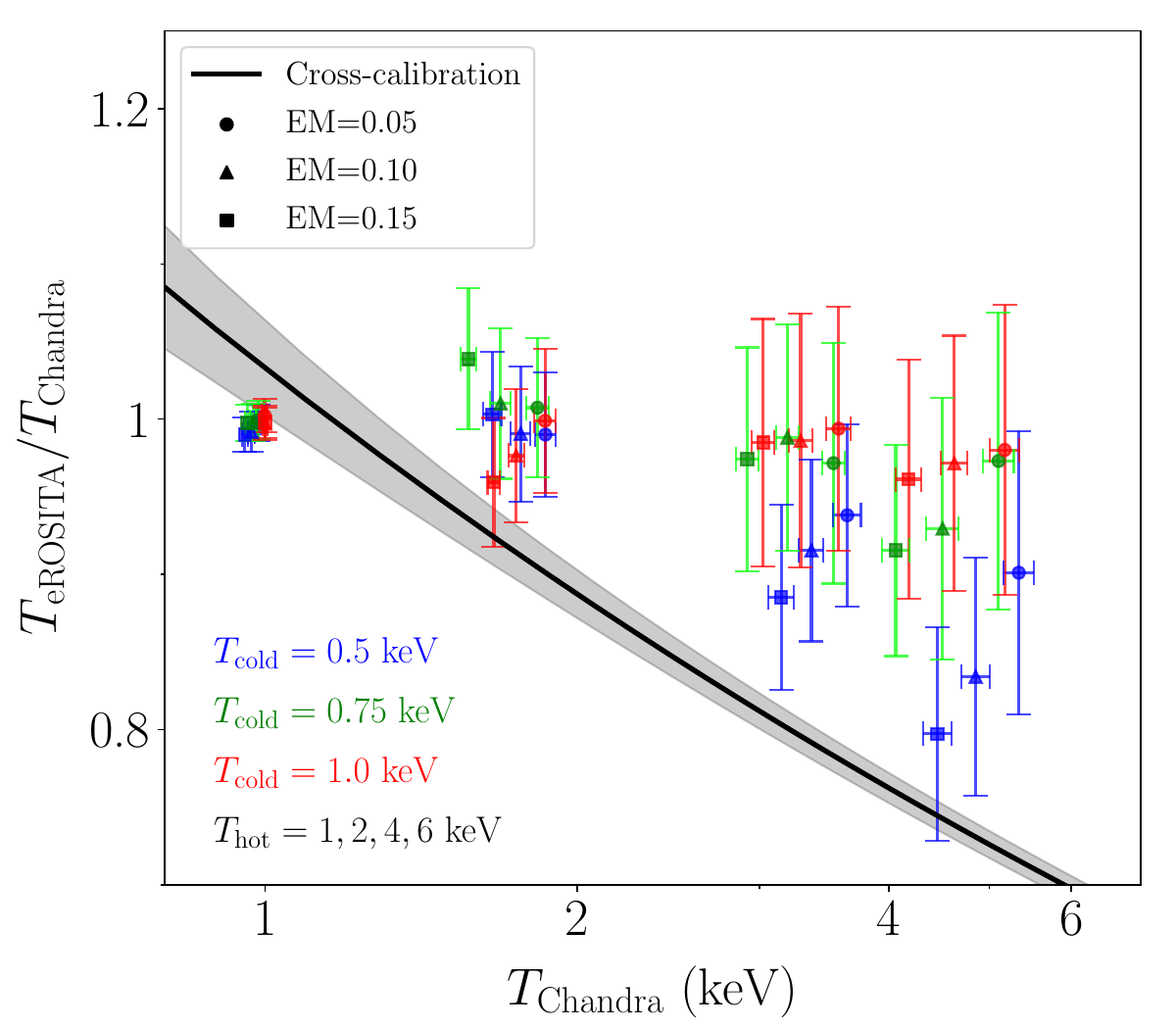}
               \includegraphics[width=0.45\textwidth]{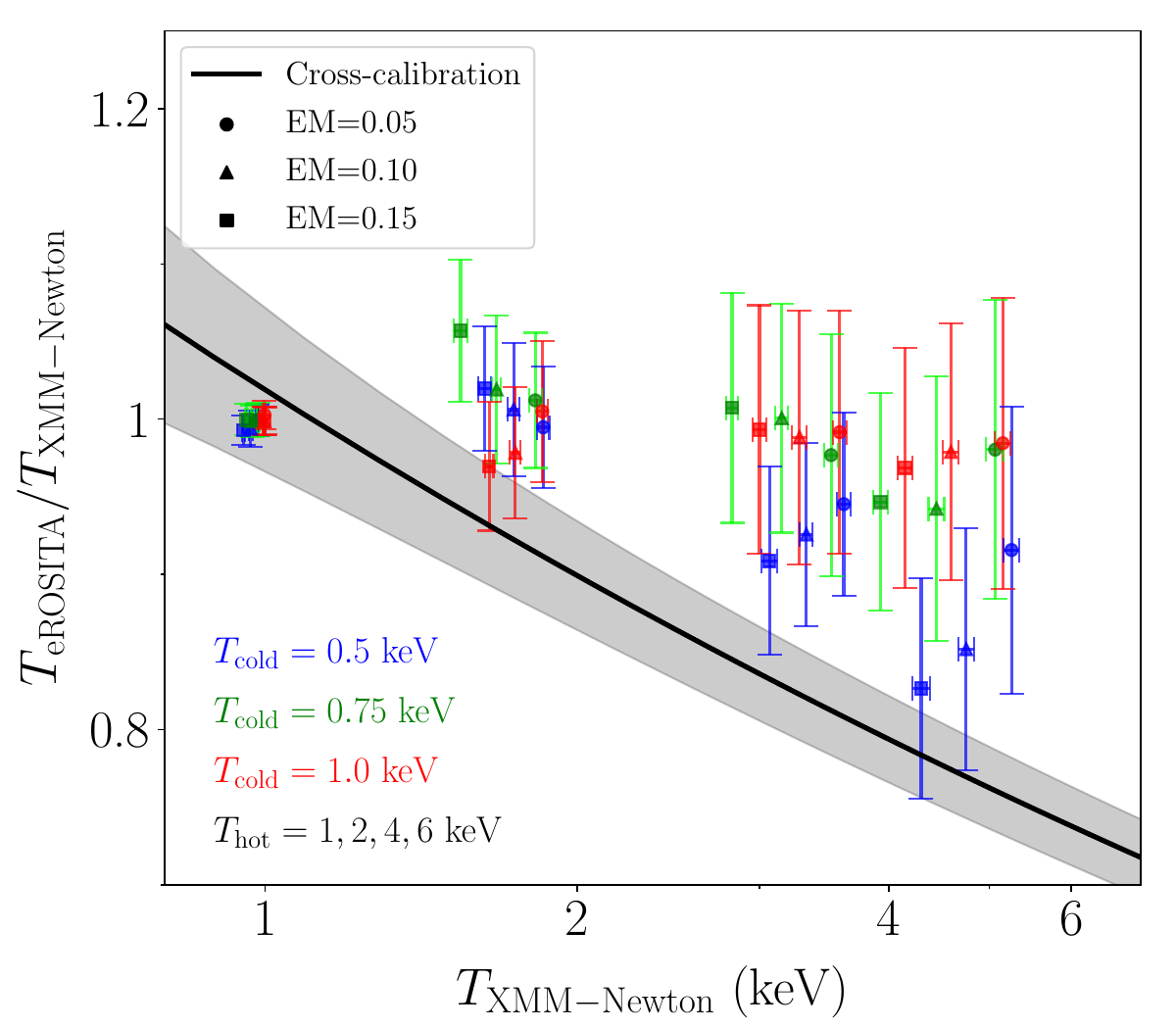}
               \caption{Ratio of the eROSITA and \textit{Chandra} (left) and eROSITA and \textit{XMM-Newton} (right) single-$T$ fits to simulated spectra with two temperature components as a function of the \textit{Chandra} and \textit{XMM-Newton} best-fit single-$T$, respectively. The blue, green, and red colors correspond to $T_{\text{cold}}=0.5, 0.75, 1$ keV, respectively. The hot component has values of $T_{\text{hot}}=1, 2, 4, 6$ keV. The circles, triangles, and squares correspond to emission measures of EM$=0.05, 0.10, 0.15$ for the cold and hot components. The black curve shows the observed $T$ difference between eROSITA-\textit{Chandra} (left) and eROSITA-\textit{XMM-Newton} (right), as presented in Table \ref{tab:comparison2}.}
        \label{2T-simul-full-band}
\end{figure*}

The average results for the full band eROSITA-\textit{Chandra} and eROSITA-\textit{XMM-Newton} comparisons are displayed in Fig. \ref{2T-simul-full-band}. It is evident that the observed cross-calibration differences are unlikely to be explained due to possible $2T$ thermal structure in the fitted spectra. although eROSITA returns slightly lower single $T$ values than \textit{Chandra} and \textit{XMM-Newton} in all cases, the $T$ difference is always $\leq 20\%$. The only combination for which the $2T$ structure could mostly explain the observed $T$ discrepancies between eROSITA and the other instruments is the most extreme case with $T_{\text{cold}}=0.5$ keV and EM$=0.15$. However, these models on average return $\chi^2\sim 1.3-1.6$, which suggests a bad fit. Such increased $\chi^2$ values are not generally seen in the real data. Subsequently, such cold, bright $T$ components in the observed spectra are disfavored as the reason for the observed $T$ discrepancy. Higher EM values return even higher $\chi^2$ values, and as such, they are not explored further. On the other hand, $T_{\text{cold}}=0.5$ keV and EM$=0.10$ models return $\chi^2\leq 1.15$, which is rather consistent with the observed $\chi^2$. Such models can alleviate most (but not all) of the $T$ tension between eROSITA-\textit{Chandra} and eROSITA-\textit{XMM-Newton}. All other combinations with $T_{\text{cold}}\geq 0.75$ keV do not show significant $T$ differences between instruments, regardless of the EM and $T_{\text{hot}}$ values. We reach the same conclusion when we only use the PN detector of \textit{XMM-Newton} to fit the spectra. Finally, we repeat the analysis using the hard band fits only. In this case, the observed cross-calibration differences cannot be explained by the $2T$ structure in the simulated data since eROSITA returns very similar single $T$ results to \textit{Chandra} and \textit{XMM-Newton}, regardless of the exact model values. The results are presented in Appendix \ref{2T-hard-section} and Fig. \ref{2T-simul-hard-band}. 

\paragraph{Conclusions on multi-temperature structure bias}
Given all the above, the bias introduced by multiphase gas and different effective areas is currently disfavored as the main cause for the observed $T$ discrepancy between eROSITA and \textit{Chandra}/\textit{XMM-Newton}. This is consistent with what past studies found for other instrument comparisons (e.g., S15). However, to accurately quantify the exact level of bias introduced due to more complicated multiphase gas structure, more work is needed.

\subsubsection{Bias from Galactic absorption}

The X-ray absorption, proxied by the $N_{\text{H}}$ parameter, has a much stronger influence on the soft part of the spectra ($\lesssim 1.5$ keV) than the hard part, for which it is almost irrelevant. Given the different energy-dependent shapes of the effective areas of the telescopes, different $N_{\text{H}}$ values can affect each telescope's $T$ measurements differently, introducing an $N_{\text{H}}$-dependent systematic bias in the $T$ comparison for the soft and full bands (while the hard band would remain unaffected from such bias). To address this possibility, we compare the full band eROSITA-\textit{Chandra} scaling relations between the $\frac{1}{3}$ of the sample with the lowest and highest $N_{\text{H}}$\footnote{The subsamples for the eROSITA-\textit{XMM-Newton} comparison are too small to return meaningfully constrained results.} ($N_{\text{H}}<2.45/$cm$^2$ and $N_{\text{H}}>5.59/$cm$^2$ respectively). The two subsamples do not show significant differences in their $T$ distributions and any difference is expected to come from the different $N_{\text{H}}$.

As displayed in the bottom panel of Fig. \ref{core-nh-contours}, the effect different $N_{\text{H}}$ values have on the eROSITA-\textit{Chandra} cross-calibration is not statistically significant, with the low and high $N_{\text{H}}$ subsamples agreeing within $\lesssim 1\sigma$. Therefore, this further supports the notion that the different effective area dependence on photon energy per telescope is not the cause of the $T$ discrepancy.

\subsection{Soft band versus hard band $T$ per instrument}\label{main_soft_vs_hard}
To better understand the cause of the $T$ discrepancy dependence on the used energy band, we explore which instrument is the most affected by the energy band change. For a self-consistently calibrated instrument, and assuming multiphase gas effects are not dominant, the different energy bands should result in statistically similar $T$. To test this, we compare the soft and hard band temperatures ($T_{0.5-4\ \mathrm{keV}}$ and $T_{1.5-7\ \mathrm{keV}}$ respectively) for all three instruments. The detailed results are given in Appendix \ref{Sect_soft_vs_hard} (see Fig. \ref{soft-vs-hard} and Table \ref{soft-vs-hard-table}). 

eROSITA shows the largest deviation between all instruments, with $T_{0.5-4\ \mathrm{keV}}$ being on average $29\pm 4\%$ higher than $T_{1.5-7\ \mathrm{keV}}$. This difference only weakly depends on $T$ at a $<1\sigma$ level. Specifically, for eROSITA:
\begin{equation}
\log_{10}{\dfrac{T_{1.5-7\ \text{keV}}}{3\ \text{keV}}}\approx -0.148+0.928\times\log_{10}{\dfrac{T_{0.5-4\ \text{keV}}}{3\ \text{keV}}}.
\end{equation}
On the other hand, \textit{Chandra} shows the highest level of agreement between $T_{0.5-4\ \mathrm{keV}}$ and $T_{1.5-7\ \mathrm{keV}}$, with the former being $8\pm 1\%$ higher than the latter on average. There is no noticeable dependence with $T$. Specifically, for \textit{Chandra}:
\begin{equation}
\log_{10}{\dfrac{T_{1.5-7\ \text{keV}}}{4.5\ \text{keV}}}\approx -0.034+0.988\times\log_{10}{\dfrac{T_{0.5-4\ \text{keV}}}{4.5\ \text{keV}}}.
\end{equation}
Finally, \textit{XMM-Newton} shows an opposite behavior compared to eROSITA and \textit{Chandra}, with $T_{0.5-4\ \mathrm{keV}}$ being on average $17\pm 2\%$ lower than $T_{1.5-7\ \mathrm{keV}}$. Again, this discrepancy is not a strong function of $T$. Specifically, for \textit{XMM-Newton}:
\begin{equation}
\log_{10}{\dfrac{T_{1.5-7\ \text{keV}}}{3\ \text{keV}}}\approx 0.068+0.936\times\log_{10}{\dfrac{T_{0.5-4\ \text{keV}}}{3\ \text{keV}}}.
\end{equation}
From all the above, it is evident that eROSITA temperatures are the most affected when one changes the used energy band. The nature of the change suggests that, at least for eROSITA and \textit{Chandra}, multi-temperature gas does not play a dominant role on $T$, otherwise one would generally expect $T_{1.5-7\ \mathrm{keV}}>T_{0.5-4\ \mathrm{keV}}$. Overall, these results indicate that the increased discrepancy in the cross-calibration between eROSITA and \textit{Chandra}/\textit{XMM-Newton} in the hard band might be mostly attributed to possible systematic calibration uncertainties at higher X-ray energies. However, this remains to be confirmed or challenged by future analysis.

\subsection{Bias introduced from eROSITA TM8 versus TM9}\label{TM8-TM9}
Next, we explore the effect that the inclusion of eROSITA TM9 data has on the $T$ measurements. As discussed in Sect. \ref{spectral}, TM9 cameras are affected by the light leak. In theory, this could potentially alter the signal at low X-ray energies while the harder energies remain unaffected, biasing the fitted $T$. This would not explain the observed $T$ discrepancies in the hard band, but it is worth checking if it has any effect on the other bands. To test the consistency of the two telescope modules, we fit the TM9-TM8 $T$ scaling relation. We find there is a general agreement between the TM8 and TM9 $T$ measurements at a $\lesssim 2.5\sigma$ level, with $\xi=-0.05$. Their scaling relation parameters for the full band fit are $A=-0.020\pm 0.015$, $B=0.921\pm 0.038$, and $\sigma_{\text{tot}}=0.193\pm 0.010$, while the intrinsic scatter is negligible (<1\%). The agreement improves to $\lesssim 2\sigma$ for the soft band, mostly due to the increase of measurement uncertainties. 

Even though it is not statistically significant, TM9 measures slightly lower $T$ than TM8 for $T\gtrsim 3$ keV clusters ($5\%$ and $10\%$ difference for $T_{\text{TM8}}=3$ keV and $T_{\text{TM8}}=6$ keV respectively). To ensure our findings are insensitive to this small TM9 effect, we refit the eROSITA-\textit{Chandra} and eROSITA-\textit{XMM-Newton} scaling relations excluding TM9 for the full and soft bands. We find very small differences with the default analysis. For the eROSITA-\textit{Chandra} and eROSITA-\textit{XMM-Newton} scaling relations both $A$ and $B$ change by $\lesssim 2\%$ and $\lesssim 2.5\%$ respectively. Consequently, TM9 does not introduce any bias to the $T$ comparisons.

\subsection{Bias from non-Gaussian scatter and asymmetric $T_{\text{eROSITA}}$ uncertainties}\label{nongaussian}

\subsubsection{Non-Gaussian scatter}
Due to the decrease of eROSITA's effective area at $\gtrsim 2$ keV, clusters with higher $T_{\text{eROSITA}}$ show larger uncertainties $\sigma_{T_{\text{eROSITA}}}$. For similar \textit{Chandra} or \textit{XMM-Newton} $T$, upscattered $T_{\text{eROSITA}}$ show higher uncertainty on average than downscattered $T$ values, making the total scatter slightly non-Gaussian. If clusters with truly high, upscattered $T_{\text{eROSITA}}$ carry larger statistical uncertainties, the fit will give more weight to lower, downscattered $T$ data points, shifting the best-fit relation away from the 1:1 relation. To test if this non-Gaussianity of $\sigma_{T_{\text{eROSITA}}}$ has a significant effect on the scaling relation constraints, we refit the eROSITA-\textit{Chandra}/\textit{XMM-Newton} scaling relations for all bands, without taking into account $\sigma_{T_{\text{eROSITA}}}$, hence, we assign the same weight to all data points. 

For the eROSITA-\textit{Chandra} cross-calibration, $A$ increases by $\approx 0.007-0.013$ dex ($1.5-3\%$ in linear space) for all bands. $B$ remains almost unchanged for the full and soft bands while it decreases by 3\% for the hard band. 

For the eROSITA-\textit{XMM-Newton} cross-calibration the effect of non-Gaussianity is stronger. $A$ increases by $0.044$ dex ($10\%$), $0.059$ dex ($13.5\%$), and $0.032$ dex ($7.6\%$) for the full, soft, and hard bands respectively. $B$ increases by $1.4\%$ and $0.3\%$ for the full and soft bands and by $24\%$ for the hard band. Non-Gaussianity has no significant effect on $B$ for the full and soft bands. However, it causes an underestimation of $A$, resulting in an overestimation of the cross-calibration differences between eROSITA and \textit{XMM-Newton}. Nevertheless, given the parameter uncertainties, this shift of $A$ is within $\lesssim 2\sigma$ from the default analysis. Of course the performed test here is quite conservative and demonstrates the maximum impact of non-Gaussianity possible. 

To conclude, the non-Gaussianity of $\sigma_{T_{\text{eROSITA}}}$ does not introduce any significant bias in the scaling relations between eROSITA and \textit{Chandra}. On the other hand, it does partially overestimate the discrepancy between eROSITA and \textit{XMM-Newton}, however, not at a statistically significant level. Moreover, this overestimation bias cannot fully alleviate the tension between $T_{\text{eROSITA}}$ and $T_{\text{XMM}}$ values. Non-Gaussianity effects become less relevant as more eROSITA counts are available for determining  $T_{\text{eROSITA}}$ and the $T_{\text{eROSITA}}$ measurement uncertainties become smaller than the intrinsic scatter.

\subsubsection{Symmetrization of asymmetric $T_{\text{eROSITA}}$ uncertainties}

In this work we adopt a widely used approach by symmetrizing $T$ uncertainties in linear-space in order to convert them to log-space (through error propagation) before we fit scaling relations. Slightly asymmetric $T$ uncertainties normally converge to  symmetry in log-space; thus, this approach is not expected to introduce significant biases most of the time. For instance, the $T_{\text{XMM}}$ and $T_{\text{Chandra}}$ uncertainties used in this work are nearly symmetric, with the average log-uncertainty difference being $\Delta\left[{\sigma_{\log{T}}}\right]^+_-\approx 0.002$ dex and $\Delta\left[{\sigma_{\log{T}}}\right]^+_-\approx 0.006$ dex for the full (or soft) and hard bands respectively, for both telescopes. Hence, their symmetrization has a nearly zero effect on log-space uncertainties. On the other hand, the $T_{\text{eROSITA}}$ uncertainties are larger and tend to be more asymmetric, that is, $\Delta{\left[{\sigma_{\log{T}}}\right]}^+_-\approx 0.096$ dex and $\Delta\left[{\sigma_{\log{T}}}\right]^+_-\approx 0.161$ dex for the full (or soft) and hard bands, respectively. Consequently, symmetrizing the $T_{\text{eROSITA}}$ uncertainties might bias our best-fit estimates for the cross-calibration scaling relations.

To test this, we refit all scaling relations without symmetrizing $\sigma_{\log{T_{\text{eROSITA}}}}$. Instead, we consider two $T_{\text{eROSITA}}$ log-errorbars: $\sigma_{\log{T}}^+=\log{T^+}-\log{T_{\text{best}}}$ and $\sigma_{\log{T}}^-=\log{T_{\text{best}}}-\log{T^-}-$, where $\log{T_{\text{best}}}$ is the best-fit $T_{\text{eROSITA}}$ and $T^+$ and $T^-$ are its upper and lower 68.3\% uncertainties respectively. For every tested $A$ and $B$ combinations in Eq. \ref{likelihood} and \ref{uncertainties-sum}, we consider $\sigma_{\log{T_{\text{eROSITA}}}}=\sigma_{\log{T}}^+$ (or $\sigma_{\log{T_{\text{eROSITA}}}}=\sigma_{\log{T}}^-$) when $T_{\text{eROSITA}}$ is downscattered (or upscattered) compared to the fitted line (i.e., when $\log{T^{'}_{\text{Y},i}}-A-B\times \log{T^{'}_{\text{X},i}}<0$ or $>0$ respectively in Eq. \ref{likelihood}). In other words, we consider only the errorbar toward the fitted line as the standard errorbar, ignoring the errorbar on the opposite side of the data point.

For the full and soft band eROSITA-\textit{Chandra} scaling relations, $A$ and $\sigma_{\text{intr}}$ change by $\leq 0.004$ dex ($\leq 1\%$) and $\leq 0.008$ dex ($\leq 1.8\%$)\footnote{Here with the percentage change, we mean that $\sigma_{\text{intr}}$ goes from, for example, 10.6\% to 12.4\% for the full band.} respectively, while $B$ decreases by $1.6\%$ and $3.5\%$. It is evident that these changes are nearly negligible compared to the overall tension of the two instruments, with only $B$ showing a $\approx 0.5-1\sigma$ shift. For the hard band, $A$ and $B$ both increase by $\approx 5.5\%$, while the scatter remains similar to the default case. These are again $\lesssim 1\sigma$ shifts for both $A$ and $B$. Therefore, it is safe to conclude that the symmetrization of temperature uncertainties has an insignificant effect on the eROSITA-\textit{Chandra} scaling relations.

For the eROSITA-\textit{XMM-Newton} scaling relations, the effect of asymmetric $\sigma_{\log{T_{\text{eROSITA}}}}$ has a somewhat larger effect, but still statistically insignificant. Both $A$ and $B$ increase by $\approx 6\%$, $\approx 3\%$, and $\approx 8.5\%$ for the full, soft, and hard band respectively. All these changes are within $< 1\sigma$. The similarity in the $A$ and $B$ shifts is noteworthy and it is attributed to some correlation between the two parameters. Additionally, $\sigma_{\text{intr}}$ increases by $<0.01$ dex ($<2.3\%$) for the full and soft bands, while it doubles for the hard band. The latter is the only parameter that changes at a $>1\sigma$ level. Overall, it is shown that the use of asymmetric temperature uncertainties only shifts the best-fit scaling relations within the statistical noise.

\subsection{Selection bias due to low $T/{\sigma_T}$}\label{selection-bias}

Due to its low sensitivity at $\gtrsim 2.5$ keV, eROSITA returned a $T/{\sigma_T}<1$ measurement for $4\%$ of the \textit{Chandra} and $12\%$ of the \textit{XMM-Newton} cluster $T$ for the full and soft bands. For the hard band, this fraction increased to $37\%$ and $58\%$ respectively. These clusters were excluded from the default analysis. With eROSITA, high $T$ clusters are more likely to show lower $T/\sigma_T$ on average than low $T$ clusters for the same counts. Consequently, excluding low $T/\sigma_T$ clusters might lead to preferentially select clusters for which $T_{\text{eROSITA}}$ was measured low. This selection effect could potentially shift the scaling relations further away from the equality line. To test this, we refit the scaling relations while considering all $T$ measurements independently of their $T/\sigma_T$ value. 

The newly added data points have nearly no effect on the eROSITA-\textit{Chandra} scaling relations for the full and soft bands. Both the new $A$ and $B$ best-fit values increase by $<1\%$ compared to the default best-fit values. This occurs since the added data with $T/\sigma_T<1$ values only marginally increase the sample size compared to the default case. Additionally, they carry very little statistical weight. For the hard band, the overall effect on $A$ and $B$ is similar, increasing only by $3\%$ and $1\%$ respectively. In all bands, the total scatter almost doubles when the low $T/\sigma_T$ values are added to the sample. 

For the eROSITA-\textit{XMM-Newton} scaling relations in the full and soft bands, both $A$ and $B$ once again change by $<1\%$ compared to the default best-fit values. For the hard band, the sample size increases significantly. Nevertheless, $A$ decreases by 4\% while $B$ increases by 9.5\%. These changes are within $0.5\sigma$ and do not practically improve the agreement between eROSITA and \textit{XMM-Newton}. 

It is evident that the data selection of $T/\sigma_T>1$ introduces no significant bias to the final results, while it reduces the total scaling relation scatter. However, it is possible that when future eRASS data are used, more high-$T_{\text{eROSITA}}$ measurements will have higher $T/\sigma_T$, and hence, a larger effect on the best-fit scaling relations. This might cause the latter to shift somewhat closer to the equality line. Nonetheless, as shown in Sect. \ref{nongaussian}, this is unlikely to alleviate the cross-calibration differences between eROSITA and \textit{Chandra}/\textit{XMM-Newton}.

\subsection{eROSITA's half-energy width and PSF effects}
%%%%%%%%
%% COMMENT BY JSS: The HEW is not around 16 arcsec!!! This is likely the on-axis value. It's around 34 arcsec when survey-averaged. See Fig A.1 in the main eRASS1 catalogue paper
%%%%%%%%
The on-axis half-energy width (HEW) of eROSITA is $\approx 15.5-16.5"$ between 1.49 keV and 8.04 keV \citep{predehl21}. For the eROSITA survey data, such as eRASS1, the average HEW increases to $\approx 26-29"$ \citep{Brunner2022}. This is only slightly larger than the \textit{XMM-Newton} on-axis HEW ($\approx 15-16"$), but significantly larger than the \textit{Chandra} on-axis HEW ($\approx 0.5-1"$). If the spectra extraction regions are not significantly larger than eROSITA's HEW then the emission from the cluster core might scatter to the core-excised annulus spectra. At the same time, \textit{Chandra} would not suffer from this due to its much lower HEW. For cool core clusters, this would bias the core-excised $T$ toward lower values. This is a potential problem for distant clusters with small apparent $R_{500}$. However, our sample consists of nearby clusters with large apparent $R_{500}$. Characteristically, the median apparent radius of the cluster cores ($<0.2R_{500}$) is $2.3'$, which is five times the eROSITA survey HEW. At the same time, only four clusters of the eROSITA-\textit{Chandra} comparison subsample ($3.6\%$ of the subsample) have a cluster core of $<65"$, that is,  $\lesssim 2.5\times$ the eROSITA survey HEW. These clusters do not show any special behavior compared to the rest of the sample. In conclusion, it is clear that the eROSITA HEW has nearly no effect on the eROSITA-\textit{Chandra} $T$ comparison. The same conclusion was reached by S15 for the \textit{XMM-Newton}-\textit{Chandra} $T$ comparison and a similar cluster population.

\subsection{Potential systematic biases from background treatment}\label{bgd-biases}
When one constrains cluster $T$, the treatment of both CXB and PIB is important. Potential biases in the background treatment can propagate to the final $T$ and the resulting scaling relations. This is particularly true for outter cluster regions with low X-ray surface brightness. In this work, we deal with central cluster regions with generally high surface brightness. The constrained $T$ of these regions are not expected to be significantly affected by mild changes in the background treatment, or even mild background biases. Nevertheless, here we discuss the possible existence of such potential biases and their impact on the final cross-calibration scaling relations.

\subsubsection{eROSITA}
For eROSITA, both the CXB and PIB were modeled together with the source spectra, after they were robustly constrained as described in Sect. \ref{spectral}. Therefore, there is no suspected bias in the background treatment of eROSITA data. Additionally, due to the large $T_{\text{eROSITA}}$ uncertainties, even (currently unknown) mild background biases would have a negligible impact on the final scaling relations.

\subsubsection{\textit{Chandra}}
For \textit{Chandra}, the CXB is modeled simultaneously with the cluster spectra, using both \textit{Chandra} and RASS data (see Sect. \ref{spectral-Chandra-xmm}). Even in the case of mild calibration differences between ROSAT and \textit{Chandra} fluxes, the inclusion of RASS data is not expected to bias the fitted $T_{\text{Chandra}}$ due to the high emissivity of the used cluster regions and the large number of resulting \textit{Chandra} counts, which eventually drive the simultaneous cluster+CXB fit. Moreover, the exact CXB values only weakly affect $T_{\text{Chandra}}$ in central cluster regions. Recently, \citet{rossetti} showed how the normalizations of the CXB components change if one uses RASS data along with \textit{XMM-Newton} data. They showed that when the CXB is constrained from cluster-free sky regions (as done for \textit{Chandra} in our work, where the CXB is constrained $>1^{\circ}$ from the cluster), the use of RASS data does not significantly change the CXB constraints. Even more importantly, they showed that the inclusion of RASS data has no effect on the final $T$ from regions where the source emission is higher than the background (which is clearly the case in our work). 

Furthermore, the \textit{Chandra} PIB is relatively stable with time and generally well-known. For the vast majority of \textit{Chandra} source spectra we analyzed, the PIB level was lower than the source emission even at the highest end of the full band range (i.e., $5-7$ keV). As such, subtracting PIB from the source spectra is followed by most studies using \textit{Chandra} data and there are no indications this causes significant biases. From all the above, it is clear that there is no reason to believe that the background treatment noticeably biases $T_{\text{Chandra}}$.

\subsubsection{\textit{XMM-Newton}}\label{XMM-bgd}
For \textit{XMM-Newton}, the CXB was constrained using only \textit{XMM-Newton} data as described in Sect. \ref{spectral-Chandra-xmm} and then fitted simultaneously with the source spectra. This is a commonly used approach and it is not expected to bias the CXB estimates. As discussed before, even if we somewhat shift the normalizations of the CXB components, $T_{\text{XMM}}$ would mostly be insensitive to these changes due to the bright cluster regions we analyze. The insensitivity of fitted $T$ to mild changes of the CXB for high surface brightness regions was also clearly shown in \citet{rossetti}. 

On the other hand, the PIB was subtracted from the source spectra, with Gaussian lines added to the model to account for any fluorescence line residuals after the PIB subtraction. However, the possible noise of the PIB normalization is not accounted for. This might introduce some bias to the fitted $T_{\text{XMM}}$, or underestimate the $T_{\text{XMM}}$ uncertainties, if the PIB estimation is biased or uncertain. To test the effect of an uncertain PIB subtraction on $T_{\text{XMM}}$ and eventually on the eROSITA-\textit{XMM-Newton} scaling relations, we perform the following. We randomly select 15 clusters (with 30 measured $T_{\text{XMM}}$) and fluctuate the PIB by $\pm 5\%$ in all EPIC detectors. We then refit $T_{\text{XMM}}$. We find that, for the cluster core regions, $T_{\text{XMM}}$ fluctuates by $<1\%$ on average for $80\%$ of the cores and $<2.5\%$ for the rest. Therefore, the effect of an uncertain (or slightly biased) PIB subtraction would be nearly negligible for the core regions, which comprise half of the sample used to constrain the eROSITA-\textit{XMM-Newton} scaling relations. For the annulus $T_{\text{XMM}}$, the average shift was $\Delta T_{\text{XMM}}\sim 3.5\%$, which is typically $<1\sigma$. Moreover, the $T_{\text{XMM}}$ shift was not systematic across all clusters when the PIB level was increased or decreased, which further decreases the impact of the PIB uncertainty on the final scaling relations. The $\Delta T_{\text{XMM}}$ results were very similar across all three bands. This test suggests that subtracting the PIB from high surface brightness cluster regions does not significantly impact the constrained $T_{\text{XMM}}$ and final eROSITA-\textit{XMM-Newton} scaling relations.

Another potential bias of the \textit{XMM-Newton} PIB treatment is that the unexposed corners of the PN detector are not fully shielded from FOV photons and soft protons, as shown in \citet{marelli}, where they mask the inner $905"$ of the FOV. They also showed that the contamination is lower as one moves to the outer edges of the unexposed corners. Recently, \citet{rossetti} showed that the $>905"$ count-rate in the $10-14$ keV band correlates very well between the PN and MOS2 detectors of \textit{XMM-Newton} for data with low solar flare contamination, that is, the PIB estimate of \textit{XMM-Newton}/PN is unbiased in this band when soft proton contamination is low. In our analysis, we mask the inner $>925"$ of the \textit{XMM-Newton}/PN detector, which ensures a slightly lower level of contamination, and use the count-rate in the $2.5-5$ keV band to determine the rescaling factor from the FWC data. To check if our PIB count-rate is overestimated due to residual contamination we perform the following. Firstly, we estimate the PIB rescaling factor based on the $10-14$ keV band from the same $>925"$ region, as done for the default analysis. We did so for 25 \textit{XMM-Newton} pointings with different levels of solar flare contamination, quantified by the IN/OUT ratio (see Appendix \ref{IN/OUT test} for details). In Fig. \ref{IN-OUT-plots} we show that our default PIB estimates are unbiased compared to the $10-14$ keV band for IN/OUT$\lesssim 1.1$, which applies to $81\%$ of our \textit{XMM-Newton} pointings (43 out of 53). At the highest-end of acceptable soft proton contamination, that is, $1.1<$IN/OUT$<1.15$, our default PIB level estimates for the \textit{XMM-Newton}/PN camera are $\sim 8\%-15\%$ overestimated compared to the $10-14$ keV case. Our results are consistent with the ones from \citet{rossetti}, although our method for probing the contamination level slightly differs.

To quantify the effect on the final eROSITA-\textit{XMM-Newton} scaling relations that the mild \textit{XMM-Newton}/PN PIB bias would have, we perform the following. We consider the ten observations with $1.1<$IN/OUT$<1.15$ and reduce the PIB level of the \textit{XMM-Newton}/PN spectra by $8\%-15\%$, according to the individual IN/OUT ratios. We then refit the full \textit{XMM-Newton} $T_{\text{XMM}}$ and the eROSITA-\textit{XMM-Newton} scaling relations. We found that the latter show completely negligible changes, $<0.3\sigma$ for all parameters (see Appendix \ref{IN/OUT test} for details). Consequently, we confirm that the residual contamination of the unexposed \textit{XMM-Newton}/PN corners has no effect on the final results of this work. 

\subsection{Consistency of spectroscopic $T_{\text{eROSITA}}$ with the eRASS1 cluster catalog}\label{eRASS1-scal-rel-section}

The eRASS1 galaxy cluster catalog \citep{bulbul23} includes cluster $T$ measurement determined by the MultiBand Projector 2D (MBProj2D) software\footnote{\url{https://github.com/jeremysanders/mbproj2d}}\citep{sanders18}. MBProj2D is a tool that uses X-ray images of galaxy clusters in different, independent energy bands, and through forward-modeling, constrains X-ray cluster properties such as the surface brightness and temperature profiles. The cluster $T$ provided in the eRASS1 galaxy cluster catalog is constrained using seven independent energy bands between 0.3-7.0 keV and within the entire $R_{500}$ of the cluster \citep[for more details see][]{bulbul23}.

To check the consistency between our spectroscopic $T_{\text{eROSITA}}$ and the official eRASS1 cluster catalog $T_{\text{eRASS1}}$, and to provide conversion functions between the latter and $T_{\text{Chandra}}$ and $T_{\text{XMM}}$, we study the scaling relations between $T_{\text{eRASS1}}$  and the other $T$ measurements. It is crucial to stress here that this comparison does not serve as a cross-calibration test between eROSITA and other instruments. In these comparisons, the cluster area used, the methodology (MBproj2D $T$ versus spectroscopic $T$), and the energy bands used for $T_{\text{eRASS1}}$ and$T_{\text{Chandra}}$/$T_{\text{XMM}}$ are all different. The sole purpose of this comparison is to test the applicability of our main results for the eRASS1 cluster catalog. Since $T_{\text{eRASS1}}$ refers to the entire $R_{500}$, we can only compare with a single $T$ per cluster for \textit{Chandra}, \textit{XMM-Newton}, and spectroscopic eROSITA; that is, with core-only or annulus-only $T$. The comparison is performed only for the full band.

The detailed results are presented in Appendix \ref{eRASS1-comparison}. In a nutshell, we find excellent agreement between our spectroscopic $T_{\text{eROSITA}}$ and $T_{\text{eRASS1}}$, especially when we use the core-only $T_{\text{eROSITA}}$, with the comparison matching the 1:1 line. When we compare $T_{\text{eRASS1}}$ to $T_{\text{Chandra}}$ and $T_{\text{XMM}}$, the results are consistent (within the uncertainties) with the main results presented in Sect. \ref{Chandra-single-pow-results} and \ref{xmm-single-pow-results}. Thus, we conclude that our results can be safely used for converting (to) $T_{\text{eRASS1}}$ as well.

\subsection{Consistency with past studies and indirect eROSITA comparison with other X-ray telescopes}
\subsubsection{\textit{Chandra}-\textit{XMM-Newton} and eROSITA-\textit{XMM-Newton}(MOS/PN only)}
The only previous eROSITA cross-calibration test using a cluster sample was performed against \textit{XMM-Newton} by \citet{turner}, who used $T$ measurements from eight common eFEDS and \textit{XMM-Newton} Cluster Survey (XCS) clusters. Despite the very small number of available data points and the different energy ranges used for eROSITA and \textit{XMM-Newton} spectral analysis, they found $T_{\text{eROSITA}}=0.75^{+0.10}_{-0.08}\times T_{\text{XMM}}$, for a fixed slope of unity. This is completely consistent with our result. As discussed in Sect. \ref{intro}, there are also a few studies that used individual cluster $T$ measurements to compare eROSITA with other instruments \citep[e.g.;][]{liu23}. However, a direct comparison of our results with these studies is not helpful since single $T$ comparisons can scatter significantly around the average behavior we present in this work.

Furthermore, based on our findings, we can 1) perform an indirect consistency test with past \textit{Chandra}-\textit{XMM-Newton} cross-calibration studies and 2) evaluate the cross-calibration of eROSITA with the individual \textit{XMM-Newton}/MOS and \textit{XMM-Newton}/PN detectors.

Using the full band eROSITA-\textit{Chandra} and eROSITA-\textit{XMM-Newton} scaling relations, we can indirectly constrain the \textit{Chandra}-\textit{XMM-Newton} $T$ scaling relation and compare it with the findings of S15 (after adjusting for the pivot points S15 used). We find
\begin{equation}
    \log_{10}{T_{\text{XMM}}}\approx 0.010+0.946\times\log_{10}{T_{\text{Chandra}}},
\end{equation}
which is entirely consistent with S15 within the $1\sigma$ uncertainties.

We now repeat this exercise to predict if using only the \textit{XMM-Newton}/MOS or \textit{XMM-Newton}/PN detectors would cause a significant shift to the eROSITA-\textit{XMM-Newton} cross-calibration scaling relations. Past studies (e.g., S15) have shown that \textit{XMM-Newton}/MOS and \textit{XMM-Newton}/PN return similar $T$ for $T\lesssim 3$ keV clusters, but \textit{XMM-Newton}/PN returns slightly lower $T$ than \textit{XMM-Newton}/MOS for hotter clusters, with the discrepancy reaching $\approx 12\%$ for $T\approx 7$ keV clusters (which is close to the maximum $T_{\text{XMM}}$ we used in this work). \citet{neval23} also showed that at higher energies ($\gtrsim 3$ keV), \textit{XMM-Newton}/PN and \textit{XMM-Newton}/MOS show cross-calibration biases. For typical $T$, \textit{XMM-Newton}/PN provides $\sim 60\%$ of the total \textit{XMM-Newton} counts. Since in this work we jointly use all the \textit{XMM-Newton} detectors to constraint $T_{\text{XMM}}$, the latter is generally expected to land somewhere between the \textit{XMM-Newton}/PN and \textit{XMM-Newton}/MOS $T$ values. All three $T$ values are expected to agree within the uncertainties for the vast majority of our clusters based on their $T_{\text{XMM}}$ distribution. Therefore, no major changes in the results should be observed if one uses individual \textit{XMM-Newton} detectors. Nevertheless, it is interesting to explore the cross-calibration of eROSITA versus \textit{XMM-Newton}/MOS or \textit{XMM-Newton}/PN alone. 

To do so, we use this work's scaling relations combined with the results provided in Table 2 of S15. For the full band eROSITA-\textit{XMM-Newton}/MOS and eROSITA-\textit{XMM-Newton}/PN scaling relations, we find
\begin{equation}
\begin{array}{l}
\log_{10}{\dfrac{T_{\text{eROSITA}}}{3\ \text{keV}}}\approx -0.106+0.802\times\log_{10}{\dfrac{T_{\text{XMM/MOS}}}{3\ \text{keV}}}, \\
\\[1ex]
\log_{10}{\dfrac{T_{\text{eROSITA}}}{3\ \text{keV}}}\approx -0.090+0.878\times\log_{10}{\dfrac{T_{\text{XMM/PN}}}{3\ \text{keV}}}.
\end{array}
\label{MOS/PN}
\end{equation}
The \textit{XMM-Newton}/PN $T$ seem to be slightly closer to $T_{\text{eROSITA}}$ than \textit{XMM-Newton}/MOS $T$ when the two EPIC detectors measure $T\gtrsim 2.5$ keV each, although the differences are minimal for most clusters. However, when \textit{XMM-Newton}/MOS and \textit{XMM-Newton}/PN measure $T\gtrsim 7$ keV each, the discrepancy with $T_{\text{eROSITA}}$ becomes $33\%$ and $26\%$ respectively. This (only slightly) better agreement of \textit{XMM-Newton}/PN with eROSITA is particularly interesting since the eROSITA telescope carries similar PN detectors. Nevertheless, such differences are still within the expected uncertainties and scatter; thus, better statistics (and a wider $T$ range) are required to draw more robust conclusions from these comparisons. 

\subsubsection{Other instruments}

Using our findings, we indirectly estimate the cross-calibration of eROSITA and other X-ray telescopes, such as \textit{Suzaku} and \textit{NuSTAR}. \citet{kettula} found that \textit{Suzaku} returned $\sim 12\%$ lower $T$ than the \textit{XMM-Newton}/PN detector. Based on that, one can predict that eROSITA and \textit{Suzaku} $T$ should be closer compared to the other instruments, especially at lower $T$ where they should be comparable. However, eROSITA is still expected to return lower $T$ than \textit{Suzaku} by $\sim 10-15\%$ for $T=7$ keV clusters. Nevertheless, it is not trivial to draw robust conclusions due to the very limited sample of \citet{kettula}, their use of \textit{XMM-Newton}/PN-only data, and the different energy bands in which they performed the \textit{Suzaku}-\textit{XMM-Newton} cross-calibration. It is interesting to note here that, if multi-temperature structure was the main cause of the observed $T$ discrepancies across different instruments, \textit{Suzaku} would be expected to return higher $T$ than \textit{XMM-Newton} (and comparable to \textit{Chandra}), which is not the case.

\citet{wallbank} reported $\approx 10\%$ lower $T$ from \textit{NuSTAR} compared to \textit{Chandra} for the 0.6-9.0 keV band and $T\approx 10$ keV clusters. As a result, one would expect eROSITA to measure much lower $T$ than \textit{NuSTAR} for massive clusters. On the $T\lesssim 2$ keV end, eROSITA is expected to return rather consistent $T$ values with \textit{NuSTAR} within $\approx 3-5\%$. As we stressed already, these predictions simply help provide a better understanding of what to expect from future studies rather than a comparison between eROSITA, \textit{NuSTAR}, and \textit{Suzaku}.

\subsection{Future improvements}
The current analysis already put tight constraints in the eROSITA cross-calibration with \textit{Chandra} and \textit{XMM-Newton} and provides a good understanding of the photon energy and cluster mass dependence of the $T$ discrepancy. These results will be highly valuable for future studies that wish to use eROSITA $T$ combined with $T$ values from other telescopes. Nonetheless, there is still room for several improvements to be made in the future. 

First and foremost, the $T_{\text{eROSITA}}$ measurement uncertainties are still large. Their decrease would result in a lower total scatter, improving the precision of the constraints. This will be achieved with the eRASS:4 data, where $\sigma_{T_{\text{eROSITA}}}$ will be reduced by a factor of $\sim 2$. Additionally, the effects of the non-Gaussian scatter will also be better studied and accounted for since higher $T_{\text{eROSITA}}$ will carry lower uncertainties and higher statistical weight compared to the present work. $T_{\text{break}}$ will also be better constrained when using a broken power law to fit the scaling relations.

The fourfold more counts in the eRASS:4 data will also allow us to use narrower energy bands for the spectral analysis, for instance, $0.5-2$ keV and $2-8$ keV. Consequently, a better characterization of the effective area cross-calibration as a function of energy will be achieved, for example, by utilizing the stacked residual ratios methodology \citep{kettula}. In this work, it was not possible to robustly constrain $T_{\text{eROSITA}}$ using narrower energy bands due to the limited available counts. As a result, the soft and hard bands inevitably overlapped. %Additionally in future work, we plan to "invert" the analysis and determine the eROSITA effective area changes required to reach an agreement with \textit{Chandra} and \textit{XMM-Newton} $T$ measurements. 

Given the higher number of counts in eRASS:4, a better characterization of the multiphase gas bias will also be achieved. Multi-temperature emission models will be fit to the spectra, which will reduce any dependence of the observed $T$ discrepancy on the different effective areas of the compared telescopes. Such fits were not possible in this study due to the limited number of counts in eRASS1. Moreover, eROSITA and \textit{XMM-Newton} have similar effective area shapes in the $0.5-2$ keV band. Therefore, when more available counts enable $T$ in this band to be constrained, eROSITA and \textit{XMM-Newton} should not show any systematic discrepancies due to the presence of multiphase gas, and any remaining discrepancies would be attributed to systematic calibration uncertainties.

Another interesting future test is the dependence of the comparison to the cluster redshift. The M20 sample used in this work is a low$-z$ sample with most clusters lying at $z<0.2$. As the $T$ comparison between eROSITA and \textit{Chandra}/\textit{XMM-Newton} depends on both the energy band and $T$ (i.e., spectral shape), the $T$ comparison might slightly change for different redshift ranges. In this work, it was not possible to perform such a test since the redshift range is small and clusters with higher $T$ are found at larger $z$. As a result, larger $z$ would deviate more than low $z$ due to the different cluster populations. The upcoming eRASS:4 data will allow the use of similar cluster samples at different $z$ in order to detect any possible redshift dependency of the eROSITA-\textit{Chandra}/\textit{XMM-Newton} cluster $T$ comparison.

Finally, the eROSITA-\textit{XMM-Newton} sample size will be increased. Due to the nature of the M20 sample, \textit{XMM-Newton} data were available mostly for low flux (i.e., low eROSITA counts on average) clusters, and only for 51 of them. This resulted in 71 independent $T_{\text{eROSITA}}$ and $T_{\text{XMM}}$ (core and core-excised annulus). Given the vast availability of \textit{XMM-Newton} cluster data and the deeper eRASS:4 data, we expect the eROSITA-\textit{XMM-Newton} $T$ scaling relations to significantly improve within the next years. A broken power law analysis will then become feasible for these scaling relations as well.

\section{Summary and conclusions}\label{conclusions} 
In this work, we provide the first-ever eROSITA-\textit{Chandra} cross-calibration in the broad, soft, and hard X-ray energy bands using the same spectral energy ranges and 186 independent galaxy cluster $T$, the largest sample used in such studies to date. In the same manner, we also provide the first eROSITA-\textit{XMM-Newton} calibration in the soft and hard bands and the first in the broad band using the same spectral energy for both instruments and a much larger $T$ sample than previously used. Our work offers robust conversion factors for the measured $T$ between eROSITA, \textit{Chandra}, and \textit{XMM-Newton}.

Using a single power law fit, we found that eROSITA shows a strong discrepancy with \textit{Chandra}, measuring $25\%$ and $38\%$ lower $T_{\text{eROSITA}}$ for $T_{\text{Chandra}}=4.5$ keV and $T_{\text{Chandra}}=10$ keV, respectively. Furthermore, we performed the first-ever (to our knowledge) broken power law fit in such a $T$ cross-calibration scaling relation. We found the $T$ where the power law breaks to be 2.7 keV and 1.7 keV for the full and soft band, respectively. At lower $T$, the values of the two telescopes were consistent within $\lesssim 5\%$, with a slope close to unity. For $T\gtrsim 6$ keV clusters, the broken power law further increased the tension between eROSITA and \textit{Chandra} by a few percent. The hard band was the one that showed the largest discrepancy between the two instruments, with eROSITA returning $\approx 25-60\%$ lower $T_{\text{eROSITA}}$ for $T_{\text{Chandra}}\approx 2-10$ keV.

Moreover, eROSITA shows lower $T$ than \textit{XMM-Newton} as well, with the discrepancy being milder than the one with \textit{Chandra}. For the full band, eROSITA measures $10-28\%$ lower $T_{\text{eROSITA}}$ for $T_{\text{XMM}}\approx 2-7$ keV clusters, while there is a slightly better agreement for cooler systems. The (dis)agreement between the two instruments improves by a few percent for the soft band. The hard band was only loosely constrained due to the large measurement uncertainties, although it is obvious that eROSITA largely underestimates $T$ compared to \textit{XMM-Newton} in the hard band. 

A wide range of possible systematics was explored. Namely, we looked for possible biases introduced by a multi-temperature gas structure, the degeneracy between the temperature and the metallicity parameters, the effect different levels of Galactic absorption have on the $T$ constraints of different telescopes, the non-Gaussianity of the scatter and the correlation between $T_{\text{eROSITA}}$ and its uncertainty, sample selection effects, and the use of eROSITA TM8 and TM9 data. We could not identify a specific systematic that had a strong effect on the eROSITA cross-calibration with \textit{Chandra} and \textit{XMM-Newton}, and more work is needed to obtain conclusive results about some of the potential biases.

Overall, the soft band demonstrated a marginally improved agreement between instruments than the full band, whereas the hard band cross-calibration revealed significant discrepancies. Similarly, clusters with softer spectra and lower $T$ showed much better agreement than clusters with harder spectra and high $T$. Subsequently, we conclude that eROSITA's effective area calibration is more comparable to \textit{Chandra} and \textit{XMM-Newton} for soft X-ray energies, though there is still a systematic bias present. Our findings further point to a larger systematic bias in eROSITA's cross-calibration with \textit{Chandra} and \textit{XMM-Newton} at harder energies. It is possible that this bias is overestimated due to the lower statistical weight that high $T_{\text{eROSITA}}$ carry given the shallow eRASS1 data. For now, in line with earlier studies, we conclude that these discrepancies are attributable to the systematic effective area calibration uncertainties.

Finally, our work offers the first robust conversion factors of spectroscopic eROSITA $T$ to \textit{XMM-Newton} and \textit{Chandra} $T$. Additionally, we offer conversion factors between the official eRASS1 cluster catalog $T$ and the \textit{Chandra} and \textit{XMM-Newton} $T$ from the core and core-excised cluster regions. All of these findings will enable the simultaneous use of eROSITA cluster $T$ with values coming from the other two telescopes. Given the unprecedented volume of cluster data eRASS1 (and eventually eRASS:4 and eRASS:8) will provide, these conversion factors are expected to be of utter importance for future cluster studies.

\section*{Acknowledgements}

We thank the anonymous referee for their constructive and insightful comments. This work is based on data from eROSITA, the soft X-ray instrument aboard SRG, a joint Russian-German science mission supported by the Russian Space Agency (Roskosmos), in the interests of the Russian Academy of Sciences represented by its Space Research Institute (IKI), and the Deutsches Zentrum für Luft- und Raumfahrt (DLR). The SRG spacecraft was built by Lavochkin Association (NPOL) and its subcontractors, and is operated by NPOL with support from the Max Planck Institute for Extraterrestrial Physics (MPE). The development and construction of the eROSITA X-ray instrument was led by MPE, with contributions from the Dr. Karl Remeis Observatory Bamberg \& ECAP (FAU Erlangen-Nuernberg), the University of Hamburg Observatory, the Leibniz Institute for Astrophysics Potsdam (AIP), and the Institute for Astronomy and Astrophysics of the University of Tübingen, with the support of DLR and the Max Planck Society. The Argelander Institute for Astronomy of the University of Bonn and the Ludwig Maximilians Universität Munich also participated in the science preparation for eROSITA. The eROSITA data shown here were processed using the eSASS software system developed by the German eROSITA consortium. K.M. acknowledges
support in the form of the X-ray Oort Fellowship at Leiden Observatory. G.S. acknowledges support through the \textit{Chandra} grant GO5-16126X. A.V. acknowledges funding by the Deutsche Forschungsgemeinschaft (DFG, German Research Foundation) -- 450861021.

\bibliographystyle{aa} %% aa.bst but adding links and notes to references
%%\raggedright              %% only for adsaa with dvips, not for pdflatex
\bibliography{49006corr}          %% XXX.bib = your Bibtex entries copied from ADS
 
\appendix
\section{Additional material and tests}

\subsection{Free-to-vary metal abundance and temperature distributions}

In Table \ref{Free-Z-table} we present the best-fit scaling relation parameters when $Z$ is left free to vary, as explained in Sect. \ref{free-z-section}. As discussed earlier, the general behavior of the scaling relations remains unaffected, with only a slight increase in the scatter being observed. Thus, the metallicity value treatment is not important for the $T$ comparison between different instruments.

In Fig. \ref{T-hist} we show the \textit{Chandra} and \textit{XMM-Newton} $T$ distributions of cluster core and core-excised regions, as discussed in Sect. \ref{multi-T}. The $T$ distributions are very similar. Hence, the core and core-excised cross-calibration results of these two subsamples are directly comparable and any observed differences should not be attributed to a different $T$ range.
\begin{table*}[h!]
\centering
\caption{Same as in Table \ref{tab:comparison2} but leaving $Z$ free to vary during the eROSITA spectral fits.}
\label{Free-Z-table}
\begin{tabular}{c | c c c c c | c}
\hline
\hline
& & & & & & \\
Comparison & Band  & $A$ & $B$ & $\sigma_{\text{intr}}$ & $\sigma_{\text{tot}}$ & $\xi$ \\
\hline
\hline\\[-0.25cm]
 &  Full & $-0.162_{-0.008}^{+0.009}$ & $0.758_{-0.016}^{+0.019}$ & $0.041\pm 0.015$ & $0.147\pm 0.007$ & -1.50 \\
 &  &  &  &  &  & \\
eROSITA-\textit{Chandra} & Soft & $-0.170\pm 0.010$ & $0.734\pm 0.018$ & $0.034\pm0.017$ & $0.143\pm 0.009$ & -1.67 \\
 &  &  &  &  & & \\
 & Hard & $-0.252\pm 0.022$ & $0.676\pm 0.079 $ & $0.132\pm 0.022$ & $0.273\pm 0.016$ & -1.51 \\
 &  &  &  &  & & \\[-0.25cm]
\hline
\\[-0.25cm]
 & Full & $-0.113_{-0.020}^{+0.023}$ & $0.734_{-0.096}^{+0.110}$ & $0.105\pm 0.024$ & $0.227\pm 0.015$ & -0.74 \\
 &   &  &  &  & & \\
eROSITA-\textit{XMM-Newton} & Soft &  $-0.077_{-0.026}^{+0.029}$ & $0.772_{-0.092}^{+0.108}$ & $0.075\pm 0.025$ & $0.193\pm 0.016$ & -0.53 \\
&  &  &  &  & & \\

 & Hard & $-0.339_{-0.041}^{+0.044}$ & $0.458_{-0.194}^{+0.261}$ & $0.160\pm 0.059$ & $0.334\pm 0.035$ & -1.80 \\

\hline
\end{tabular}
\end{table*}

\subsection{$\chi^2$-statistic bias compared to C-statistic}\label{T-shift-sect}

The $\chi^2$-statistic assumes Gaussian distribution of spectral counts and can return biased $T$ values during spectral fitting, while the C-statistic is based on Poissonian count distributions and it is known to return nearly unbiased $T$ results. However, the $\chi^2$ bias reduces in amplitude as the number of available counts increases per bin (and in total) since the counts distribution can be well approximated by a Gaussian distribution. Due to that, using the $\chi^2$-statistic for fitting X-ray spectra is commonly used in X-ray astronomy. In this work, the \textit{XMM-Newton} and \textit{Chandra} spectra we use typically contain $\gtrsim 10,000$, $\gtrsim 8,000$, and $\gtrsim 5,000$ counts for the full, soft, and hard bands respectively. As a result, adopting the $\chi^2$-statistic (to be consistent with the full band results from M20) should not introduce strong biases in the best-fit $T$ compared to the C-statistic. We confirm this by fitting 20 randomly selected full band \textit{XMM-Newton} spectra by using the \texttt{cstat} option in \texttt{XSPEC}. The average shift in the best-fit $T$ is $\approx 0.7\sigma$, or $\approx 2.9\%$. This is consistent with what was shown in \citet{veronica22}, where they observed similar $T$ shifts by changing between the above-mentioned statistics.

\begin{figure}[h]
               \includegraphics[width=0.45\textwidth]{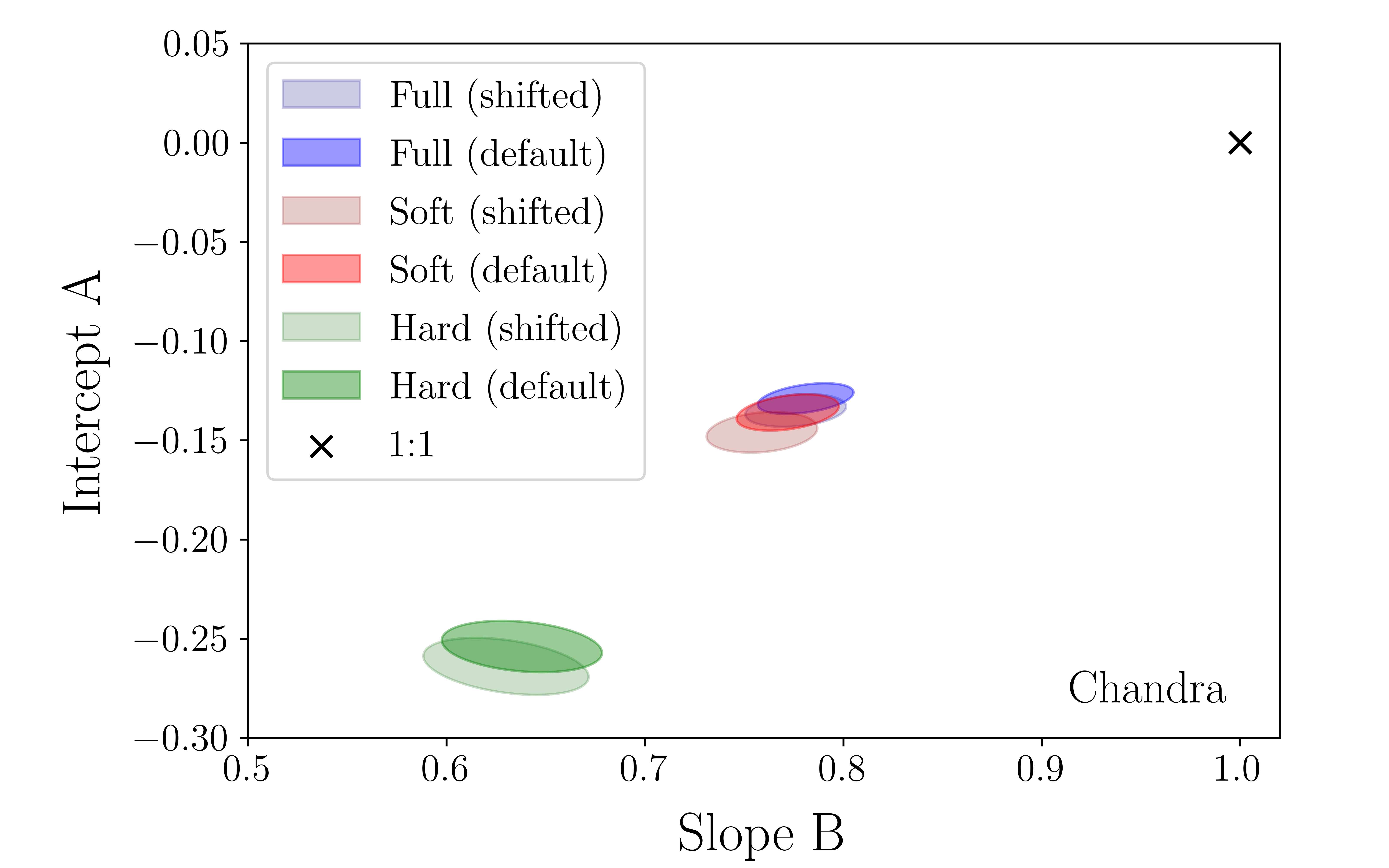}
               \includegraphics[width=0.45\textwidth]{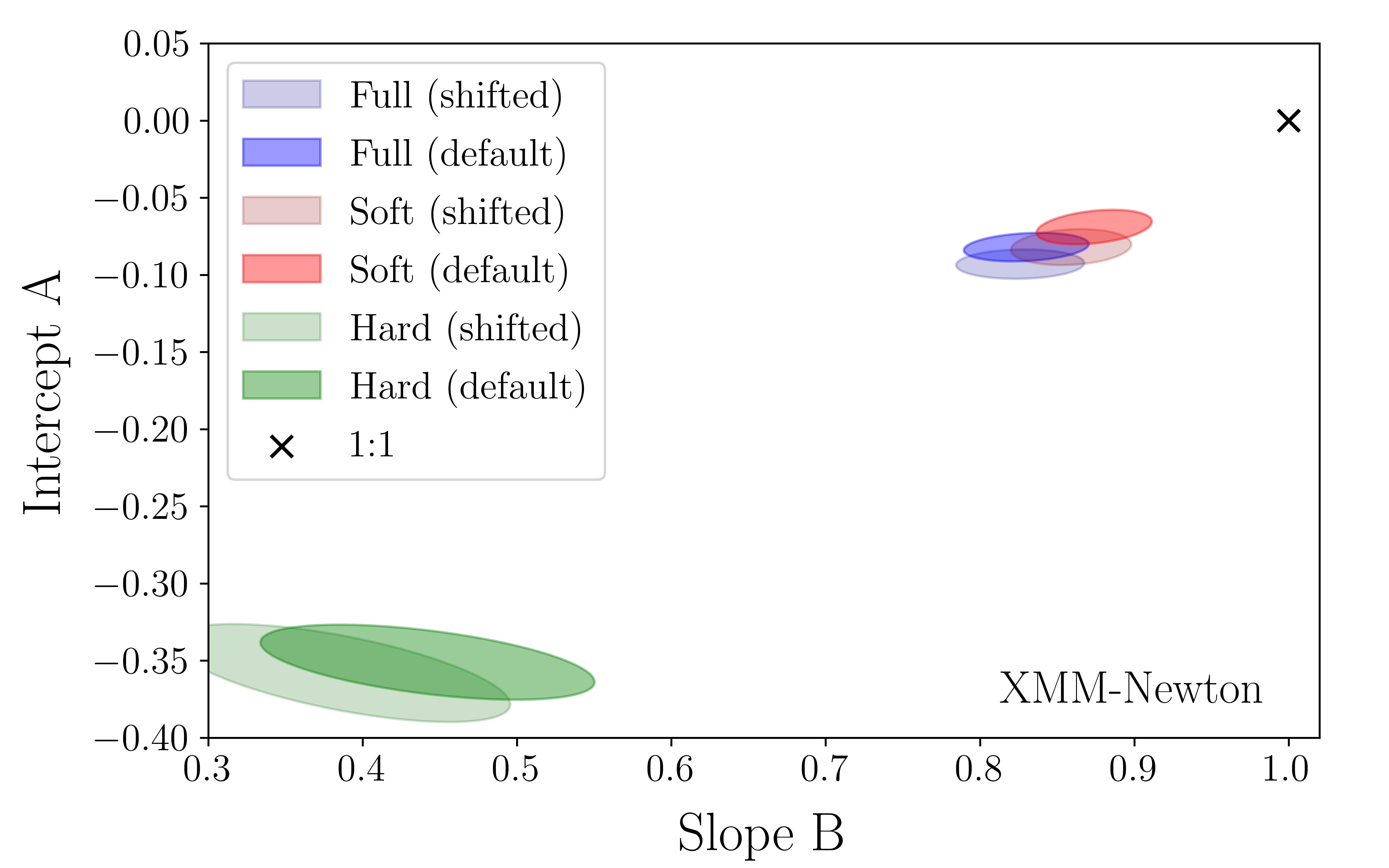}
               \caption{$1\sigma$ confidence levels for the eROSITA-\textit{Chandra} (top) and eROSITA-\textit{XMM-Newton} (bottom) scaling relations for the full (blue), soft (red), and hard (green) bands. The more opaque ellipses correspond to the default results, while the more transparent ellipses correspond to the best-fit results when $T_{\text{Chandra}}$ and $T_{\text{XMM}}$ are increased by $\sim 0.5-1\sigma$, as described in Sect. \ref{T-shift-sect}}.
        \label{ellipses-T-shifted}
\end{figure}

\begin{figure}[h]
               \includegraphics[width=0.45\textwidth]{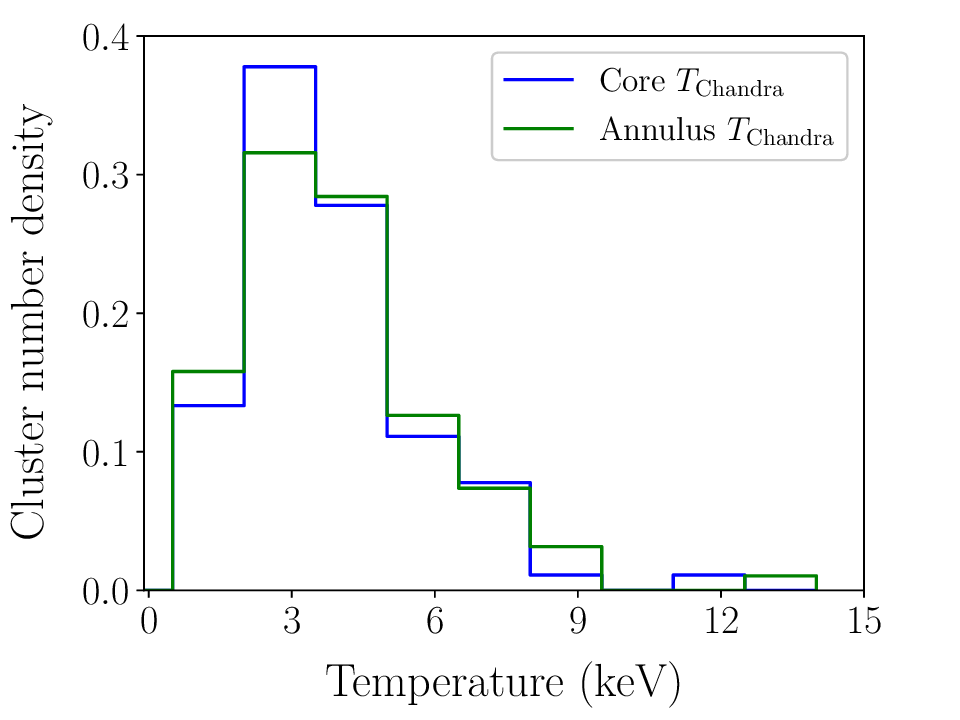}
               \includegraphics[width=0.45\textwidth]{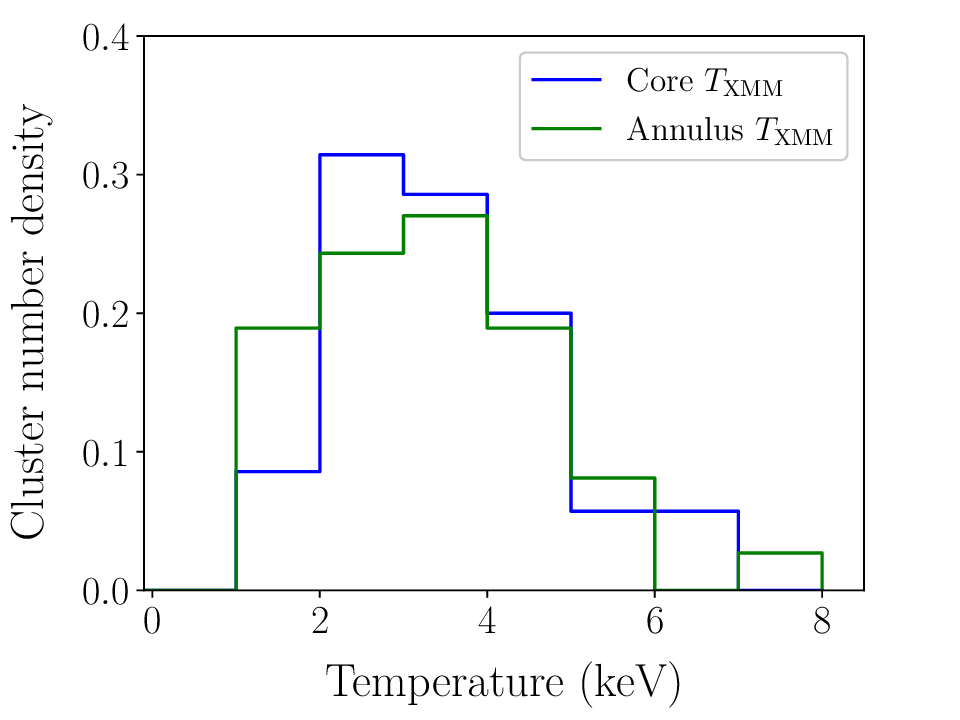}
               \caption{Histogram of core (blue) and core-excise (green) temperature values for \textit{Chandra} (right) and \textit{XMM-Newton} (right). The two distributions are very similar for both instruments.}
        \label{T-hist}
\end{figure}

\citet{humphrey} explored this issue in detail. They used simulated thermal plasma emission spectra and folded them with the \textit{Chandra} response to determine the bias of the $\chi^2-$ and C-statistics as a function of the available counts and thermal plasma properties. They found that, while the amplitude of the bias decreases with increasing counts, the statistical significance of the $T$ shift increases until $\sim 10^5$ counts (Fig. 2 in that paper). This occurs because, as counts increase, $\sigma_{T}$ drops faster than the absolute $T$ bias. For our number of counts in the \textit{XMM-Newton} and \textit{Chandra} spectra, \citet{humphrey} predicts a $T$ bias of $\sim 0.5-1\sigma$. To test the effect such a bias would have on the constrained scaling relations, we refit the latter in all bands for both the eROSITA-\textit{Chandra} and eROSITA-\textit{XMM-Newton} comparisons, but we increased $T_{\text{Chandra}}$ and $T_{\text{XMM}}$ by $\sim 0.5-1 \sigma_{T}$.\footnote{The exact factor depends on the average number of counts in that band per instrument, and it is based on Fig. 2 of \citet{humphrey}.} The shift in the $1\sigma$ contours of all scaling relations is displayed in Fig. \ref{ellipses-T-shifted}. The best-fit parameters change by $\leq 1.1\sigma$ for all scaling relations. The largest change is found for the two soft band comparisons, where $A$ changes by 0.01 dex (2.3\%) and 0.013 dex (3\%) for the eROSITA-\textit{Chandra} and eROSITA-\textit{XMM-Newton} scaling relations respectively. The slope $B$ changes by 1.7\% for both relations. For the full band comparisons, the best-fit parameters change by $\lesssim 1\sigma$.

The minimal effect of this temperature shift on the scaling relations occurs because small changes in $T_{\text{XMM}}$ and $T_{\text{Chandra}}$ are negligible compared to the uncertainties of $T_{\text{eROSITA}}$ for most clusters. Additionally, the slope of the relations ($B\sim 0.65-0.85<1$) further minimizes the effect that small $T_{\text{XMM}}$ and $T_{\text{Chandra}}$ changes have on the results. Consequently, the exact statistic used to fit the \textit{XMM-Newton} and \textit{Chandra} spectra does not affect the conclusions of this work. However, we note that for future similar studies with decreased $T_{\text{eROSITA}}$ uncertainties and total scatter, a consistent analysis using the C-statistic would be preferable since the mild $\chi^2$-statistic bias would have a larger impact on the final scaling relations.

\subsection{Intrinsic scatter dependency on temperature}\label{scatter-dependence}

\begin{figure}[h]
               \includegraphics[width=0.45\textwidth]{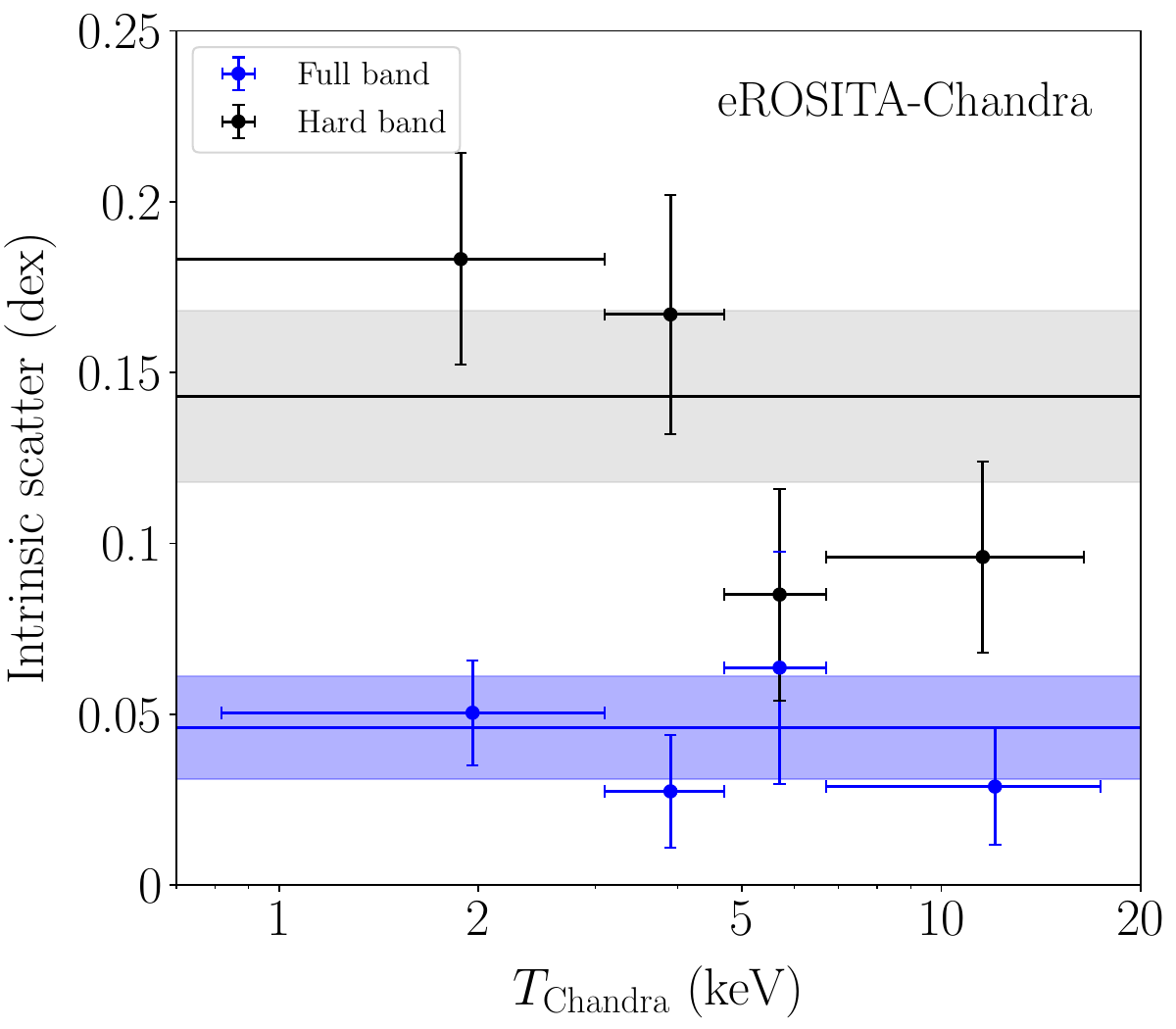}
               \includegraphics[width=0.45\textwidth]{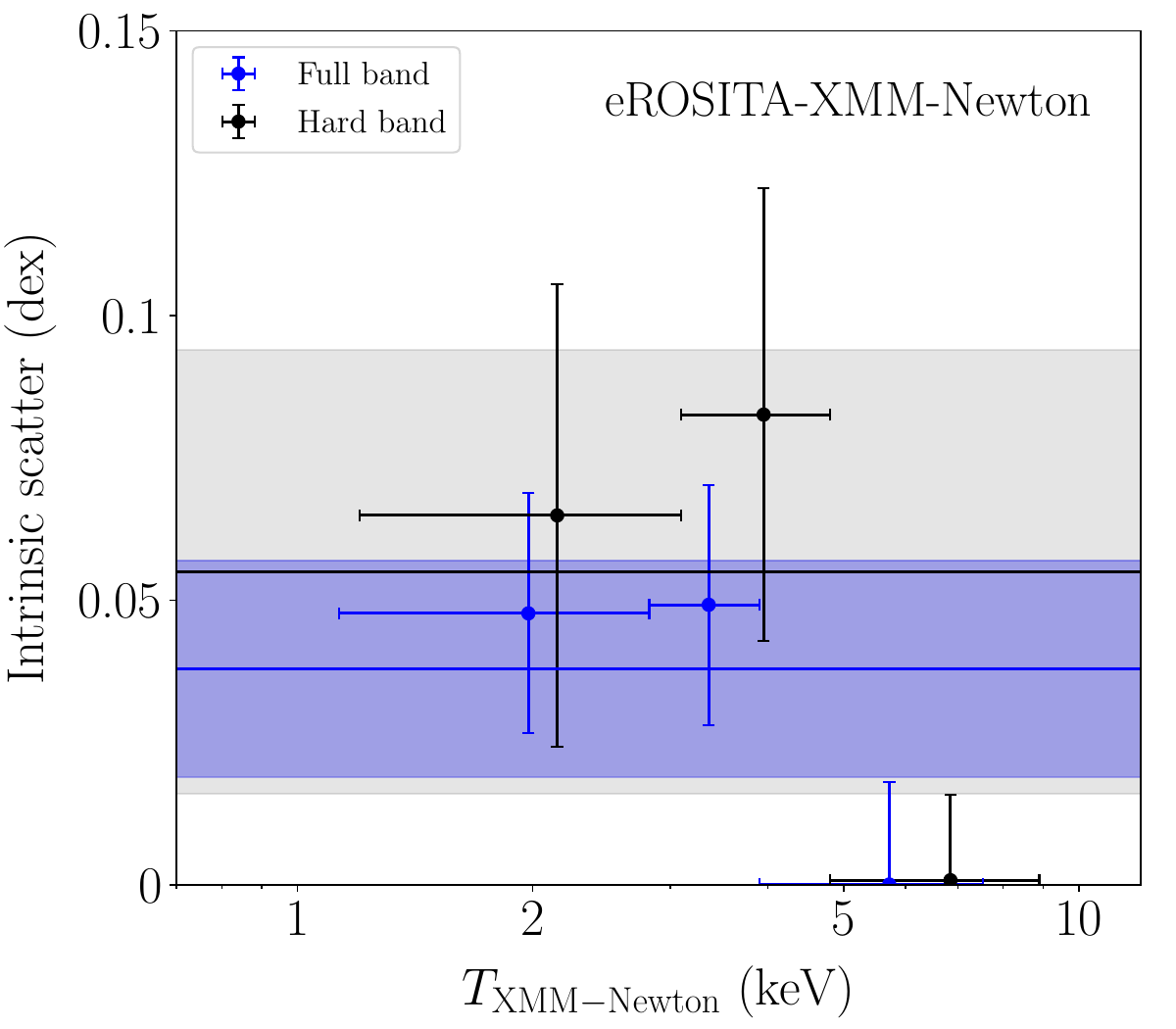}
               \caption{Dependence of the intrinsic scatter on $T_{\text{Chandra}}$ and $T_{\text{XMM}}$ for the eROSITA-\textit{Chandra} (top) and eROSITA-\textit{XMM-Newton} (bottom) scaling relations, respectively. Full band results are displayed in blue, while the hard band results are shown in black and gray. The shaded bands represent the best-fit, full sample $\sigma_{\text{intr}}$ with its $1\sigma$ uncertainties.}
        \label{scatter-dependency}
\end{figure}

The intrinsic scatter $\sigma_{\text{intr}}$ plays an essential role on the propagated statistical uncertainty when one converts a measured $T$ value from one telescope to the equivalent $T$ of another telescope. If $\sigma_{\text{intr}}$ is not relatively constant with $T$, then one might need to adopt different values for such conversions, depending on the used $T$. To test this, we fix $A$ and $B$ to their full sample best-fit values and fit only $\sigma_{\text{intr}}$ in four (three) independent, increasing-$T$ bins for the eROSITA-\textit{Chandra} (eROSITA-\textit{XMM-Newton}) scaling relations. All bins have the same number of clusters per scaling relation. We display the results for the full and hard bands in Fig. \ref{scatter-dependency}.

For the eROSITA-\textit{Chandra} full and soft band comparisons, $\sigma_{\text{intr}}$ does not depend on $T_{\text{Chandra}}$. For the hard band, $T_{\text{Chandra}}<5$ keV clusters show $\sigma_{\text{intr}}\approx 0.18$ dex while hotter clusters return $\sigma_{\text{intr}}\approx 0.09$. This suggests that soft-spectrum clusters show a more noisy behavior when the hard band is used since eROSITA struggles more to constrain the low $T$ of these clusters. However, all bins are within $<2\sigma$ from the full sample average $\sigma_{\text{intr}}$. Consequently, until lower statistical uncertainties are achieved with future data, we cannot argue in favor of a $T$-dependent scatter in the hard band.

For the eROSITA-\textit{XMM-Newton} comparison, all three bands show similar behavior. Clusters with $T_{\text{XMM}}<5$ keV show a constant $\sigma_{\text{intr}}$, consistent with the full sample average. Hotter clusters (which represent 33\% of the full sample) show $\sigma_{\text{intr}}\approx 0$ dex, deviating by $\approx 1.5\sigma$ from the average. Nevertheless, this can again suggest that hotter clusters have a lower $\sigma_{\text{intr}}$ than cooler clusters. For now, the currently poor statistics do not allow us to draw robust conclusions on the scatter's dependency on $T$.

\subsection{Outliers in the eROSITA-\textit{XMM-Newton} hard band comparison}\label{Xmm-hard-outliers}

There are two strong outliers in the eROSITA-\textit{XMM-Newton} hard band comparison: the core of A0602 ($T_{\text{XMM}}=2.850^{+0.158}_{-0.122}$) and the annulus of A2721 ($T_{\text{XMM}}=6.781^{+0.703}_{-0.553}$). Both of these regions return $T_{\text{eROSITA}}<0.3$ keV with $T/\sigma_T>2$. Both hard band spectra have very low eROSITA counts and higher $T$ models return rather consistent fits as well; thus, their obtained $T_{\text{eROSITA}}$ uncertainties seem to be underestimated. Including these two data points, the LMM slope decreases to $B=0.324^{+0.361}_{-0.314}$ while the intrinsic scatter increases by a factor of five. Thus, these outliers have a strong effect on the best-fit results. The core of A0602 has $T_{\text{eROSITA}}\approx 1.9$ keV in the full and soft bands while the annulus $T_{\text{eROSITA}}$ of A2721 could not be constrained in these bands. Consequently, based on all the above, their hard band $T$ seem to be problematic and they were excluded. 

\subsection{Multi-temperature fits with single $T$ models for the hard band}\label{2T-hard-section}

\begin{figure}[hbtp]
               \includegraphics[width=0.45\textwidth]{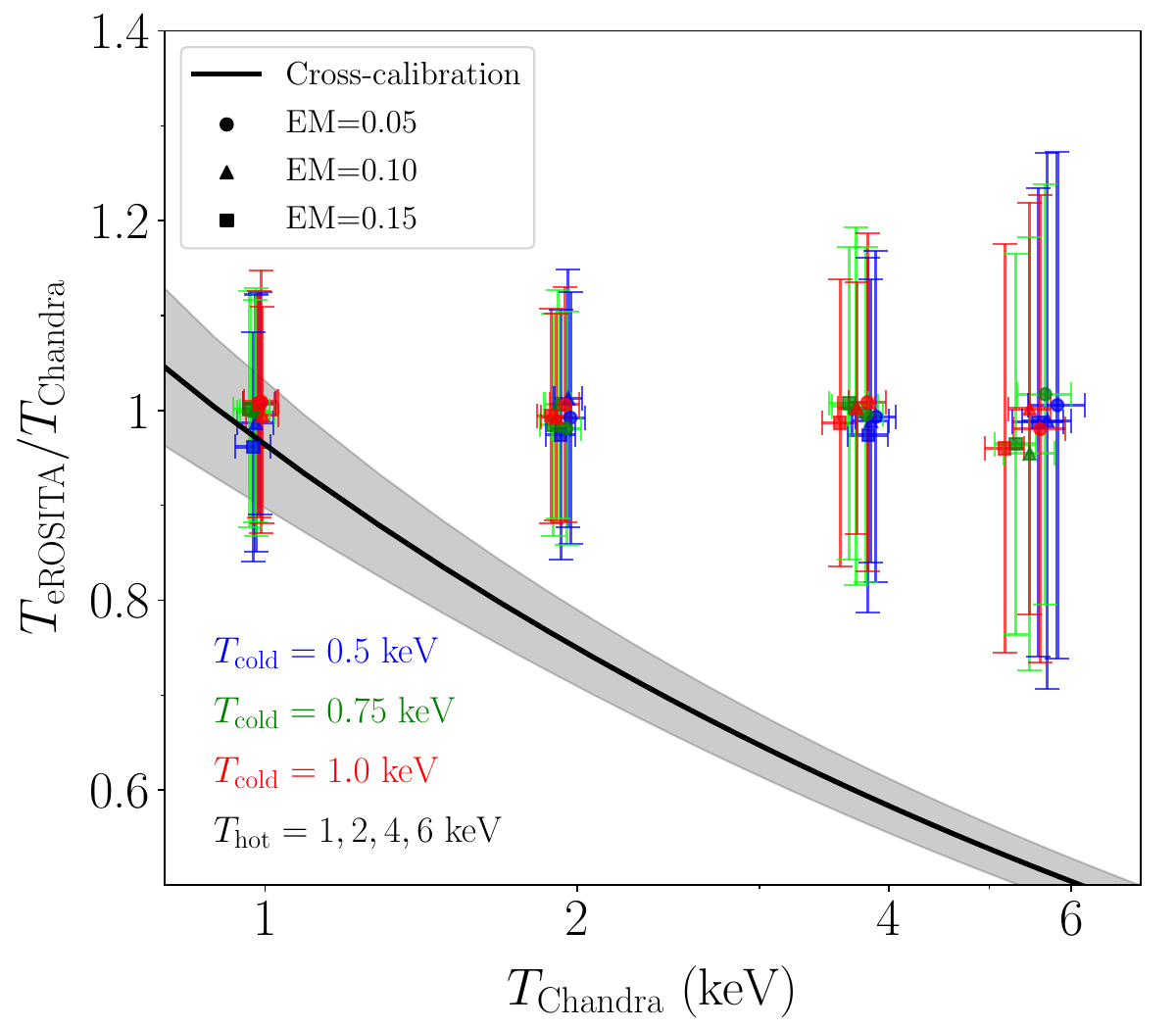}
               \includegraphics[width=0.45\textwidth]{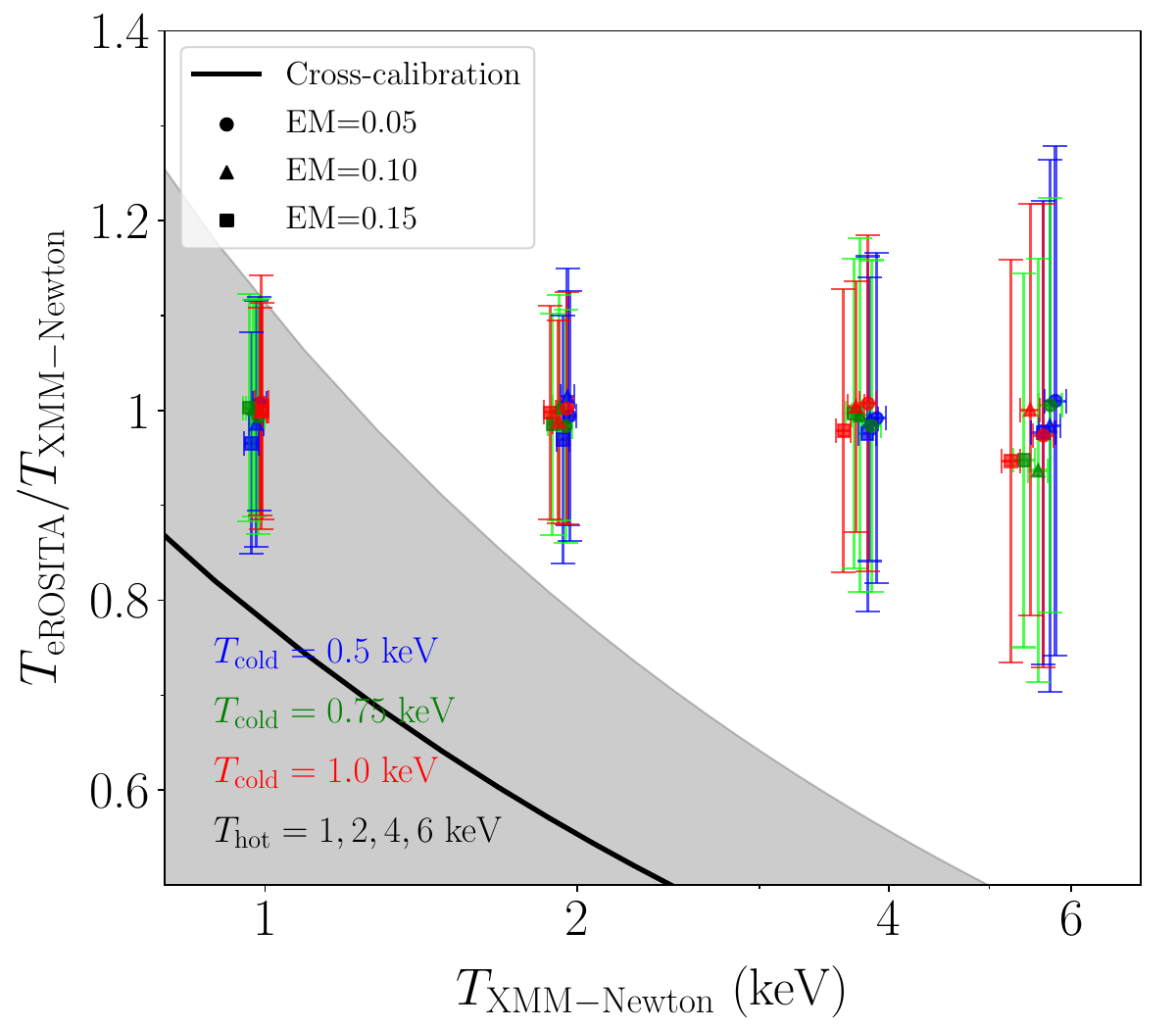}
               \caption{Same as in Fig. \ref{2T-simul-full-band} but for the hard band fits.}
        \label{2T-simul-hard-band}
\end{figure}

We repeat the analysis of Sect. \ref{2T-full-section}, focusing on the hard band this time. As shown in Fig. \ref{2T-simul-hard-band}, the three instruments return very similar single $T$ constraints for all $2T$ combinations. Although the $T_{\text{eROSITA}}$ uncertainties are very large and only $<2\sigma$ away from the observed cross-calibration differences, the general trend does not seem to explain the observed $T$ tension, especially between eROSITA and \textit{Chandra}.

\subsection{Soft versus hard band $T$ comparison}\label{Sect_soft_vs_hard}
In Sect. \ref{main_soft_vs_hard} we explored the self-consistency of $T$ measurements when one uses only the soft or hard band of the same telescope. To do that, we compared $T_{0.5-4\ \mathrm{keV}}$ and $T_{1.5-7\ \mathrm{keV}}$ for all available cluster regions per instrument. We used 132 cluster regions for eROSITA, 189 for \textit{Chandra}, and 78 for \textit{XMM-Newton}. The pivot point values $T_{\text{piv}}$ used in Eq. \ref{uncertainties-sum} are 3 keV, 4.5 keV, and 3 keV for eROSITA, \textit{Chandra}, and \textit{XMM-Newton} respectively. $T_{0.5-4\ \mathrm{keV}}$ and $T_{1.5-7\ \mathrm{keV}}$ are not fully independent since they overlap within the $1.5-4$ keV energy range. Here, the partial covariance of their uncertainties is ignored since this test aims to provide indicative results rather than numerically precise ones.

\begin{figure*}[hbtp]
               \includegraphics[width=0.33\textwidth]{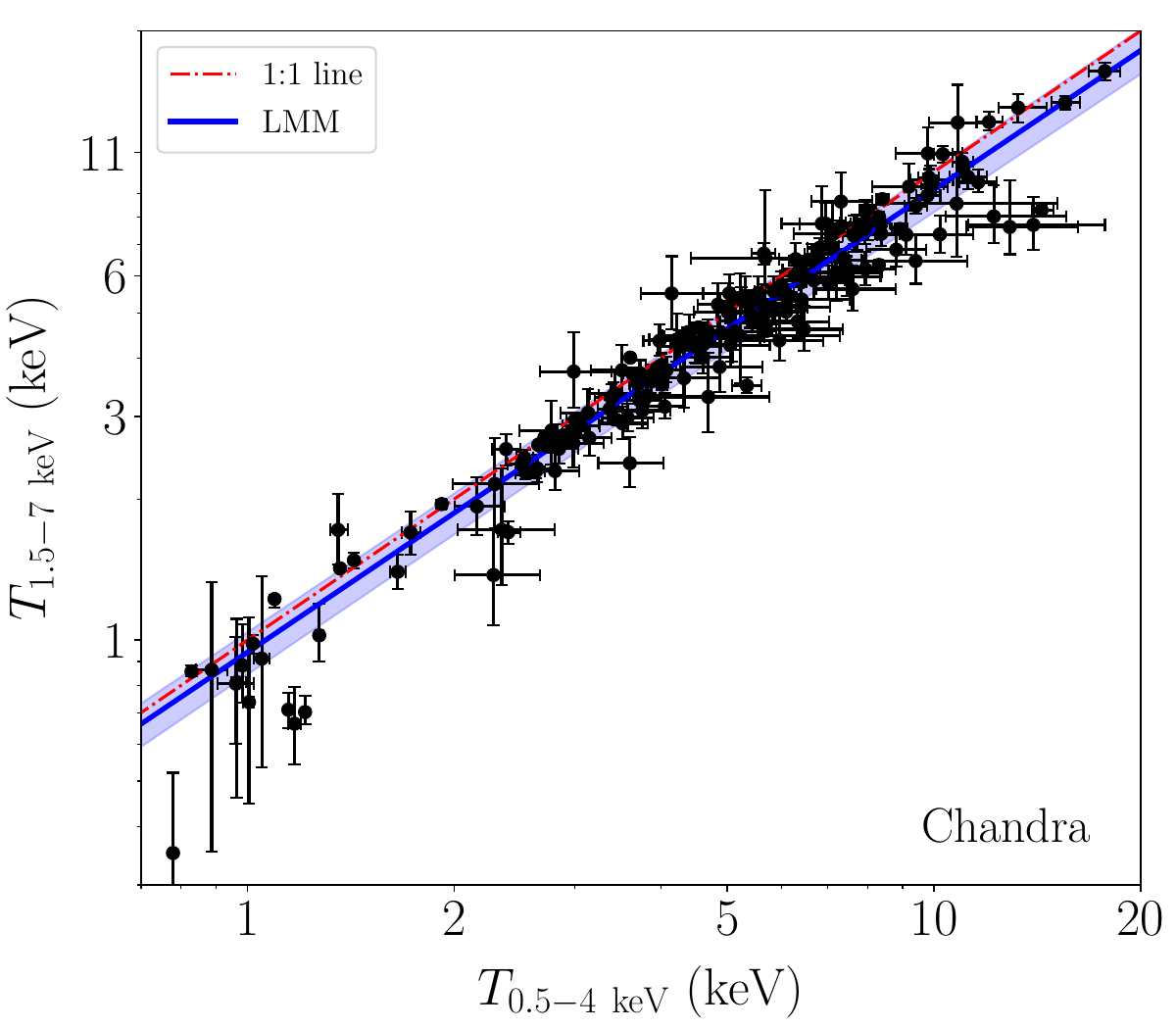}
               \includegraphics[width=0.33\textwidth]{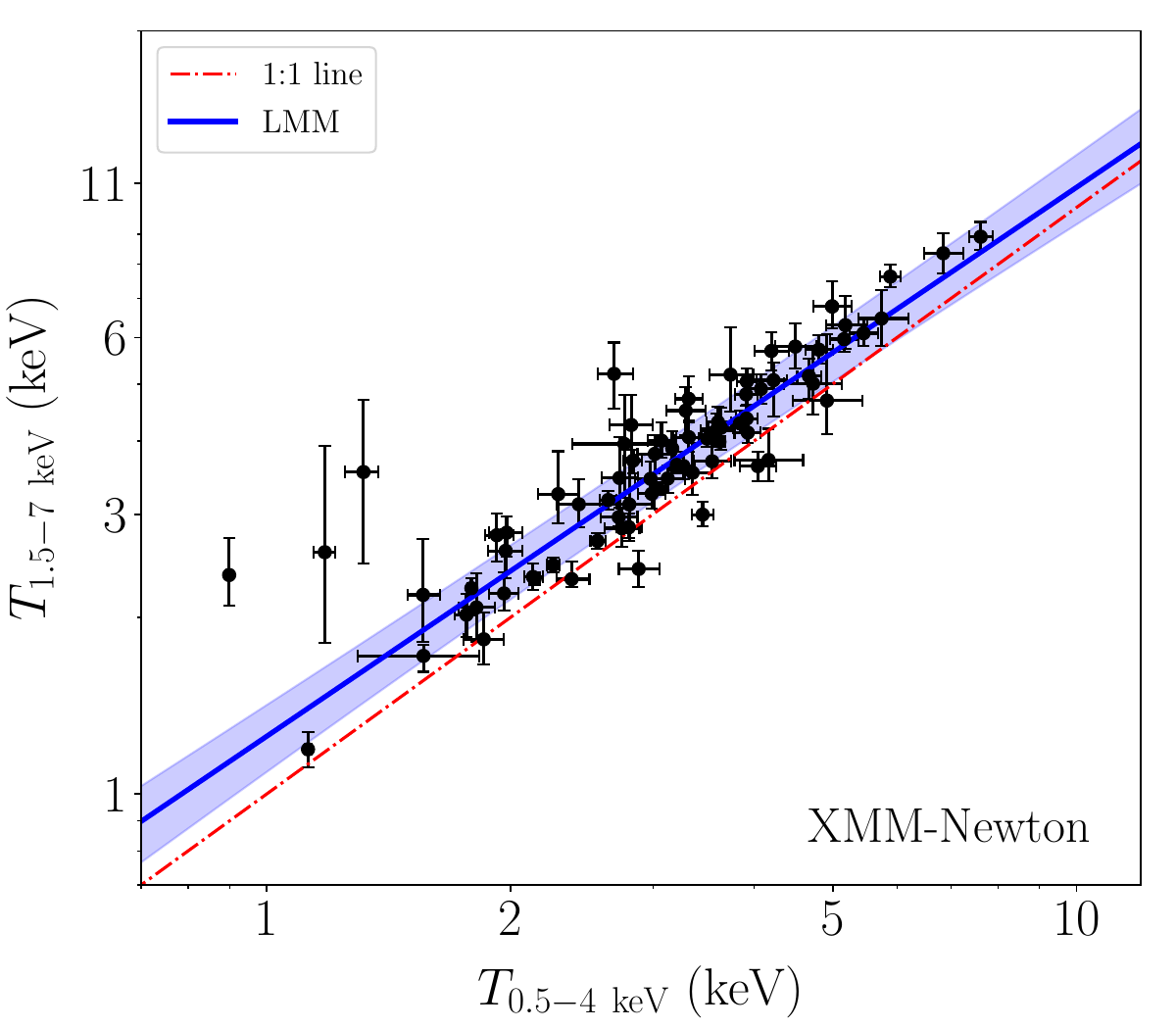}
               \includegraphics[width=0.33\textwidth]{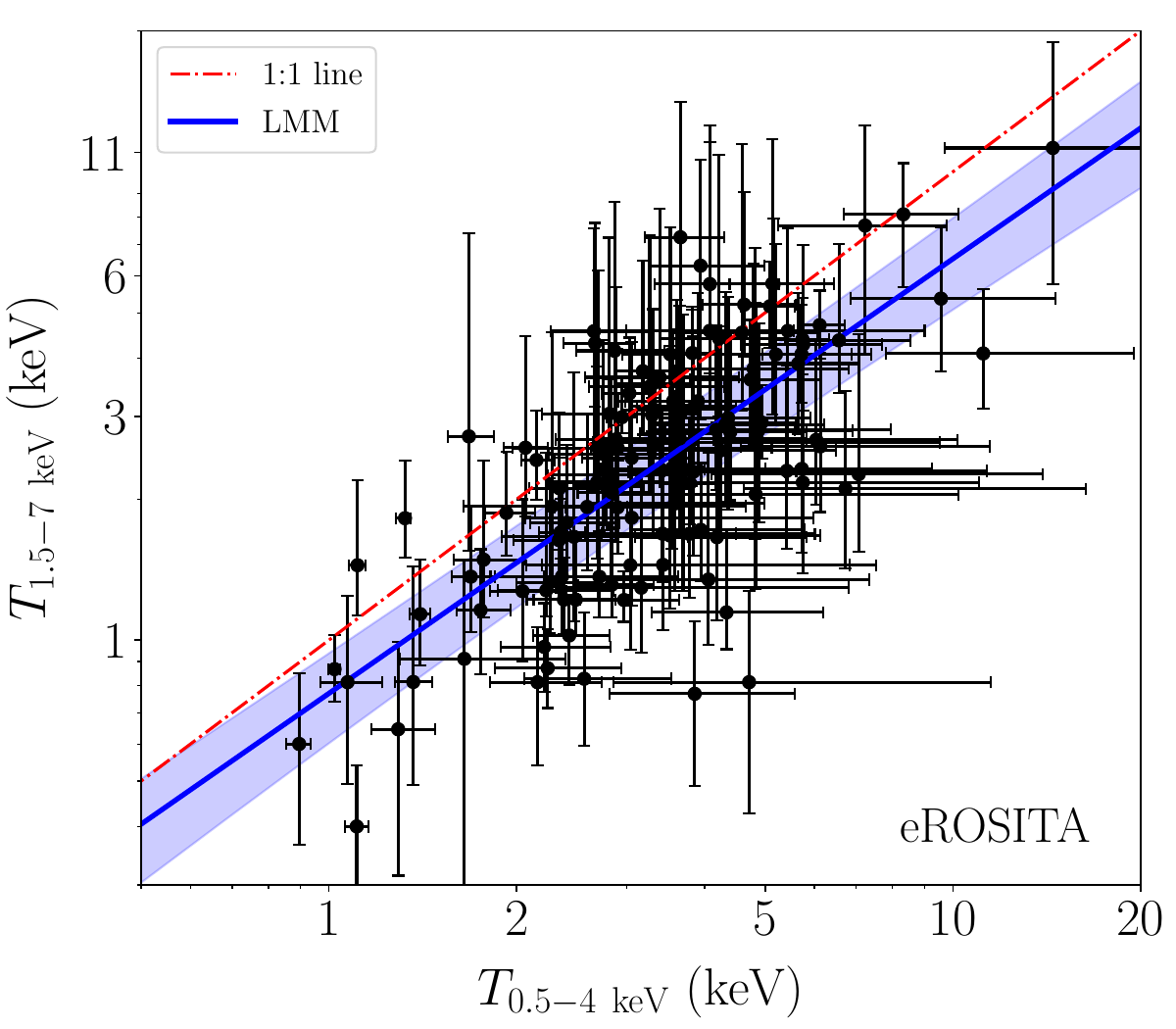}
               \caption{Comparison between $T$ measured in the soft (0.5-4 keV) and hard (1.5-7 keV) bands for \textit{Chandra} (left), \textit{XMM-Newton} (middle), and eROSITA (right). The best-fit scaling relation line by LMM (blue) is displayed. The equality 1:1 line is shown in red (dash dot). The blue shaded area represents the LMM statistical error plus the intrinsic scatter.}
        \label{soft-vs-hard}
\end{figure*}

As shown in Table \ref{soft-vs-hard-table} and Fig. \ref{soft-vs-hard}, any discrepancy between $T_{0.5-4\ \mathrm{keV}}$ and $T_{1.5-7\ \mathrm{keV}}$ is almost constant with increasing $T$, for all instruments. The intrinsic scatter is $\approx 10\%$ for the \textit{Chandra} and \textit{XMM-Newton} $T_{1.5-7\ \mathrm{keV}}-T_{0.5-4\ \mathrm{keV}}$ comparisons, while it almost doubles for the eROSITA comparison. In the left and middle panels of Fig. \ref{soft-vs-hard}, one sees that, for low $T_{0.5-4\ \mathrm{keV}}$, there is some non-Gaussianity of the scatter. However, excluding these measurements does not significantly alter the results since their statistical weight is generally low due to the increased $T_{1.5-7\ \mathrm{keV}}$ uncertainties.

\begin{table*}[htbp]
\centering
\caption{Best-fit parameters for the $T_{1.5-7\ \mathrm{keV}}-T_{0.5-4\ \mathrm{keV}}$ scaling relations for all three instruments based on using the parametrization in Eq. \ref{scal_rel} and Eq. \ref{scatterYX}. For \textit{Chandra}, \textit{XMM-Newton}, and eROSITA, $T_{\text{piv}}=4.5, 3, 3$ keV respectively.}
\label{soft-vs-hard-table}
\begin{tabular}{c | c c c c | c}
\hline
\hline
 & & & & & \\
Soft-hard band comparison & $A$ & $B$ & $\sigma_{\text{intr}}$ & $\sigma_{\text{tot}}$ & $\xi$ \\
\hline
\hline\\[-0.25cm]
\\[-0.25cm]
\textit{Chandra}  & $-0.034\pm 0.005$ & $0.988\pm 0.024$ & $0.043^{+0.008}_{-0.006}$ & $0.082\pm 0.005$ & -0.80 \\
%\hline
\\[-0.25cm]
\textit{XMM-Newton} & $0.068\pm 0.008$ & $0.936\pm 0.070$ & $0.043^{+0.013}_{-0.011}$ & $0.076\pm 0.008$ & 1.75 \\
\\[-0.25cm]
eROSITA  & $-0.148\pm 0.018$ & $0.928\pm 0.081$ & $0.083^{+0.025}_{-0.017}$ & $0.267\pm 0.012$ & -0.72 \\
%\hline

\hline
\end{tabular}
\end{table*}

\subsection{Effects of residual contamination in the \textit{XMM-Newton}/PN camera}\label{IN/OUT test}

In Sect. \ref{XMM-bgd}, we discuss the effect that the residual contamination in the \textit{XMM-Newton}/PN unexposed corners has on the eROSITA-\textit{XMM-Newton} scaling relations. In this section, we provide more details for this test.

To characterize the soft proton contamination we use the IN/OUT diagnostic as defined in \citet{miriam} \citep[see also][]{deluca}. Briefly, this refers to the ratio of the surface brightness within the FOV after masking the central $10'$ and the surface brightness of the unexposed corners. Both values are measured in the $5-7$ keV plus $10-14$ keV bands for \textit{XMM-Newton}/PN and the $6-12$ keV band for \textit{XMM-Newton}/MOS. We estimate the IN/OUT ratio for the actual observation (IN/OUT$_{\text{obs}}$) and for the FWC data (IN/OUT$_{\text{FWC}}$). The final IN/OUT value is given by the ratio of IN/OUT$_{\text{obs}}$ and IN/OUT$_{\text{FWC}}$. \textit{XMM-Newton}/EPIC detectors with IN/OUT$<1.15$ are considered sufficiently filtered from solar flares and they are used in this work.

\begin{figure}[hbtp]
               \includegraphics[width=0.45\textwidth]{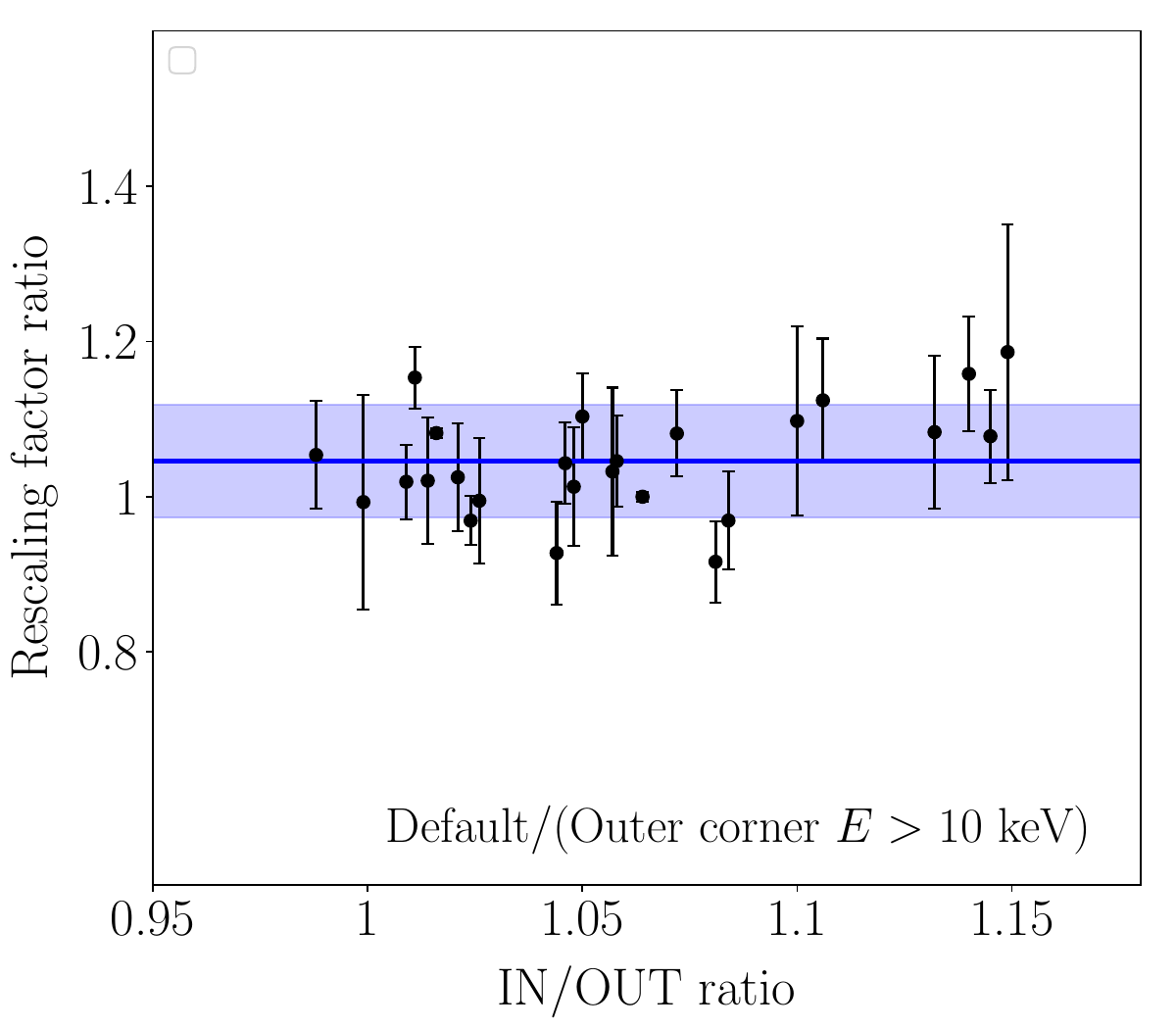}
               \includegraphics[width=0.45\textwidth]{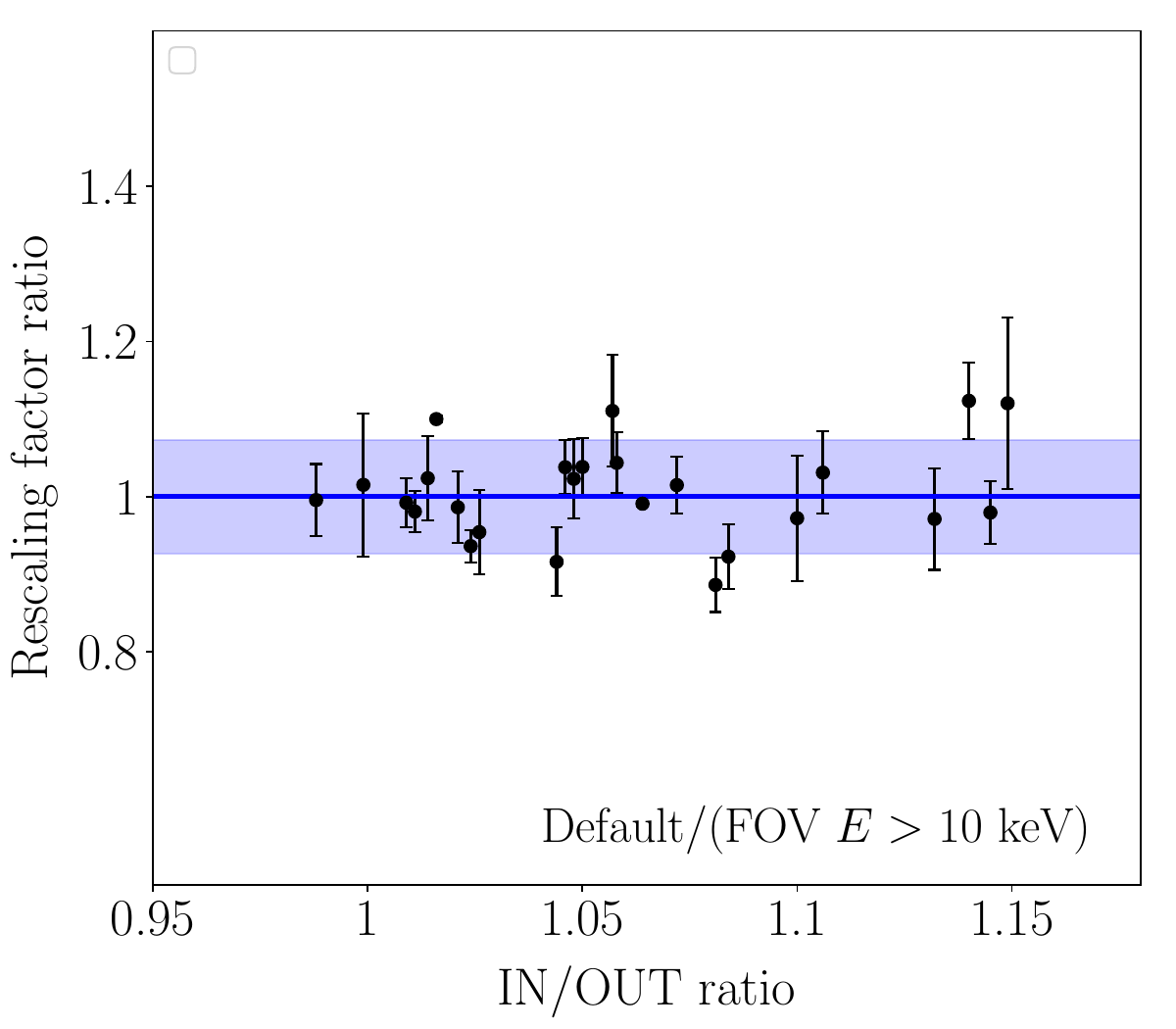}
               \caption{Ratio of PIB rescaling factors as they occur from the $2.5-5$ keV (default) and the $10-14$ keV count rates of the unexposed \textit{XMM-Newton}/PN corners (top) and the entire FOV (bottom). The blue shaded region corresponds to the standard deviation of the distribution.}
        \label{IN-OUT-plots}
\end{figure}

In Sect. \ref{XMM-bgd} we discussed our methodology for checking the bias level in our default PIB estimation of \textit{XMM-Newton}/PN as a function of the latter's IN/OUT ratio. The results are shown in Fig. \ref{IN-OUT-plots}. This test clearly shows that there is no obvious bias for the majority of our \textit{XMM-Newton} data. For \textit{XMM-Newton}/PN data with $1.1<$IN/OUT$<1.15$ our method indeed overestimates the rescaling factor by $<15\%$, that is, we subtracted more photons to correct for the PIB than we should have. We then proceed to correct the PIB subtraction in the \textit{XMM-Newton}/PN camera for the ten clusters in our sample with $1.1<$IN/OUT$<1.15$ and refit their $T_{\text{XMM}}$. We found that the core $T_{\text{XMM}}$ are negligibly affected by this mild PIB bias ($\Delta T<2\%$ typically), due to the very high surface brightness of the clusters at these regions. For the core-excised $T_{\text{XMM}}$, we found that the shift was more significant ($\Delta T\sim 3-15\%$) and depended on the IN/OUT ratio of \textit{XMM-Newton}/PN. Moreover, the $T_{\text{XMM}}$ shifts were not systematic for all clusters (i.e., not all $T_{\text{XMM}}$ increased). Overall, only 11\% (8 out of 71) of the $T_{\text{XMM}}$ used in the eROSITA-\textit{XMM-Newton} scaling relations show a $\Delta T_{\text{XMM}}>4\%$ shift, which is the average $1\sigma$ uncertainty for these values.

To quantify the effect of these $T_{\text{XMM}}$ changes to the eROSITA-\textit{XMM-Newton} scaling relations, we replace the default $T_{\text{XMM}}$ with the newly fitted ones that account for the contamination in the \textit{XMM-Newton}/PN unexposed corners. As expected, the effect is minimal. For the full and soft band eROSITA-\textit{XMM-Newton} scaling relations, $A$ increased by 0.005 dex ($1\%$) while $B$ increased by $\approx 2-2.5\%$. These correspond to $<0.3\sigma$ changes for both parameters. For the hard band, the effects were even less statistically significant. Subsequently, we show that the default PIB treatment does not introduce any noticeable bias to the final scaling relations.

Finally, we repeat the estimation of the rescaling factor using the whole FOV in the $10-14$ keV band instead of the unexposed corners. The comparison with the default rescaling factor is shown in the bottom panel of Fig. \ref{IN-OUT-plots}. Interestingly, the full FOV method returns a similar rescaling factor to our default methodology for all IN/OUT values.

\subsection{Comparison of our measurements with the official eRASS1 cluster catalog $T$}\label{eRASS1-comparison}

To compare $T_{\text{eRASS1}}$ to the eROSITA, \textit{Chandra}, and \textit{XMM-Newton} $T$ presented in this work, we matched all the clusters from M20 with all the eRASS1 cluster catalog systems that have available $T$ measurements. Since $z$ correlates with $T$ during spectral fittings, we excluded six clusters that showed a redshift difference of $\left|\dfrac{\Delta z}{1+z_{\text{M20}}}\right|>0.01$\footnote{Alternative filtering based on a 10\% or 20\% redshift difference results in similar cuts and have a completely negligible effect on the analysis.} between M20 and the eRASS1 cluster catalog. This resulted in 78 eRASS1-\textit{Chandra}, 33 eRASS1-\textit{XMM-Newton}, and 96 eRASS1-spectroscopic eROSITA clusters.

\begin{table*}[htbp]
\centering
\caption{Best-fit parameters for the eROSITA-\textit{Chandra} and eROSITA-\textit{XMM-Newton} scaling relations for all energy bands and fitting methods using the parametrization in Eq. \ref{scal_rel} and Eq. \ref{scatterYX}. }
\label{erass1-table}
\begin{tabular}{c | c c c c c | c}
\hline
\hline
 & & & & & \\
Comparison & Cluster region & $A$ & $B$ & $\sigma_{\text{intr}}$ & $\sigma_{\text{tot}}$ & $\xi$ \\
\hline
\hline\\[-0.25cm]

\\[-0.25cm]
& Core-excised & $-0.107\pm 0.012$ & $0.831\pm 0.030$ & $0.037\pm0.010$ & $0.136\pm 0.008$ & -1.03 \\
eRASS1-\textit{Chandra}  &  &  &  &  & \\
& Core & $-0.103\pm 0.011$ & $0.880\pm 0.024$ & $0.003^{+0.007}_{-0.003}$ & $0.118\pm 0.008$ & -0.95 \\
\hline
\\[-0.25cm]
& Core-excised & $-0.086\pm 0.029$ & $0.738\pm 0.088$ & $0.077^{+0.041}_{-0.032}$ & $0.210\pm 0.016$ & -1.45\\
eRASS1-\textit{XMM-Newton}  &  &  &  &  & \\
& Core & $-0.113\pm 0.027$ & $0.823\pm 0.113$ & $0.033^{+0.039}_{-0.023}$ & $0.174\pm 0.015$ & -0.45 \\
\hline
\\[-0.25cm]
& Core-excised & $-0.003\pm 0.014$ & $0.934\pm 0.049$ & $0.027^{+0.017}_{-0.011}$ & $0.194\pm 0.011$ & -0.05 \\
eRASS1-eROSITA spec-$T$  &  &  &  &  & \\
& Core & $-0.003\pm 0.011$ & $1.004\pm 0.026$ & $0.000^{+0.002}_{-0.000}$ & $0.178\pm 0.010$ & -0.01 \\

\hline
\end{tabular}
\end{table*}

\begin{figure*}[hbtp]
               \includegraphics[width=0.33\textwidth]{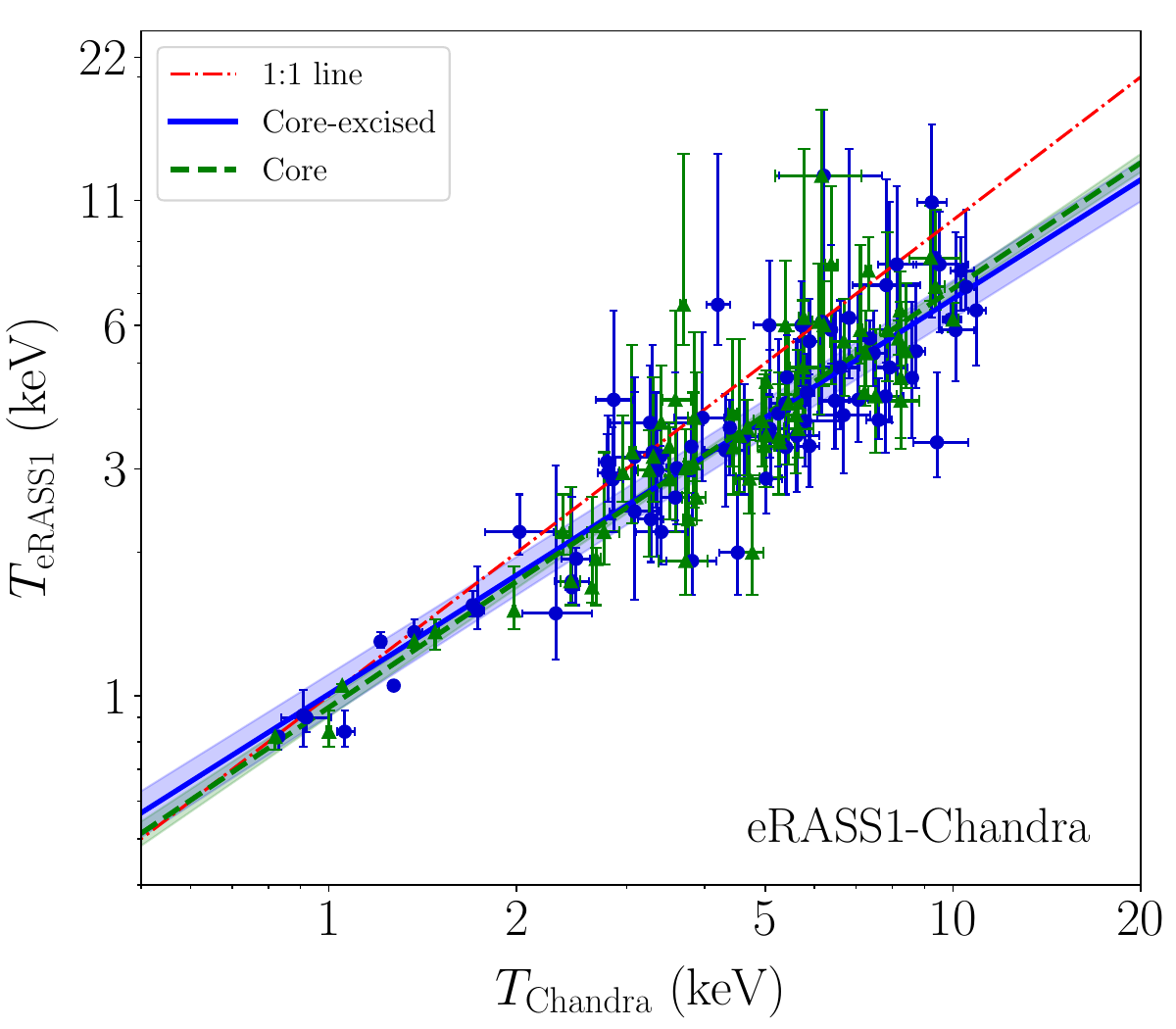}
               \includegraphics[width=0.33\textwidth]{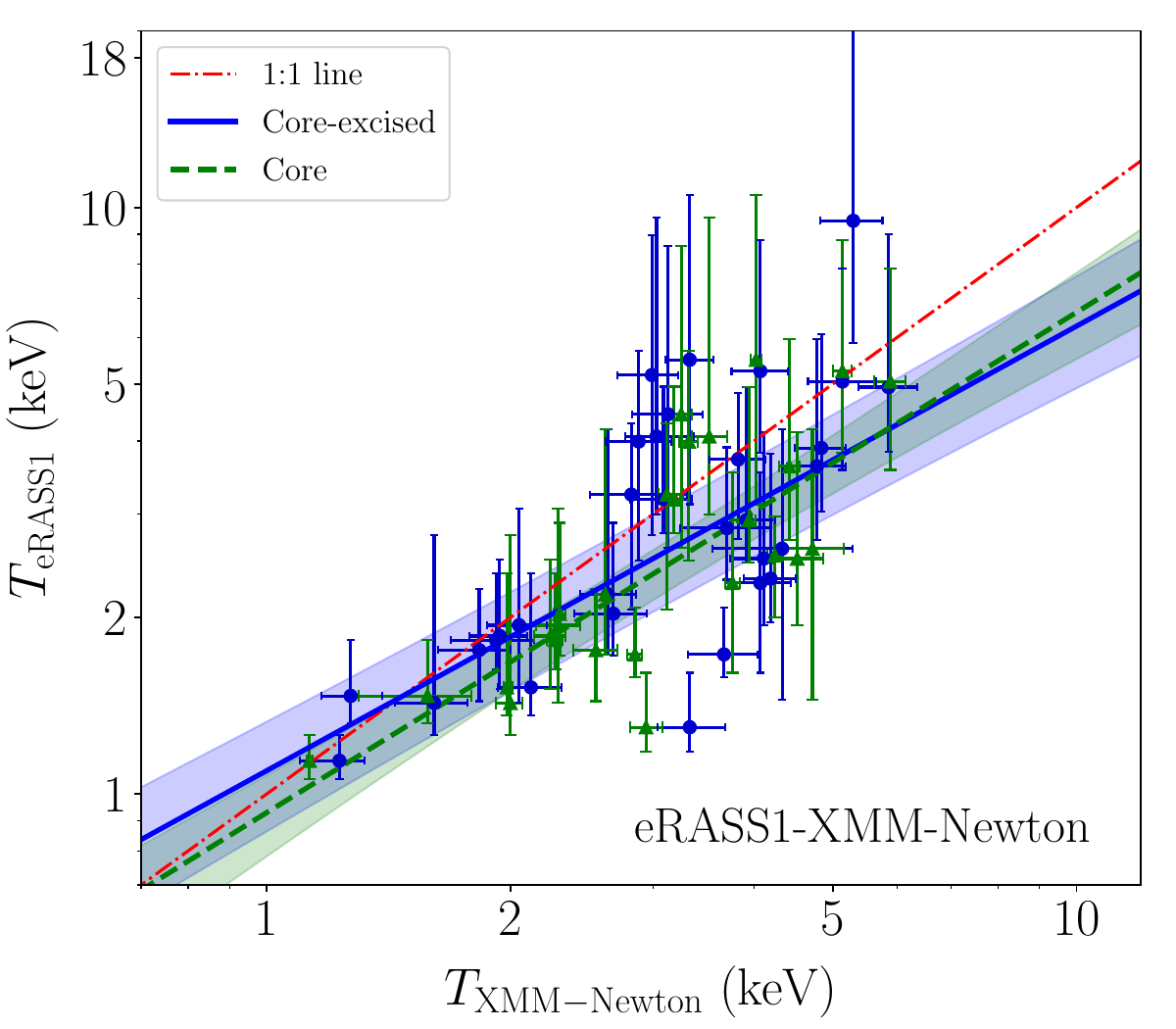}
               \includegraphics[width=0.33\textwidth]{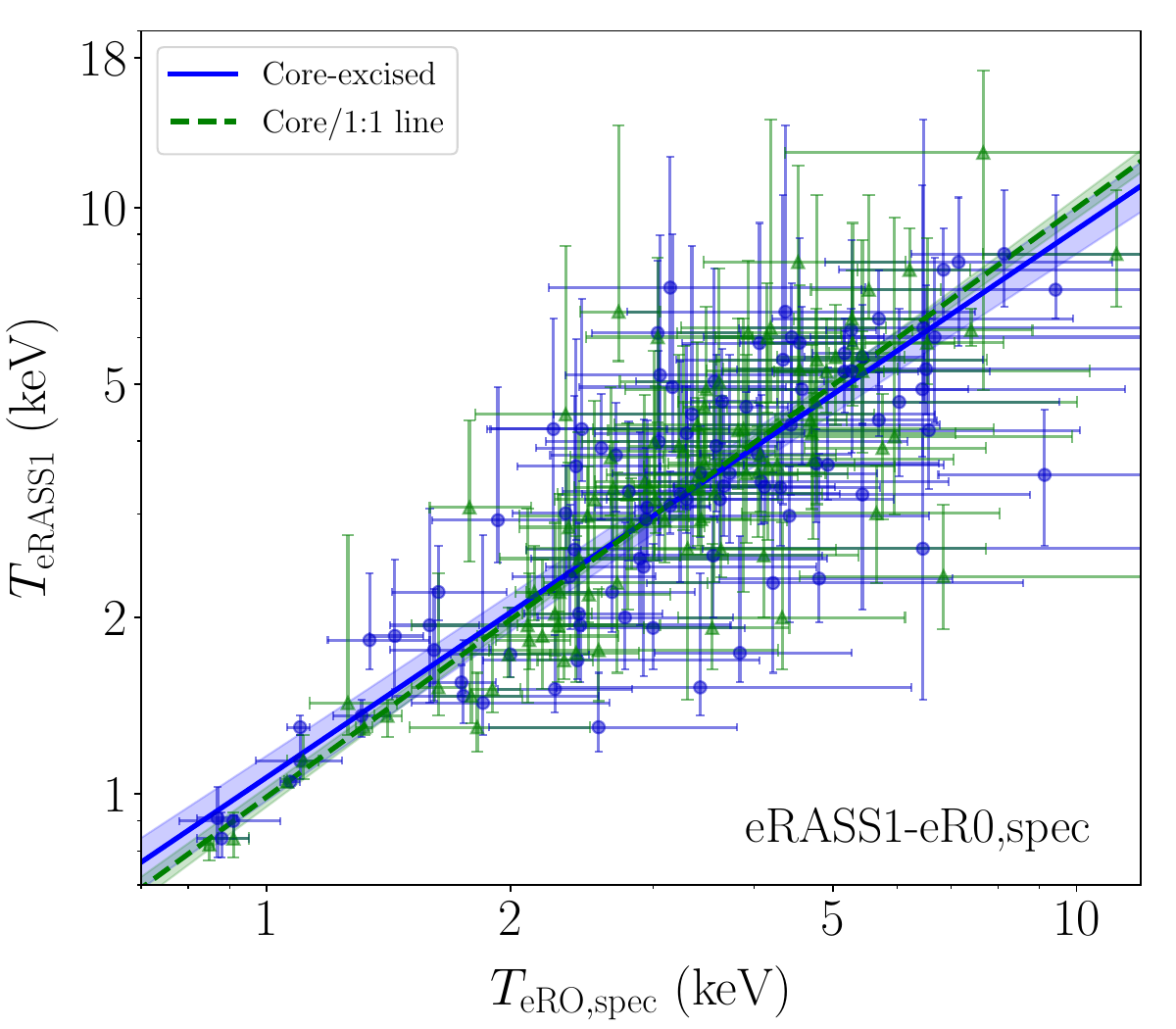}
               \caption{Comparison between eRASS1 and \textit{Chandra} (left), \textit{XMM-Newton} (middle), and spectroscopic eROSITA (right) temperatures. The comparison between the core-excised (core) $T_{\text{Chandra}}/T_{\text{XMM}}/T_{\text{eRO,spec}}$ and the single $T_{\text{eRASS1}}$ from the entire $R_{500}$ is displayed in blue (green). The best-fit scaling relation lines are displayed with the same colors. The equality 1:1 line is shown in red (dash dot), except for the right panel, where it coincides with the green best-fit line. The blue and green shaded areas represent the statistical error plus the intrinsic scatter for the core-excised and core scaling relations respectively.}
        \label{eRASS1-scaling-rel}
\end{figure*}

The best-fit scaling relation parameters for the $T$ comparison between the eRASS1 cluster catalog and the other spectroscopic $T$ measurements are presented in Table \ref{erass1-table}.

\subsubsection{eRASS1 versus \textit{Chandra} temperatures}

The eRASS1 versus \textit{Chandra} $T$ comparison is displayed in the left panel of Fig. \ref{eRASS1-scaling-rel}. Both the core and core-excised $T_{\text{Chandra}}$ show similar deviations from $T_{\text{eRASS1}}$. The main difference is found in the negligible intrinsic scatter of the relation when the core $T_{\text{Chandra}}$ is used. $T_{\text{eRASS1}}$, which come from the full $R_{500}$ area of the clusters, are expected to be mostly influenced by the cluster core since this is where most of the emission originates. The eRASS1-\textit{Chandra} scaling relations agree slightly better than the results presented in Sect. \ref{Chandra-single-pow-results}, but within the statistical uncertainties. The eRASS1-\textit{Chandra} relations still deviate from the equality line by $>10\sigma$. $T_{\text{eRASS1}}$ are $22\%$ ($21\%$) lower than the core-excised (core) $T_{\text{Chandra}}$ for $T_{\text{Chandra}}=4.5$ keV. For hotter plasma with $T_{\text{Chandra}}=10$ keV, the deviation rises to $32\%$ ($ 28\%$). Cooler gas values of $T_{\text{Chandra}}=2$ keV disagree by $10\%$ ($21\%$) between the eRASS1 catalog and \textit{Chandra}. However, these differences come with non-negligible uncertainties and individual clusters can deviate significantly from these averages. Overall, the conclusions from the eRASS1-\textit{Chandra} $T$ comparisons do not significantly change compared to the results presented in Sect. \ref{Chandra-single-pow-results}.

\subsubsection{eRASS1 versus \textit{XMM-Newton} temperatures}

The eRASS1 versus \textit{XMM-Newton} $T$ comparison is displayed in the middle panel of Fig. \ref{eRASS1-scaling-rel}. Due to the limited sample and the significant scatter, no statistical significant differences can be found between the core and core-excised $T_{\text{XMM}}$ compared to $T_{\text{eRASS1}}$, although the core-excised $T_{\text{XMM}}$ offers a best-fit relation slightly closer to the equality line. The eRASS1-\textit{XMM-Newton} scaling relations are consistent with the results presented in Sect. \ref{xmm-single-pow-results} within the uncertainties. However, the best-fit functions deviate more than the equality line compared to the comparison with the spectroscopic $T_{\text{eROSITA}}$ in \ref{xmm-single-pow-results}, while the scatter here is also larger. For cool gas with $T_{\text{XMM}}=1.5$ keV, $T_{\text{eRASS1}}$ is found within $2\%$ (13\%) when the core-excised (core) scaling relation is considered. Respectively, for $T_{\text{XMM}}=3$ keV the difference is $18\%$ (23\%), while it rises to $34\%$ ($33\%$) for $T_{\text{XMM}}=7$ keV.

\subsubsection{eRASS1 versus spectroscopic eROSITA temperatures}

The comparison between the eRASS1 and the spectroscopic eROSITA $T$ measured in this work is displayed in the right panel of Fig. \ref{eRASS1-scaling-rel}. Both the core-excised and core scaling relations show excellent agreement with the equality line (the core relation practically coincides with the 1:1 line). The scatter of the relations is almost exclusively driven by the $T$ measurement uncertainties with no significant intrinsic scatter being present.

\end{document}